\documentclass{elsart}

\usepackage{graphicx}
\usepackage[latin1]{inputenc}
\usepackage{amssymb}
\usepackage[sort&compress]{natbib}
\journal{Progress in Particle and Nuclear Physics}

\newcommand{\bi}[1]{\mathbf{#1}}
\newcommand{\AM}{MeV/nucleon}
\newcommand{\Zmoy}{\langle Z \rangle}

\bibpunct{[}{]}{,}{n}{}{}
\begin{document}

\begin{frontmatter}
\title
{Nuclear Multifragmentation and Phase Transition for hot nuclei}
\author{B.~Borderie\corauthref{cor}},
\corauth[cor]{Corresponding author - borderie@ipno.in2p3.fr}
\author{M.~F.~Rivet\ead{rivet@ipno.in2p3.fr}},

\address{Institut de Physique Nucl\'eaire, CNRS/IN2P3, Université Paris-Sud 11,
 F-91406 Orsay Cedex, France}

\begin{abstract}
That review article is focused on the tremendous progresses realized 
during the last fifteen years in the understanding of multifragmentation 
and its relationship to the liquid-gas phase diagram of nuclei and 
nuclear matter. The explosion of the whole nucleus, early predicted by 
Niels Bohr~\cite{Boh36}, is a very complex and rich subject which
continues to fascinate nuclear physicists as well as theoreticians who
extend the thermodynamics of phase transitions to finite systems.
\end{abstract}


\end{frontmatter}

\tableofcontents



\normalsize

\section{Introduction}

  The nuclear multifragmentation phenomenon was predicted and studied 
  since the early 80's. It is however only with the advent of powerful 
  4$\pi$ detectors~\cite{Sou06} that real advances were made. Such arrays 
  allow the detection of a large amount of the many fragments and light
  particles produced in nuclear collisions at intermediate and high 
  energies. Indeed it now appears that further progresses are linked to 
  the knowledge of many observables and the possibility to study
  correlations inside the multifragment events. This paper is a snapshot 
  of what is known at the present time, making use of a large panel of the
  available data from heavy-ion collisions at intermediate energies, and
  from hadron-nucleus collisions. A comparison
  between fragmentation properties of quasi-projectiles formed in
  semi-peripheral collisions and of ``fused'' systems produced in the more
  violent ones will be particularly developed. The formation of the
  fragmenting systems (or sources) in the course of nuclear collisions
  makes it necessary to model the collisions and several transport codes 
  have been developed for two decades~\cite{Fuc06,Ono06}, which account for 
  many properties experimentally observed. On the other hand statistical
  descriptions based on phase space occupation also account well for the
  static properties (partitions) of multifragmenting systems. The 
  connection between both descriptions will be discussed. 
  The Equation of State describing nuclear matter, similar to the van der
  Waals equation for classical fluids, foresees the existence of a 
  liquid-gas type phase transition; multifragmentation was 
  long-assimilated to this transition ( or to a ``liquid-fog''
  transition~\cite{Sie83}).
  Nuclear physicists are however dealing with finite systems -~nuclei 
  feeling nuclear and Coulomb forces~- and not with infinite nuclear matter.
  Following the concepts of statistical physics, a new definition of phase
  transitions for such systems was recently proposed, showing that 
  specific phase transition signatures could be expected. Different
  and coherent signals of phase transition have indeed been evidenced in
  a few cases.
   
  After a general survey of the multifragmentation phenomenon in
 section~\ref{survey}, the necessity of sorting events and  the different 
 ways to proceed are presented in section~\ref{sorting}. The properties of 
 the emitted fragments  are detailed in section~\ref{frag}.
 Section~\ref{models} introduces statistical and 
 dynamical models whose results are commonly compared to experimental data.
 The reconstruction of the multifragmenting systems connected with
 calorimetric and thermometric measurements (section~\ref{CaloTh})
 is  followed by a
 study of the properties of these systems at the freeze-out stage in
 section~\ref{FO}. Finally the view of multifragmentation in terms of the
 phase transition of a finite system and the experimental signatures 
 evidencing the transition are developed in sections~\ref{FinitePT} 
 and~\ref{SignPT}.

\section{A general survey}\label{survey}
One can come to the multifragmentation concept and its relation to 
a phase transition from two different starting points. Firstly by using 
kinetic models which show that nucleus-nucleus collisions at intermediate 
energies produce matter at subnormal density which breaks and where 
a phase transition is predicted to occur; it is a pure theoretical 
starting point. The other one consists in studying both
experimentally and theoretically the evolution of decay mechanisms with
increasing excitation energy.

\subsection{Nuclear matter at subnormal density and phase transition}

\subsubsection{The nuclear liquid-gas phase transition}
\begin{figure}[htb]
\begin{center}
 \includegraphics[scale=0.75]{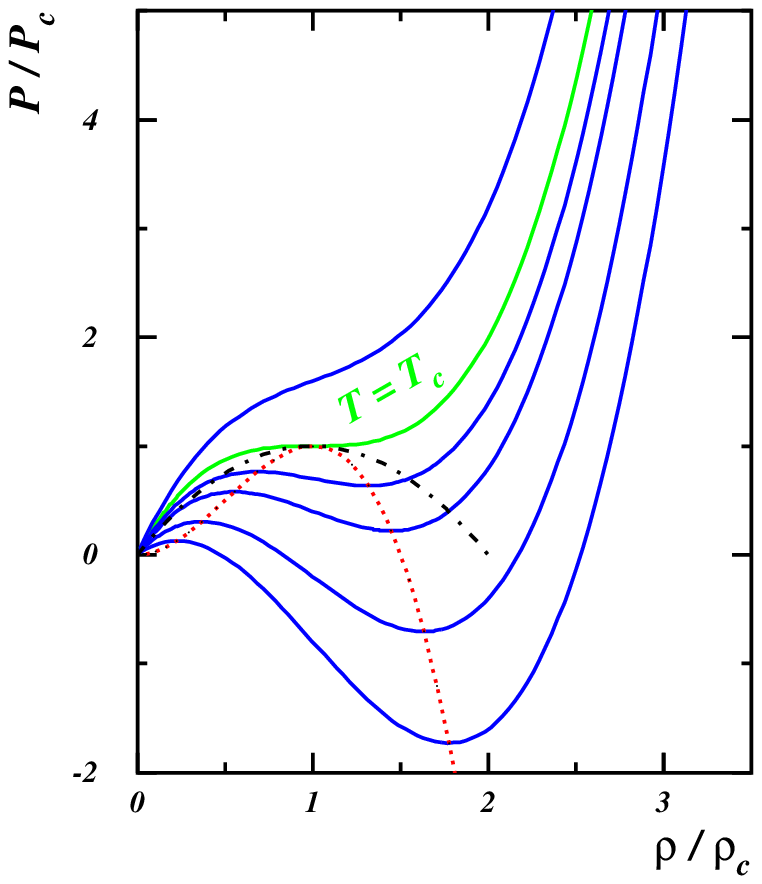}
 \includegraphics[scale=0.75]{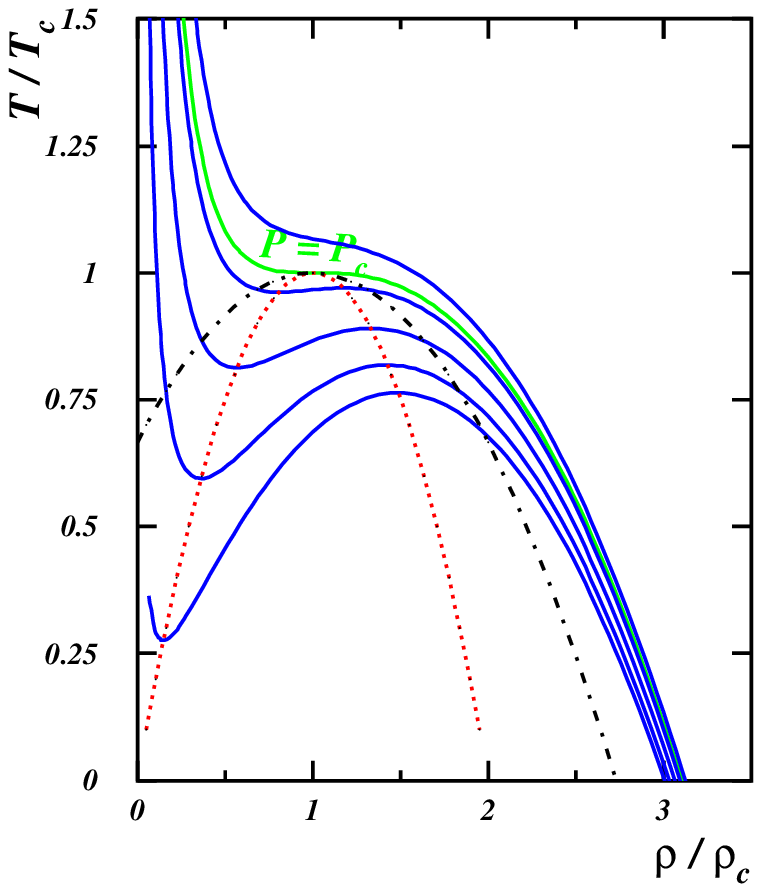}
\end{center}
\caption{Equation of state relating the pressure (left) or the temperature
(right) and the density (normalised to critical values) in nuclear matter.
The curves represent isotherms (left) and isobars (right). The dashed-dotted
lines are the coexistence lines and the dotted lines the spinodal lines
(from~\protect\cite{I46-Bor02}).}
\label{fig:eos}
\end{figure}

Nuclear matter is an idealized macroscopic system with an equal number
of neutrons and protons. It interacts via nuclear forces, and Coulomb
forces are ignored due to its size. Its density $\rho$ is  spatially uniform.
The nucleon-nucleon interaction is constituted by two components according
to their radial interdistance : a very short-range repulsive part
which takes into account the compressibility of the medium and a long-range
attractive part. Changed by five orders of magnitude the nuclear interaction is
similar to van der Waals forces acting in molecular medium. In a sense
the phase transition in nuclear matter resembles  the liquid-gas
phase transition in classical fluids. 
However, as compared to classical fluids the main difference comes from the
gas composition:
for nuclear matter the gas phase is predicted to be composed not only of
single nucleons, neutrons and protons, but also of complex particles and
fragments depending on temperature conditions~\cite{Bug00,Bug01}.\\
A set of isotherms for an
equation of state (pressure versus density) corresponding to
 nuclear forces (Skyrme effective interaction and finite temperature
Hartree-Fock theory~\cite{Jaq83}) is shown
in figure~\ref{fig:eos}.
It exhibits the maximum-minimum structure typical of van der Waals
equation of state.
Depending on the effective interaction chosen and on the
model~\cite{Jaq83,Jaq84,Cse86,Mul95}, the nuclear
equation of state
exhibits a critical point at $\rho_{c}\approx$-0.3-0.4$\rho_{0}$ and
$T_{c}\approx$16-18~MeV. $\rho_{0}$ and $V_{0}$ refer to normal density/volume.
The region below the dotted line in figure~\ref{fig:eos}
corresponds to a domain of negative compressibility: at constant temperature
an increase of density is associated to a decrease of pressure. Therefore
in this region a single homogeneous phase is unstable and the system breaks
into a liquid phase and a gas phase in equilibrium. It is the so called
spinodal region,
and spinodal instability corresponds to the breaking into the two phases.
Such instability has been proposed, for a long time, as a possible
mechanism responsible for multifragmentation~\cite{Ber83,Hei88,Lope89}. 
It will be discussed in section~\ref{FinitePT}.
The spinodal region constitutes the major part of the coexistence region
(dashed-dotted line in figure~\ref{fig:eos}) which also contains two
metastable regions: one at density below
$\rho_{c}$ for the nucleation of drops and one above $\rho_{c}$
for the nucleation of bubbles (cavitation). 

\subsubsection{From nuclear matter to hot nuclei}
Evidently the hot piece of nuclear matter produced in any nuclear
collision  has at more a few hundreds nucleons and so is not adequately
described by the properties of infinite nuclear matter; surface
and Coulomb effects can not be ignored. These effects have been
evaluated and lead to a sizeable reduction of
the critical temperature~\cite{Jaq83,Jaq84,De99}.
Finite size effects have been found to reduce the temperature by
2-6 MeV depending on the size of nuclei while the Coulomb force is responsible
for a further reduction of 1-3 MeV. However large reductions due to small
sizes are associated with small reductions from Coulomb. Consequently,
in the range A=50-400 a total reduction of about 7 MeV is calculated leading
to a ``critical'' temperature of about 10 MeV for nuclei or hot pieces of
nuclear matter produced in collisions between very heavy nuclei. 
The authors of reference~\cite{Jaq84}  indicate
that, due to some approximations, the derived values
 can be regarded as upper limits. Finally we can recall that,
in infinite nuclear matter, the binding energy per particle is 16 MeV 
whereas it
is about 8 MeV in a finite nucleus. Clearly these values well compare
with the  $T_{c}$ values for infinite nuclear matter and nuclei just
discussed. 

\subsection{Evolution of decay mechanisms: from evaporation 
to vaporization}\label{sec:vapo}

\subsubsection{Evaporation}
The behaviour of nuclei at excitation energies around 1~\AM{} has
been  extensively studied and rather well understood using statistical
models~\cite{Col00}. The theoretical treatment of particle emission 
involves the estimation of microstate densities defined for equilibrium 
states. Such a treatment is justified only when there is enough time between
successive emissions for the relaxation of the emitting nucleus to a new
equilibrium state. At such excitation
energies the density stays very close to normal density of cold nuclear
matter and the earliest evaporation model rests  on the basic idea:
an emitted particle can be considered as originally situated somewhere on 
the surface of the emitting nucleus with a randomly directed velocity, 
it is why we
use the term evaporation. Moreover particles are emitted sequentially and
independently without any correlation. Note that in this excitation energy
domain fission is also a deexcitation mechanism and can compete with
evaporation.

\subsubsection{The multifragmentation domain}

\begin{figure}[htb]
\begin{center}
 \includegraphics[width=9cm]{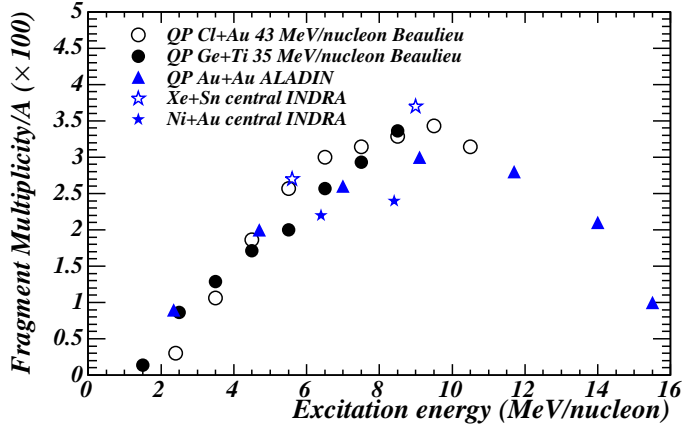}
\end{center}
\caption{Average fragment multiplicity (normalized to the number of incident 
nucleons) as a function of the excitation energy per nucleon
(data from~\protect\cite{Beau96,Sch96,T25NLN99,I37-Bel02}).} \label{fig:v1}
\end{figure}

\begin{figure}[htb]
\begin{center}
 \includegraphics[width=9cm]{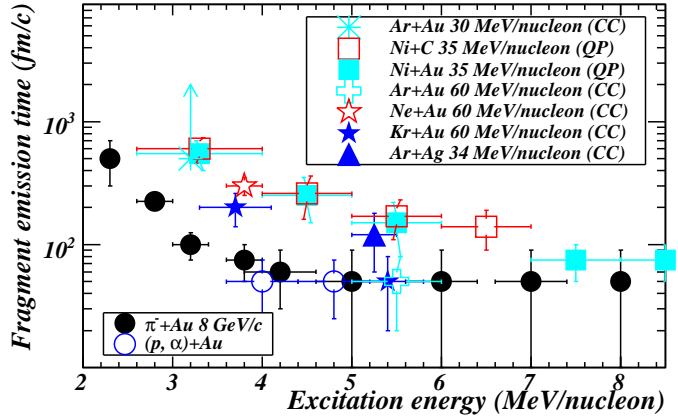}
\end{center}
\caption{Average fragment emission time as a function of the excitation 
energy per nucleon (data from~\protect\cite{Bou89,Lou94,Lou95,Gel95,He00,%
He01,Beau01,Kar06}. CC stands for central collisions, QP for
quasi-projectile sources).} \label{figdtf}
\end{figure}

At excitation energies comparable with the binding energy of nuclei
 a copious emission of particles and
fragments is experimentally observed and the name ``multifragmentation''
was introduced.
 A first attempt to connect multifragmentation to a possible phase 
 transition is to derive the excitation energy domain where this
 type of deexcitation is experimentally observed. Figure~\ref{fig:v1}
 displays, for different reactions, the evolution of the 
 average fragment multiplicity, normalized to the size of the
 multifragmenting system,  as a function of the
 excitation energy per nucleon deposited into the system. All 
 points fall on a single bell shape curve. The onset of
 multifragmentation takes place for excitation energies around 3~\AM{} 
 and the maximum for fragment production is found around 9~\AM{}, 
 i.e. close to the binding energy of nuclei. At higher excitation
 energy, due to the opening of the vaporization channel,
 the fragment production reduces.
On the other hand, average time intervals between successive emissions have
been estimated by analysing space-time correlations between fragments,
taking advantage of proximity effects induced by Coulomb repulsion.
Figure~\ref{figdtf}
displays those average time intervals measured as a function of excitation
energy deposited into the emitting system. A strong decrease of measured
times with the increase of excitation energy is observed up to around 5~\AM{}.
Then a saturation appears around 50-100 fm/$c$ which reflects
the limit of sensitivity of the method. For such short times fragments can
be considered as emitted quasi simultaneously and fragment emissions can not
be treated  independently. Note that, correlatively, sequential statistical
models fail in reproducing the observed emission characteristics~\cite{Cha90}.

\subsubsection{Vaporization and identification of a gas phase}
\begin{figure}[htb]
\begin{minipage}[t]{0.52\textwidth}
\includegraphics*[trim=0 0 52 0,width=\textwidth]{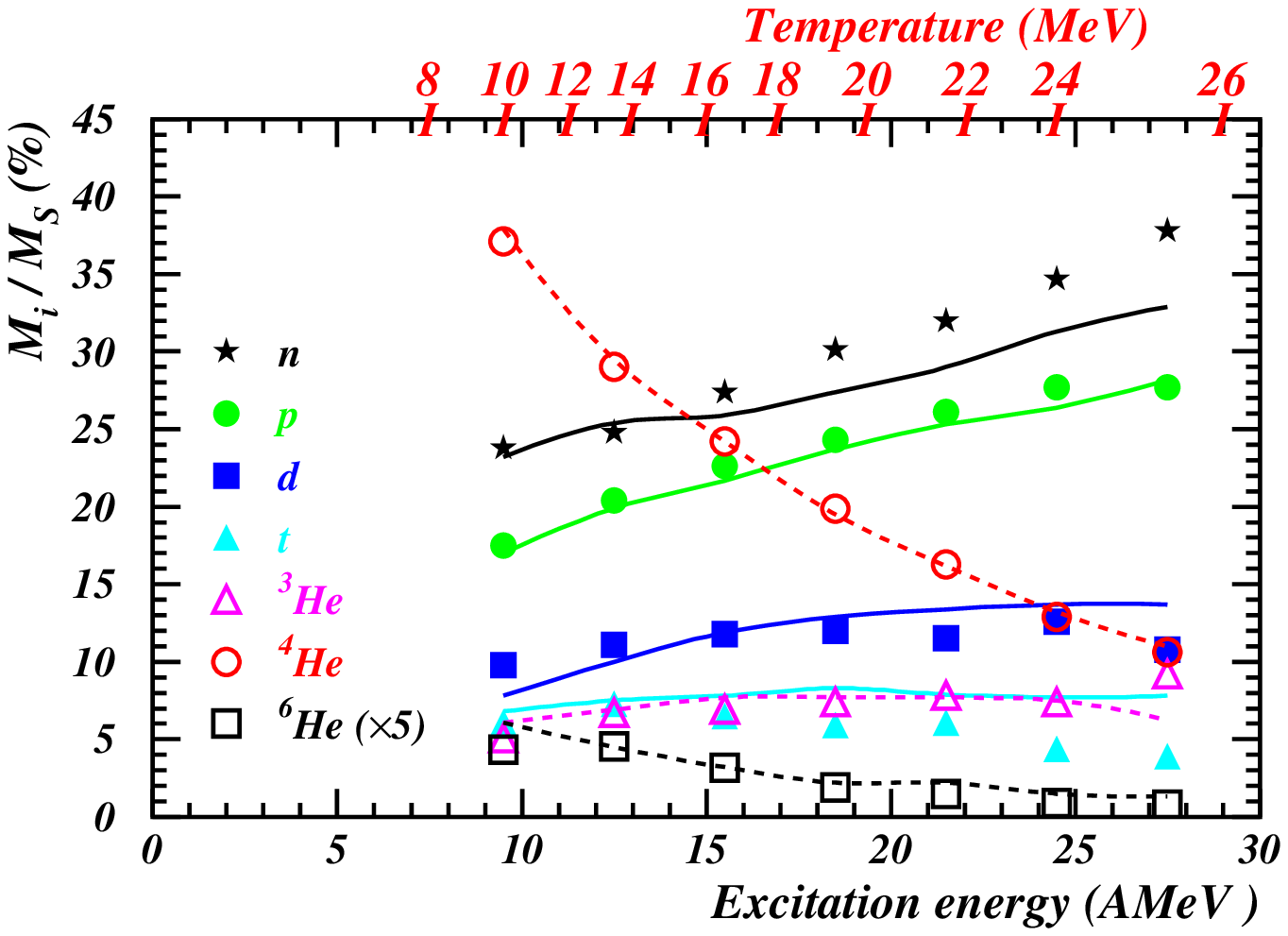}
\caption{Composition of vaporized quasi-projectiles, formed in 95~\AM{} 
$^{36}Ar$+$^{58}Ni$ collisions, as a function of their excitation
energy per nucleon. Symbols are for
data while the lines (dashed for He isotopes) are the results of the model.
The temperature values used in the model are also given. 
(from~\protect\cite{I15-Bor99}).} \label{fig:b2}
\end{minipage}%
\hspace*{0.03\textwidth}
\begin{minipage}[t]{0.45\textwidth}
\includegraphics*[trim= 15 17 52 50,width=\textwidth]{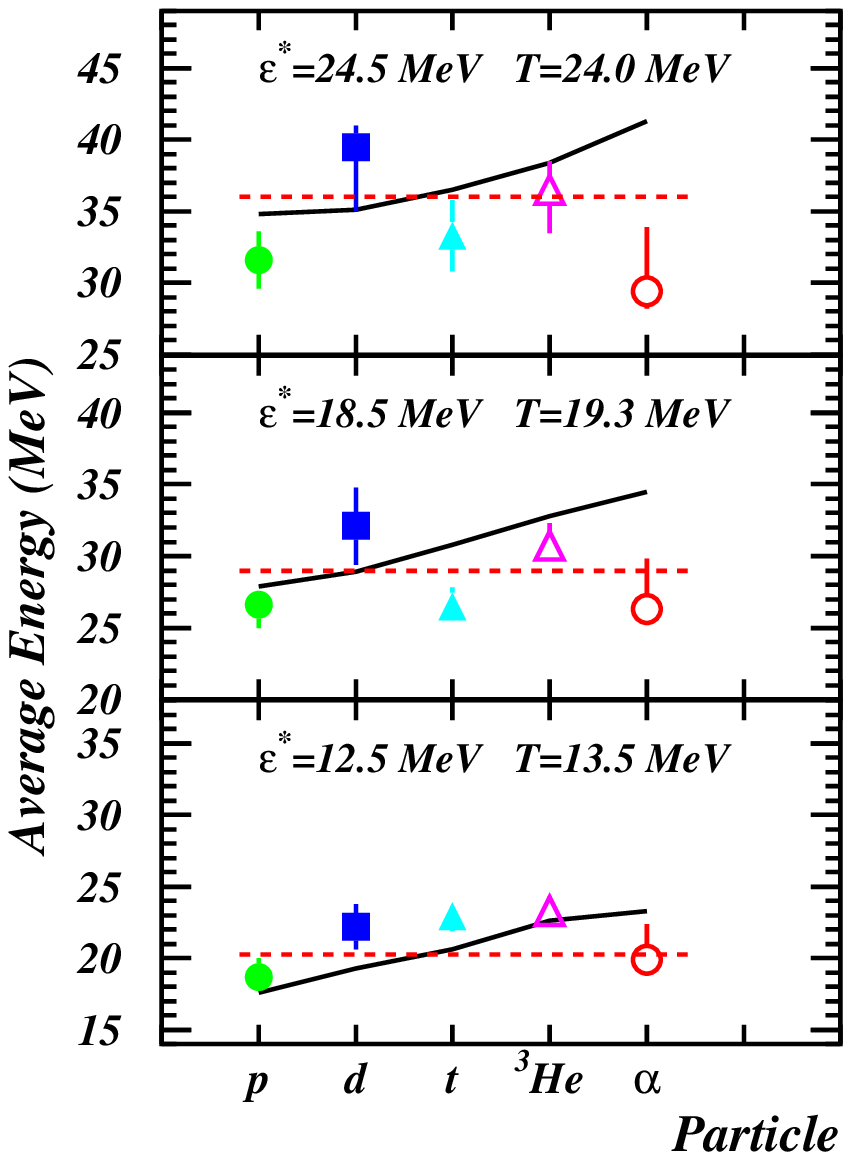}
\caption{Average kinetic energies of particles emitted by vaporized
quasi-projectiles at different excitation energies and formed in 95~\AM{}
$^{36}Ar$+$^{58}Ni$ collisions. Symbols are for data while full lines 
are the results of the model. The dashed lines refer to average kinetic 
energies of particles expected for an ideal gas (3T/2) at temperatures 
derived from the model. (from~\protect\cite{I15-Bor99}).} \label{fig:b3}
\end{minipage}%
\end{figure}

At excitation energies around 10~\AM{},
first experimental indications for the onset of vaporization (disintegration
into light particles with Z $\leq$ 2) were reported in 
references~\cite{Tsa93,I2-Bac95,I6-Riv96,I7-Bor96,Pie00}. Moreover  
a gas phase was identified by comparison with a model by studying the
deexcitation properties of vaporized quasi-projectiles
produced in $^{36}$Ar+$^{58}$Ni reactions~\cite{I15-Bor99}. 
Chemical composition
and average kinetic energies of the different
particles are well reproduced by a quantum statistical model 
(grandcanonical approach) describing a real gas of fermions and bosons
in thermal and chemical equilibrium. The
evolution with excitation energy of the composition of vaporized
quasi-projectiles is shown in figure~\ref{fig:b2}. Nucleon production
increases with excitation energy whereas emission of alpha particles, 
dominant at lower excitation energies, strongly decreases. 
The regular behaviour observed is a strong 
indication that a rapid change of
phase does not occur in the considered excitation energy range.
Note that an excluded volume correction due 
to finite particle size~\cite{Gul97} (van der Waals-like behaviour) was
found decisive to obtain the observed agreement.
In the model, the experimental range in
excitation energy per nucleon of the source (9.5 to 27.5 MeV) was covered
by varying the temperature from 10 to 25 MeV and the only free parameter,
the free volume, was fixed at 3$V_{0}$ in order to reproduce the
experimental ratio between the proton and alpha yields at 18.5~\AM{} 
excitation energy. The average kinetic energies of the different
charged particles are also rather well reproduced over the whole excitation
energy range (fig.~\ref{fig:b3}) but the model fails to accurately follow
the dependence on the different species especially for alphas. The dashed
lines in fig.~\ref{fig:b3} indicate the average kinetic energies, 
3T/2, expected for an ideal gas, which appear as a rather good 
approximation; this is due to the low density of the emitting source. 
We are in presence of a quantum weakly-interacting gas.

\section{Experimental event sorting}\label{sorting} 

\begin{figure}
\centering \includegraphics*[trim= 0 0 0 27,scale=0.7]{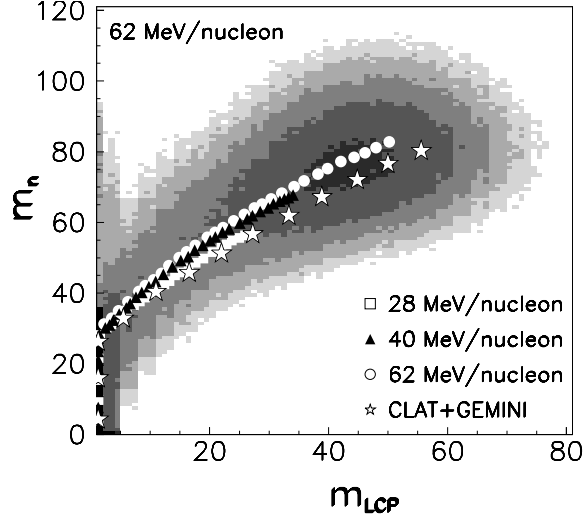}
\caption{Neutron vs light charged particle multiplicities for Xe+Bi 
reactions at various incident energies. Larger multiplicities correspond to
more violent collisions. From~\cite{Gal05}} \label{fig3.1}
\end{figure}

Multifragmentation is by essence associated to the emission of several 
fragments. Any study of the phenomenon requires a coincident and efficient
detection of these fragments and of the associated particles (Z$\leq$2). 
This is why,
in the recent years, multifragmentation studies were performed with $4\pi$
detectors. These do not however guarantee a full detection in
all reaction configurations. The combined geometrical efficiency 
(due to dead zones between detection
cells and the necessary free space close to 0 and 180$°$ to allow beam
propagation) and detection/identification thresholds of charged
particle (CP) arrays for instance,
strongly disfavour any correct detection of the most peripheral collisions 
for energies above $\sim$20~\AM{}; indeed the projectile residues, 
emitted very close to 0$°$, escape detection, while the 
low energy of the target residues makes them unidentifiable, except 
if their time of flight is measured. 
Moreover for heavy systems 
a very large number of neutrons is emitted before the release of any 
light charged particle as shown in fig.~\ref{fig3.1}. Coupling a 4$\pi$ CP 
array to a neutron ball partly remedies this last drawback. \\
For an efficient study of multifragmentation, the notion of complete event
was proposed: a high enough part of the reaction products should be
correctly detected and identified. For central collisions, a condition 
on the total detected charge (more than 70-90\% of the system charge)
is set. For quasi-projectile, one
requires the detection, in the forward centre of mass hemisphere, of a large
fraction of the projectile charge, or momentum. Owing to the non-detection
of neutrons, a pseudo momentum, $ZV$,  is sometimes used, replacing the mass 
of each product by its charge. In any case, the representative character 
of the selected events should be verified. \\
Studies on multifragmentation should apply to homogeneous 
samples of events, which requires an appropriate sorting; 
this is mandatory for thermodynamical studies for
instance. In hadron-nucleus collisions all events have similar 
topological properties independently of the impact parameter, as a 
single hot nucleus is created after a more or less abundant 
preequilibrium emission. Conversely, in heavy-ion collisions, the 
outgoing channel is different depending on the masses and asymmetry 
of the incident partners, the incident energy and the
impact parameter. At intermediate energies 
residual interactions (NN collisions) strongly compete with mean
field effects; the number of NN collisions largely fluctuates, leading to
different final reaction channels for the same initial conditions. The
weakening of the mean field hinders, on average, full stopping above 
about 30~\AM{};
the large fluctuations mentioned above allow however the observation of
 "fusion" (one outgoing heavy fragment) at higher energies, although with 
 small cross sections~\cite{I60-Lau06}.
Most of the collisions end-up in two remnants coming from the projectile 
and the target - accompanied by some evaporated particles -, and some 
fragments and particles with velocities intermediate between those of the
remnants: these are called mid-velocity products. They may have several
origins, e.g. direct preequilibrium emission from the overlap region 
between the incident partners, or a neck of matter between them which
may finally separate from QP or QT, or both.
Whatever the type of reaction is, a fraction of the incident translational
energy is transformed into ``excitation energy'', $E^*$, which may be 
shared into thermal energy (heat) and collective energies (rotational, 
expansion{\ldots}). While several experimental methods give access to  
$E^*$, the knowledge of the fraction allotted to thermal or collective 
energies relies on models.
The sorting of events, is generally done through 
global variables, which condensate the information measured on each 
event. 
Two philosophies guide the methods used for event sorting:  
the impact parameter dependence, and the event topology. 

\subsection{Impact parameter selectors}\label{IPS} 
Sorting events with respect to an ``experimental'' impact parameter, 
$b_{exp}$ is appropriate for studying global properties of collisions 
versus their violence. These methods are also useful for comparisons 
with transport 
codes, provided their outputs are filtered and sorted accordingly.
Some observables are strongly connected to the impact parameter and
are thus commonly used as sorting variables. One of the most popular
impact parameter selector (IPS) is the charged product  
multiplicity~\cite{Bow91}, sometimes reduced to the -~barely smaller~- 
light charged particle (lcp, Z=1,2) multiplicity. The fragment (Z$\geq$3)
multiplicity is in no way a good selector, because its small value, 
less than 10, induces too large fluctuations. IPS based on coupled 
neutron and lcp multiplicities are used with neutron 
balls~\cite{Tok01,Ma05}. Other IPS are the total charge
bound in charged products, excluding hydrogen isotopes, 
$Z_{bound}=\sum(Z \geq 2)$, introduced by the ALADIN 
group~\cite{Kre93}, or the value of the largest charge forward emitted
in the centre of mass~\cite{I30-Dor01}, $Z_{max}^{AV}$. This last variable 
requires arrays able to identify products in a very large 
range of atomic number and angles.
\begin{figure} \centering 
\begin{minipage}[t]{0.565\textwidth}
\includegraphics[width=8cm,height=6cm]{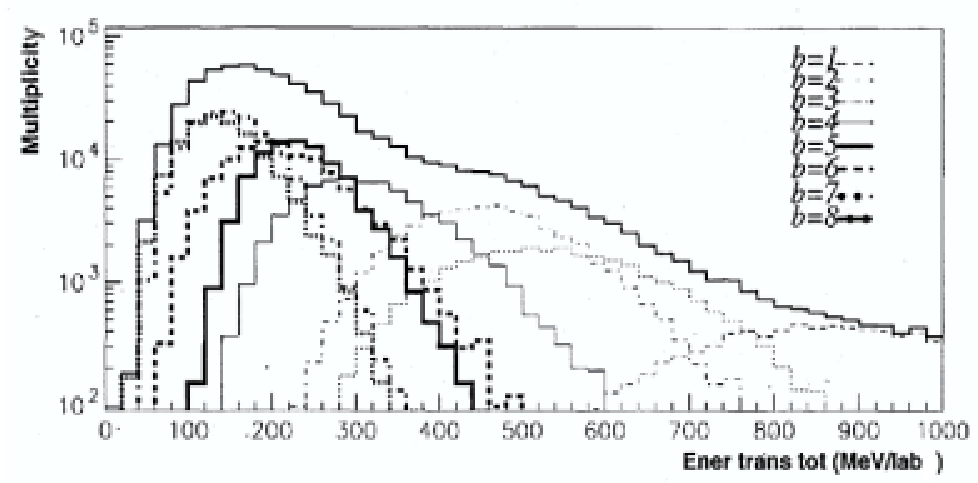}
\caption{Total transverse energy distributions for various impact parameters,
calculated in the DYWAN simulation, for the  95~\AM{} Ar+Ni system.
(private communication from F. Sébille).} \label{fig3.2}
\end{minipage}%
\hspace*{0.035\textwidth}
\begin{minipage}[t]{0.4\textwidth}
\includegraphics[scale=0.6]{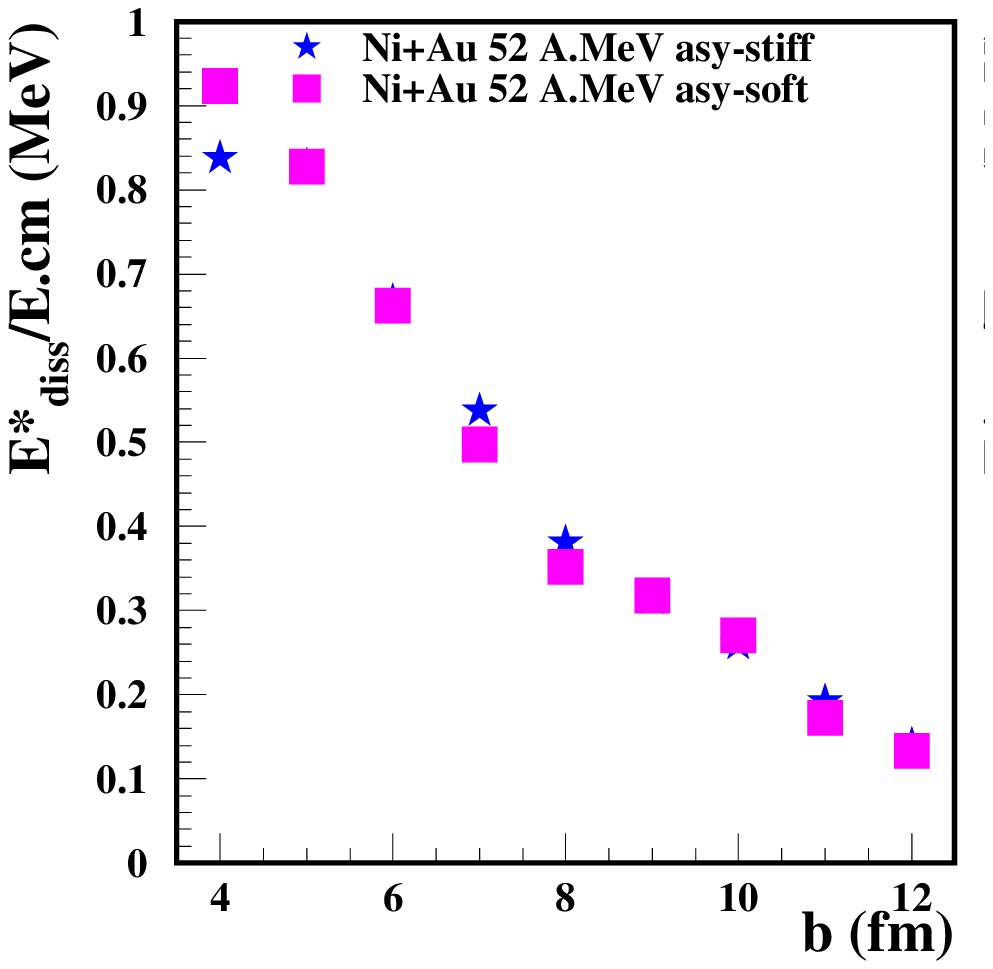}
\caption{Relation between the dissipated energy, $E_{diss}$ (see text),
and the impact parameter for the Ni+Au reactions at 52~\AM{} modelled 
with a stochastic mean field simulation. From~\cite{Gal07} }
\label{fig3.3}
\end{minipage}%
\end{figure}
Implicit in the previous IPS is the notion of dissipation, namely the
part of the initial translational energy transformed into excitation 
energy of the system. Several IPS reflect more directly this
connection, and first of all the transverse (perpendicular to the beam) 
component of the energy, either of all products~\cite{Mor97}, or of lcp 
only~\cite{I10-Luk97}. The latter choice is due to
the generally better efficiency of the arrays to lcp than to fragments.
At intermediate energies, 
a narrow zone of $b_{exp}$  generally covers a
broad range of true impact parameters, as shown in fig.~\ref{fig3.2}, 
obtained with the DYWAN code~\cite{Jou98}.
for the Ar+Ni system at 95~\AM{}. 
If $Z_{max}^{AV}$ is identified, its velocity, $V_{Z_{max}^{AV}}$, 
can be used to quantify dissipation~\cite{Yan03}. A related variable 
is the energy dissipation calculated as if the reaction was purely 
binary, without mass transfer\cite{Pia06}: 
$ E_{diss}=E_{cm}-1/2 \,\mu \,V_{Z_{max}^{AV}} \times (A_{P}+A_{T})/A_{T}$.
Fig.~\ref{fig3.3} shows for instance the good correlation between the impact
parameter and $ E_{diss}$, calculated in stochastic mean field
simulations for the Ni+Au reactions at 52~\AM{}, despite the presence of
mid-rapidity emission in the simulation; 
in this figure stars and squares
correspond to simulations with different isospin terms in
the EOS~\cite{Bar02}. \\
Finally correlations between the studied variables and the sorting 
variables must be avoided as much as possible. A detailed study on this 
subject was published in~\cite{Llo95}, where the average and the variance 
of the fragment multiplicity distributions obtained when selecting central
collisions with different IPS (upper 10\% of the IPS distribution) 
were examined. The authors found that $M_{IMF}$ was positively 
auto-correlated (small variance and 
normalized variance) with the total detected charge and negatively 
auto-correlated (small variance and large normalized variance) with variables
related to lcp or proton multiplicity. They conclude that the more neutral
selectors were the charged product multiplicity and transverse energy
(fig.~\ref{selIPS}).
\begin{figure}
\centering \includegraphics[scale=0.8]{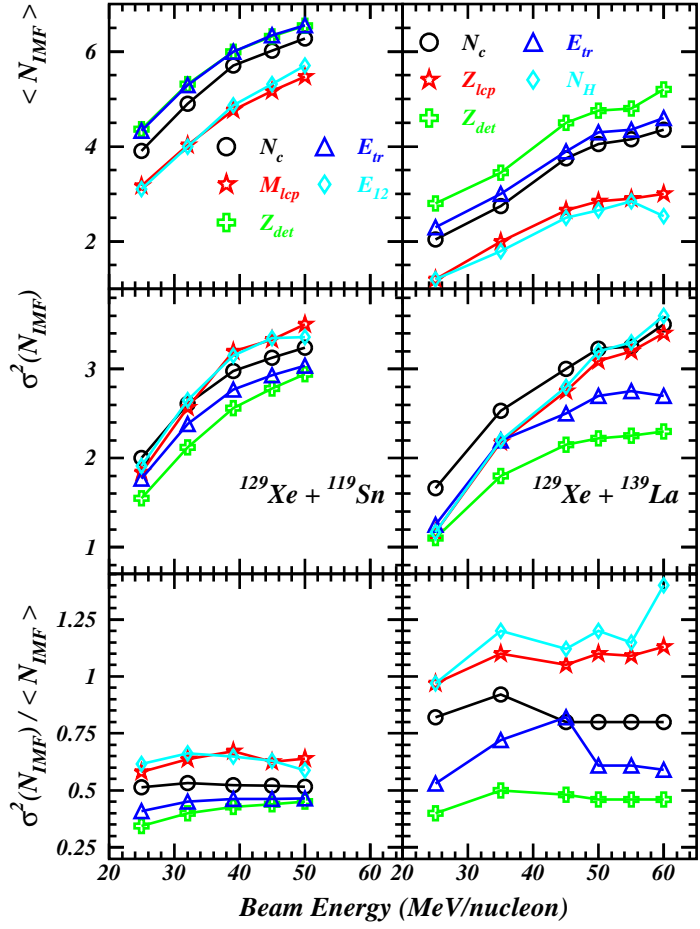}
\caption{Means, variances and normalized variances of the fragment
(Z$\geq$3, left) or IMF (3$\leq$ Z $\leq$ 20, right) multiplicities
versus the beam energy for central collisions between \nuc{129}{Xe} and
\nuc{nat}{Sn} (INDRA data) and \nuc{129}{Xe} and \nuc{139}{La}
(from~\cite{Llo95}), using different IPS: multiplicity of charged products,
$N_c$, of lcp, $M_{lcp}$, of hydrogen, $N_H$, total detected charge, 
$Z_{det}$, charge bound in lcp, $Z_{lcp}$, transverse energy total, $E_{tr}$,
or of lcp, $E_{12}$.} \label{selIPS}
\end{figure}
These conclusions on the best IPS are  detector dependent: the same 
study, on a similar system but using the INDRA array, shows that the 
negative correlation with lcp related variables weakly persists.
 The total detected charge (complete events) is, for 
this system, an IPS as neutral as
the charged product multiplicity and total transverse energy.

\subsection{Topology selectors}\label{topo} 
IPS classify events with respect to the violence of the collisions, 
related to the energy transformed into thermal energy, irrespective 
of the kind of reaction which happened. Indeed, at intermediate
energies (10-100~\AM{}), there are large fluctuations in the exit channel
associated to a given impact parameter, as shown in stochastic transport
models~\cite{Colon98}. For this reason, and particularly when it 
appeared desirable to work on a rather well defined ``source'', sorting 
variables related to the topology of the events, 
\emph{namely to their shape in velocity space}, 
were proposed. Most of the following variables are
borrowed from high energy physics. \\
The flow angle, $\theta_{flow}$, is the angle between the beam axis and 
the main axis of the kinetic energy-flow tensor~\cite{Buc84},
$Q_{ij} = \sum_{n=1}^{M_{tot}} p_n ^i p_n ^j / 2 m_n $,
with $m_n$ the mass and $p_n ^i$ the $i^{th}$ Cartesian component of 
the momentum of particle $n$. The eigenvectors of the tensor define an
ellipsoid, which corresponds to a rotated
reference frame with respect to the centre of mass.
The flow angle is connected with the impact 
parameter~\cite{Sch93}. By analogy with the evolution of
low energy deeply inelastic collisions, it is expected
that this angle increases with the violence of the collision while
the exit channel is dominantly binary; around the Fermi energy the
rotation angle of the system remains small so the bulk of
the cross section is associated to small flow angles. 
If ``fusion'' occurs no privileged direction is expected, so this type 
of events is better isolated for large flow angles. This selection
was largely used in the INDRA collaboration for the study of 
compact single source events. \\
Among shape global variables are the isotropy ratio, $R_{iso}$,
and the energy ratio~\cite{Kam93}, $E_{rat}$:
$$R_{iso}=\frac{2}{\pi} \frac{\sum_{n} |\bi{p}_n| \sin \theta_n}
{\sum_{n} | \bi{p}_n| \cos \theta_n}  \hspace{2cm}
E_{rat}= \left. \frac{\sum_{n} E_n \sin ^2 \theta_n}
{\sum_{n} E_n \cos ^2 \theta_n} \right| _{y \geq y_{cm}},$$
where $\theta_n$ is the emission angle of particle $n$ 
in the centre of mass (or in the ellipsoid) frame,
and $E_n$ its kinetic energy. 
Both $R_{iso}$ and $E_{rat}$ quantify the part of the momentum or 
energy transferred from the beam to the transverse direction; they are 
thus equal to zero for peripheral collisions and take large values
for the more central collisions, associated to more spherical shapes.
Large flow angles, and $R_{iso}$ or $E_{rat}$ values close to 1 
equally select events with compact shapes.
A very powerful variable, which would deserve a larger utilisation,
is the charge density,
$\rho_z (k) = \sum Z_i (k) / \sum Z_i$, where k is the centre of mass
velocity of particle $i$ (with charge $Z_i$)~\cite{I18-Lec00}. 
With this variable it is possible to isolate rare events which are 
either binary - without mid-rapidity emission - or monosource, from the 
majority of events which exhibit a mid-rapidity component between 
quasi-projectile and quasi-target remnants.
All these selections implicitly assume the isotropy - or at least the
forward-backward symmetry - of any source emission. 

\subsection{Multivariate analysis techniques}\label{multivar}
 Event selections with both IPS and topology 
selectors are made \emph{via} a sharp cut-off in the distribution of one 
or two variables. While it has the advantage to be simple and transparent,
the physical implications might be more difficult to evaluate due to the 
large fluctuations of most of the global variables. 
A generalisation to more variables is
given by the multivariate analysis techniques~\cite{Des95}. 
 Principal Component Analyses allow reduction of the dimensionality of the
information, and are based on a set of global variables. The
multidimensional space is rearranged on axes which are linear combinations
of the initial global variables, and carry a more or less important part of
the information. Projecting the events on the plane defined by the two axes
bearing maximum information allows to separate classes of events with a
rather close topology~\cite{Des96,I37-Bel02}. Discriminant Analyses, 
neuronal networks~\cite{Dav95,Bas96,Had97} aim at discriminating types 
of events. They need a learning phase before being applied to a physical
problem, which is often performed with the help of a 
model~\cite{I22-Des00,I50-Lau05}, but can as well  directly utilize
a sample of real
events~\cite{T39Mou04,H4Lau05}. Indeed if the model used does not 
correctly consider and weigh all the possible reaction channels, the 
final results may be biased. The advantage of these methods is that 
they generally provide samples of events with a higher statistics than 
the sharp-cut ones. The properties of the samples must however be 
studied in detail in order to be sure to have
selected an homogeneous ensemble of events.

\subsection{Cross sections}
The knowledge of absolute cross sections is of great interest, whether one
sorts events with an IPS or isolates one emission source. In the first case,
for a comparison of experimental results with those of a transport code, the
knowledge of the correct zone of impact parameter is mandatory. 
Indeed an experimental impact parameter, $b_{exp}$, is estimated with the 
help of a variable $\Phi$ assumed to vary monotonously with $b$, using a
geometric prescription~\cite{Cav90}:

$$b_{exp} (\Phi) = \frac{b_{max}}{\sqrt{N_{ev}}}
\sqrt{\int_{\Phi_1}^{\Phi_{max}} \frac{\mathrm d N}{\mathrm d \Phi} \mathrm
d \Phi}$$

In the above formula, $\Phi$ decreases for increasing values of $b$, 
from a value $\Phi_{max}$ for $b$=0 to 0 for $b=b_{max}$. Obviously 
$b_{exp}$ can be compared to the true impact parameter only if $b_{max}$
is associated to the reaction cross section, or if the total measured 
absolute cross section is known.
In the case of the study of a source properties, the interest is to find 
out whether the observed phenomenon is dominant or rare.

\subsection{Conclusion}
It was shown in this section that, when dealing with the results of 
powerful 4$\pi$ arrays, it is necessary to sort the data. Several methods 
are used by different groups, based on global variables which condensate
the information.  
The adequate sorting differs depending on the type of analysis which is 
foreseen, either an evolution with the violence of the collision and a 
comparison with dynamical simulations, or a thermodynamical study of the
properties of a source. As an example, for central collisions, it is known
that topology selectors and IPS do not select the same events, due to 
fluctuations in the reaction mechanism associated to a given impact 
parameter; 
while the former isolate "compact shape events", IPS
favour events with elongated shapes in velocity space.      
It must also be stressed  that \emph{any} sorting is 
detector dependent, even if the ``same'' global variable is adopted; 
this must be kept in mind when comparing data obtained with different 
experimental set-up.

\section{Fragment properties}\label{frag}

In this section several fragment properties will be discussed, 
either static like multiplicities or charge distributions, or dynamical. 
The ``references'' are chosen  as raw as possible, namely the total 
system or the projectile masses, and the available centre of mass 
energy, to facilitate comparisons with dynamical simulations. 
The main trends displayed in this section are however representative of 
those found for the properties of well isolated emitting sources
versus the excitation energy, as it will appear in the next sections.

\subsection{Multiplicities} \label{MulF}

\begin{figure}
\begin{minipage}[t]{0.49\textwidth}
\includegraphics[width=\textwidth]{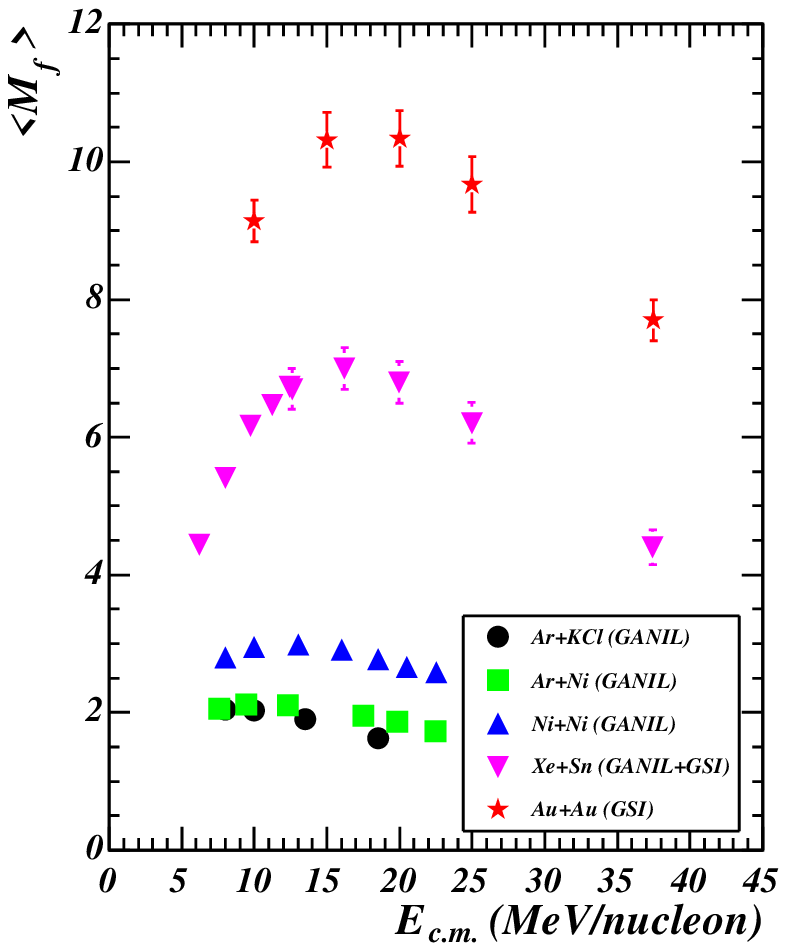}
\caption{Average fragment (Z$\geq$3) multiplicities as a function of the 
centre of mass energy for central collisions and systems of various 
sizes, measured with INDRA. Statistical error bars are smaller than the 
size of the symbols. Systematic errors are reported for 
the GSI data.}\label{fig4.1}
\end{minipage}%
\hspace*{0.02\textwidth}
\begin{minipage}[t]{0.49\textwidth}
\includegraphics[width=\textwidth]{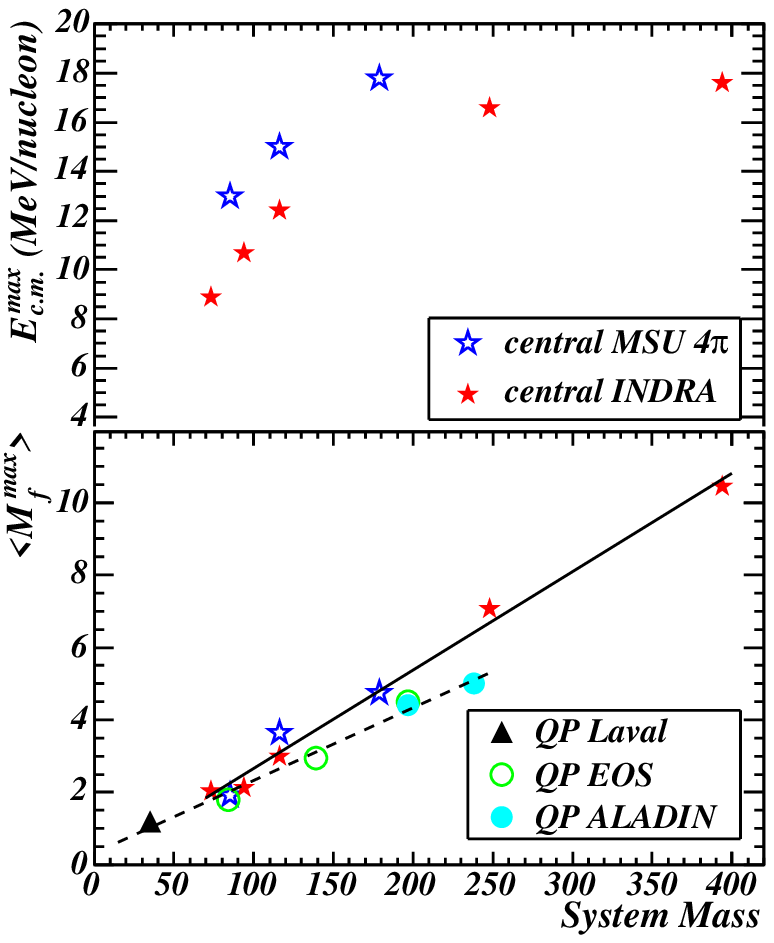}
\caption{Maximum average fragment  multiplicities (bottom panel) and the
centre of mass energy of its occurrence (top panel) for systems of various 
sizes. Central collisions (stars) measured with INDRA and the MSU 4$\pi$ 
array; quasi-projectiles (circles) measured by EOS, ALADIN and
Laval~\cite{Beau96}. The lines are
linear fits of central (full) and QP (dashed) data.}\label{fig4.2}
\end{minipage}%
\end{figure}

The fragment multiplicities in multifragmentation were abundantly studied 
in the last years. The most prominent observation is a rise and fall 
of the average fragment multiplicity, observed since the early 90's 
for Au quasi-projectiles~\cite{Ogi91} and for Au+Au central 
collisions~\cite{Tsa93}:
below 100~\AM{} the fragment multiplicity increases when going from
peripheral to central collisions, while above that beam energy 
the maximum fragment multiplicity is observed in semi-central collisions.
The absolute maximum value is however reached in central collisions at 
rather moderate energy.
A first study of the evolution of the IMF (3$\leq$Z$\leq$20) multiplicity 
versus the bombarding energy and the size of the system was performed for 
central collisions in~\cite{Sis01}. In the same line, we show in 
fig.~\ref{fig4.1} similar results obtained with the INDRA array 
operating at GANIL and GSI. 
Central collisions were selected, as in~\cite{Sis01}, by a cut of the top 
10\% of the total transverse energy distribution; we chose here to 
present fragment multiplicities, $M_f$, fragments being defined as
 all fully identified products with a charge 
Z$\geq$3. The identification
thresholds are respectively 0.8, 1.1, 1.3, 1.7~\AM{} for Z=3, 10, 20, 50. 
Lowering all thresholds below 0.8~\AM{} for all species would increase the
average multiplicities by less than 3\% for symmetric systems and up to a
maximum of 10\% for Ar+Ni at 32~\AM{}. Larger multiplicities are found when
the mass of the system increases. Systematic errors, essentially due to
efficiency, can be estimated by requiring additionally a high completeness 
of the events; they lie between 0.5 and 1.5 units, increasing with the 
incident energy, and are larger for the GSI data due to the dysfunction
of some INDRA modules. 
With increasing bombarding energy, the
rise and fall of the average fragment
multiplicity is visible in all cases (but perhaps for the lighter system 
Ar+KCl, where there is no rising part in the measured energy range); 
$M_f$ rises rapidly at low energies up to its maximum value,
$M_{f}^{max}$, but decreases smoothly on the high energy side;
indeed sizeable fragment multiplicities persist
in central collisions at much higher energies~\cite{Ins00,Rei04}. 
The c.m. energy at which the fragment multiplicity is the largest also 
increases with the size of the system. In order to quantify these effects, 
the data of fig.~\ref{fig4.1} were fitted with a third degree polynomial; 
for comparison, the data of~\cite{Sis01}
were fitted in the same way. Figure~\ref{fig4.2} shows, as a function 
of the total mass of the systems, the maximum fragment multiplicity, 
$M_{f}^{max}$, and the energy at which this maximum occurs. 
The maximum multiplicities measured with the MSU 4$\pi$ array and with INDRA
exhibit a remarkably coherent behaviour, despite the differences in the
detection devices: $M_{f}^{max}$ has a value around 2 for masses 90-100, 
and grows further on proportionally to the total mass of the system 
(2.7 units of multiplicity for 100 incident nucleons). 
The consistency of the results indicates that $M_{f}^{max}$ is little 
sensitive to the highest charge limit included in $M_{f}$ - indeed this 
limit only affects the tail of the largest fragment distribution at 
low energies; conversely it strongly depends on the low charge threshold 
chosen, as shown in~\cite{I40-Tab03}.
No saturation of $M_{f}^{max}$ at the highest masses, as quoted 
in~\cite{Sis01}, due to the Kr+Au data of~\cite{Pea94}, is visible. 
The energy where $M_{f}^{max}$ is observed first rapidly increases with the 
system mass, then tends to level off beyond mass 150 for the INDRA data. 
Note that $E_{c.m.}^{max}$  is systematically higher for the MSU data, 
although its evolution with A$_{sys}$ is parallel to that of the INDRA data. 
$E_{c.m.}^{max}$ is also sensitive to the low charge limit included in
 $M_{f}$, decreasing when this limit is raised. 
The increase of the degree of fragmentation with the available energy
is expected if thermal or radial flow energy put in the system increases. 
When it becomes very high, more hydrogen and helium isotopes are formed at
break-up, and during the deexcitation stage, leading to a fall of the 
fragment multiplicity. Fragments being smaller for lighter systems 
can be a first reason of the early
drop of fragment multiplicity for such systems. \\
\noindent \begin{figure}
\begin{minipage}{0.49\textwidth}
\includegraphics[width=\textwidth]{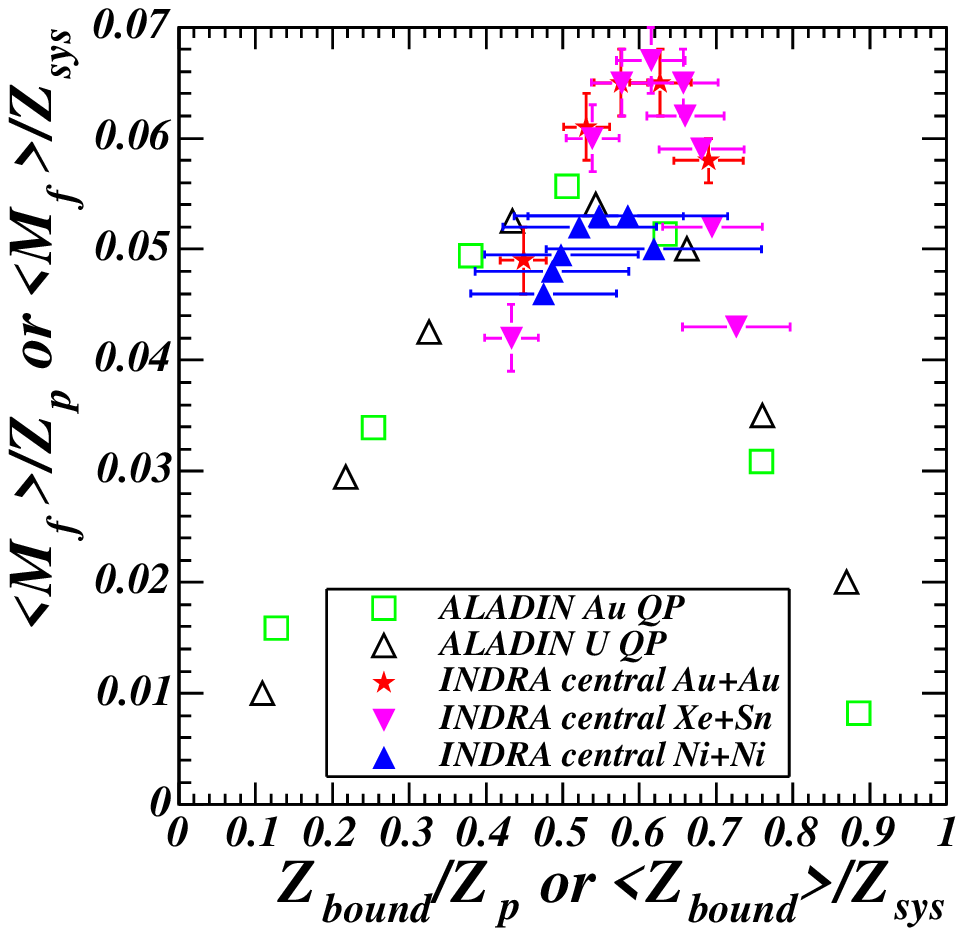}
\caption{Normalized average fragment  multiplicities  versus 
normalized $Z_{bound}$ (Z$\geq$2) for central collisions measured 
with INDRA and for Au and U quasi-projectiles detected with ALADIN 
(adapted from figs.~6 and 10 of ref.~\cite{Sch96}) The horizontal 
error bars for  the INDRA
data represent the RMS of the $Z_{bound}$ distributions.}\label{fig4.3}
\end{minipage}%
\hspace*{0.02\textwidth}
\begin{minipage}{0.49\textwidth}
\includegraphics*[trim= 0 0 38 24,width=\textwidth]{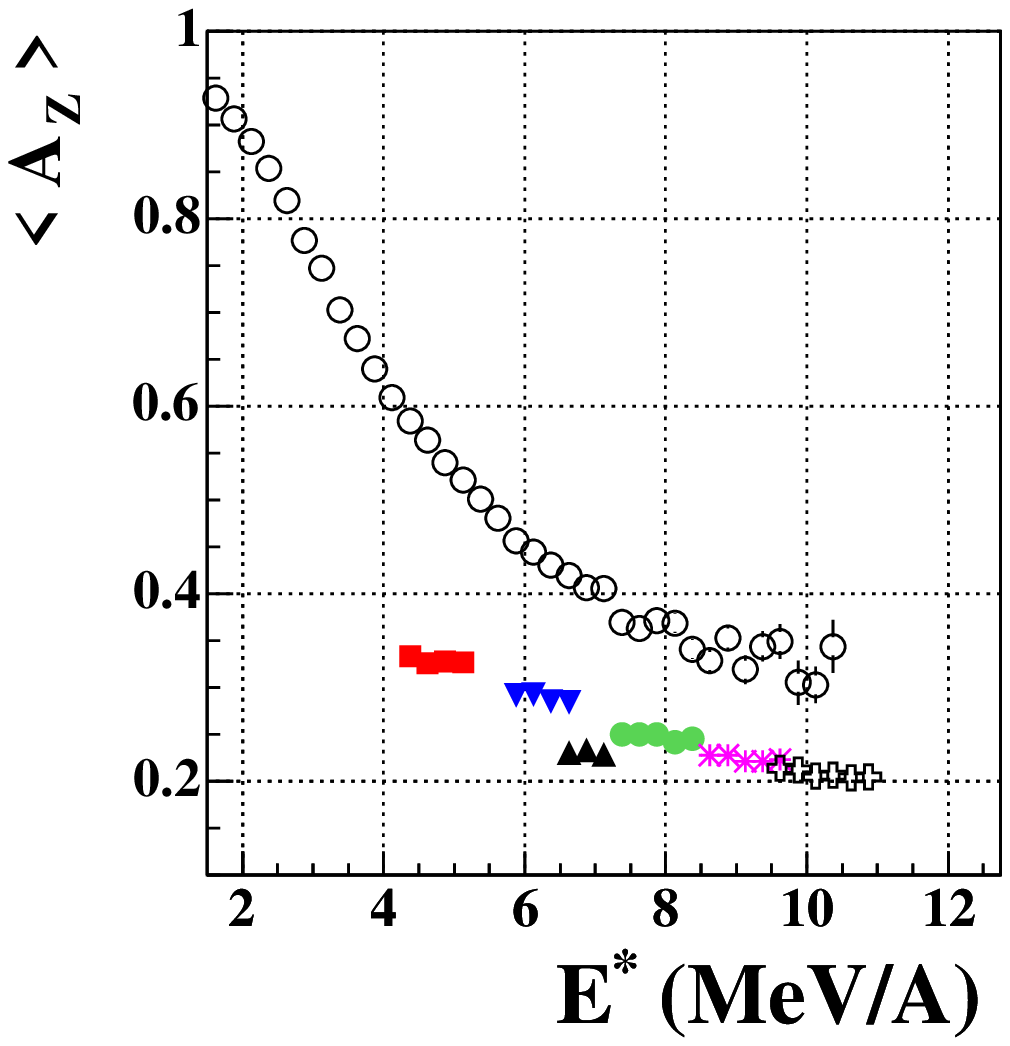}
\caption{Charge asymmetry of fragments as a function of the excitation energy
of the emitting source, for Au quasi-projectiles (open circles), monosources
from central Gd+U collisions (triangles) and Xe+Sn from 25 to 50~\AM{}
(other symbols). From~\cite{T41Bon06}}\label{fig4.3b}
\end{minipage}%
\end{figure}
Finally it is also instructive to compare the fragmentation of 
quasi-projectiles, as those studied by the Laval group, the ALADIN and 
EOS collaborations, with that of medium mass to heavy systems formed 
in central collisions, to get a first hint on the similarities
or differences between the fragmentation mechanisms. 
The maximum QP 
fragment multiplicity (circles in fig.~\ref{fig4.2}) rises linearly with 
the projectile mass; the values remain however smaller than those reached 
in central collisions.
The values of $M_f^{max}$ for quasi-projectiles are independent of both 
the target mass and the incident energy~\cite{Sch96}; 
this is an indication that multifragmentation is mainly driven
by the thermal energy deposited in the system, as noted in~\cite{Tam06}.
The difference between the maximum multiplicities for 
QP's and central collisions might then come from the larger
expansion energy found in the latter (see next subsection).
Figure~9 of ref.~\cite{Sch96} showed a scaling of the average fragment 
multiplicity versus $Z_{bound}$ when both
quantities were normalized to the projectile charge (Xe, Au or U). 
This scaled result is summarized in fig.~\ref{fig4.3}; $Z_{bound}$ was in
that case taken as an IPS. For central collision samples measured with 
INDRA and used in this section, the average multiplicity 
and $Z_{bound}$ (and its RMS) were normalized to the total charge of 
the system: a scaling is also observed for the Xe+Sn and Au+Au systems,
but fails for Ni+Ni (Z$_{tot}$=56), whereas it persists for Xe 
quasi-projectile (Z$_p$=54)~\cite{Sch96}.
The fragment multiplicity-$Z_{bound}$ scaling is thus a property of 
multifragmentation and not just a geometrical property when $Z_{bound}$
is taken as an IPS.
As already observed, for heavy systems, the scaled maximum multiplicity 
is higher in central collisions and occurs at higher scaled 
$Z_{bound}$, meaning that the multifragmentation partitions 
are different in both types of collisions. This is confirmed in
fig.~\ref{fig4.3b}, where the average charge asymmetry of fragments,
$A_Z=\sigma _Z/ (\langle Z \rangle \times \sqrt{M_f - 1})$ is plotted 
versus the excitation energy of QP sources or monosources~\cite{T41Bon06}
(see sect.~\ref{CaloTh} for the calculation of $E^*$): the asymmetry
is smaller for monosources formed in central collisions, which
means that the system is more fragmented.

\subsection{$Z_{bound}^f$ and charge distributions in central collisions}

\begin{figure}
\begin{minipage}[t]{0.47\textwidth}
\includegraphics[width=\textwidth]{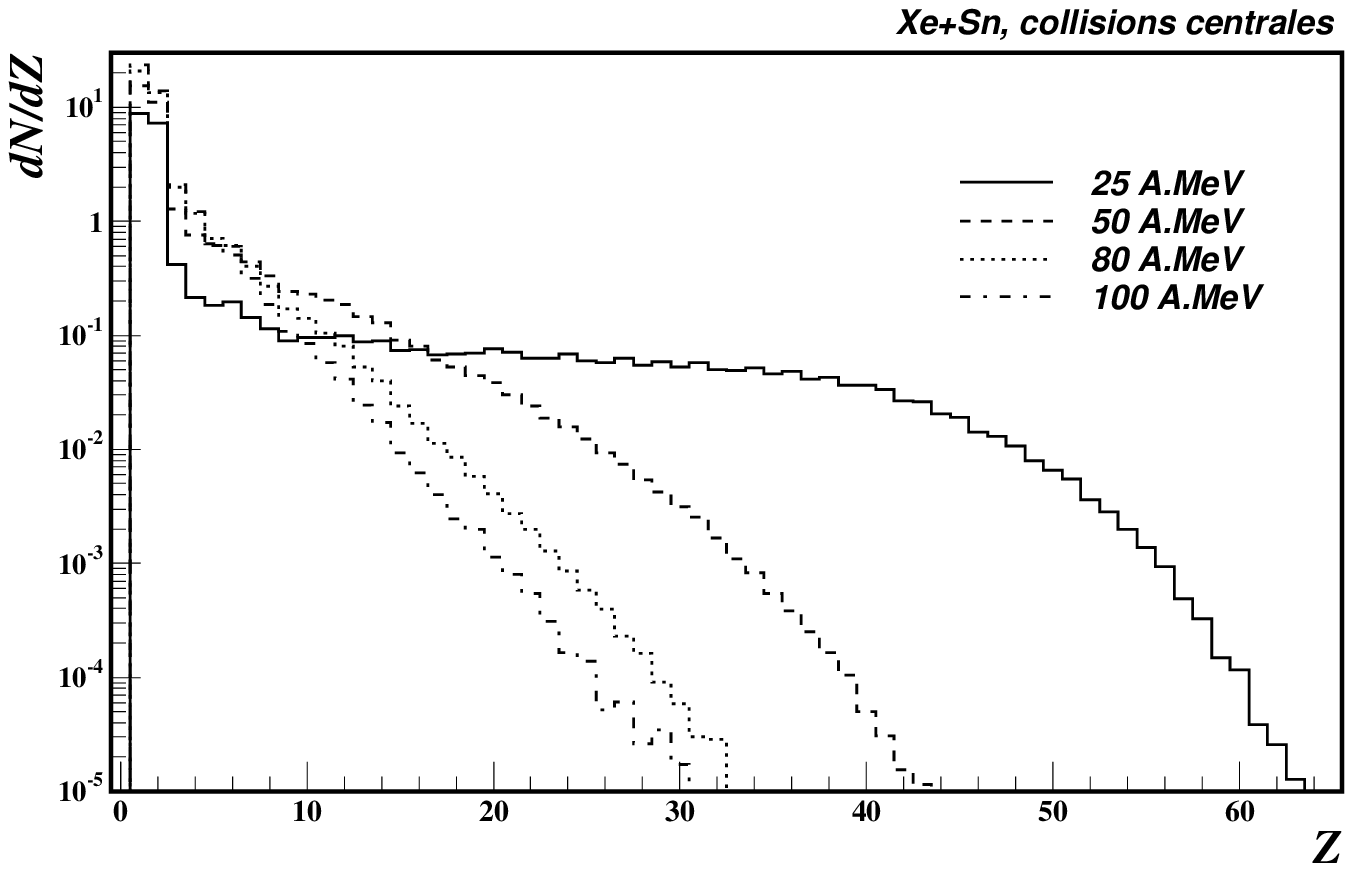}
\caption{Charge distributions measured for central Xe on Sn collisions 
between 25 and 100~\AM{} incident energies. 
From~\cite{T32Hud01}}\label{fig4.4}
\end{minipage}%
\hspace*{0.05\textwidth}
\begin{minipage}[t]{0.47\textwidth}
\includegraphics[width=\textwidth]{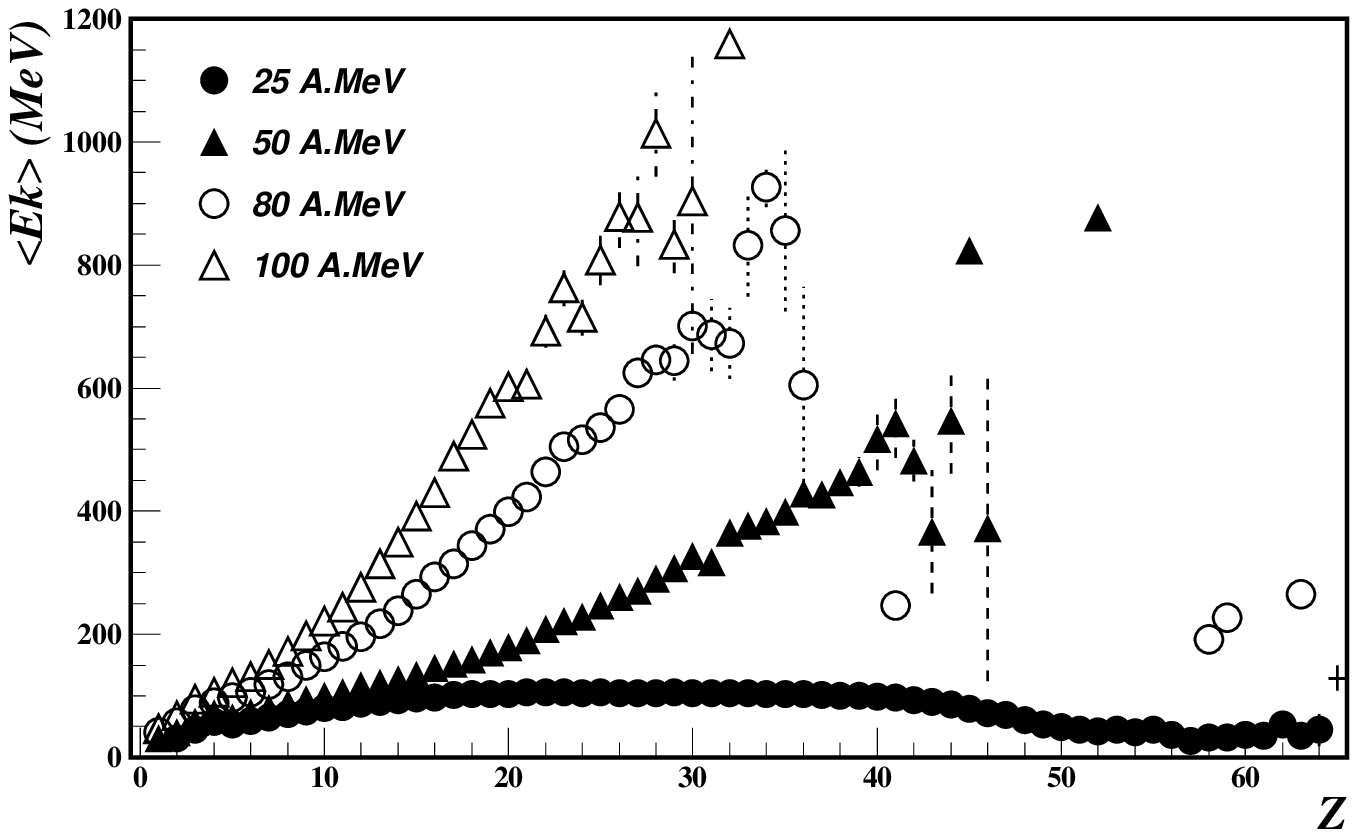}
\caption{Average centre of mass kinetic energy of fragments as a function of
their charge for central Xe on Sn collisions (selected with the lcp
transverse energy) between 25 and 100~\AM{} incident energies. 
From~\cite{T32Hud01}}\label{fig4.5}
\end{minipage}
\end{figure}

While the fragment multiplicity exhibits a rise and fall with the incident
energy, the total charged product multiplicity keeps increasing, indicating
that the formed fragments become smaller. Indeed the total charge bound 
in fragments, $Z_{bound}^f=\sum(Z \geq 3)$, monotonously diminishes with 
increasing energy~\cite{I40-Tab03}; for heavy systems (Xe+Sn, Au+Au) the 
$Z_{bound}^f$ distributions shift to smaller values and become narrower 
when raising the incident energy from 30 to 150~\AM{}. For the Ni+Ni 
system it appears that the high Z tail of these distributions barely  
evolves above 80~\AM{}. The distributions are no longer symmetric, but 
raise for low $Z_{bound}^f$, indicating that the system goes toward 
vaporization. The charge distributions show  the same
evolution, extending over a narrower charge range at higher energies, as
shown in fig.~\ref{fig4.4} for the Xe+Sn system. The shape of the 
distribution comprises a broad plateau at moderate energies, and evolves 
towards  an exponential decrease above 100~\AM{} (see also~\cite{I40-Tab03}). 
The yield of fragments  with charges Z=3-10 is almost
independent of the incident energy, between 39 and 150~\AM{}, for this
system. An interesting property of the charge distributions 
is their independence, when scaled by the fragment multiplicity, 
with respect to the total charge of the (heavy) system 
(104 and 156 in~\cite{I12-Riv98}), 
provided that the available energy is similar.

\subsection{Energy distributions - expansion} \label{EneF}

Energy spectra in the c.m. frame of fragments emitted by compact 
multifragmenting sources were shown in~\cite{I57-Tab05} for the two 
heavy systems 32~\AM{} Xe+Sn and 36~\AM{} Gd+U. 
They have asymmetric shapes for the lighter
fragments, and tend to become more symmetric for Z$\geq$15. These spectra
bear information about the Coulomb repulsion,
the radial expansion and the temperature of the source from which 
they originate. Indeed in~\cite{Viol04} the Coulomb barrier deduced from the
maximum of some fragment energy spectra was used to infer the density of the
emitting source; a warning was made on this method by~\cite{Rad05}, where
it was shown that the same shift of the maximum of the spectrum may
also be due to the total Coulomb energy which increases faster than the 
fragment multiplicity. In any case fragment energy spectra are precious
probes to test dynamical multifragmentation models~\cite{Wad04,I57-Tab05}.
For compact source selection, fragment emission is roughly isotropic in the 
centre of mass. This is not the case for selections based on transverse 
energy, ACP~\cite{T29Lav01} or neural networks~\cite{T30Bou01}. 
In this case either the sources are shown to be elongated 
along the beam axis in coordinate space, and the fragment energy 
is larger in this direction, or the selection includes collisions with a
binary character in the exit channel (see section~\ref{sorting}). \\
The average kinetic energy of fragments as a function of their charge
is thus dependent on the event selection and in some cases on the emission 
angle, except that of the lighter fragments which is more stable against 
experimental selections. For compact sources formed in central collisions
between 25 and 50~\AM{} Xe and Sn, the average c.m. kinetic energy is rising 
with the fragment charge, then saturates and even decreases for charges 
Z$\geq$20-25. The decrease is essentially due to the properties of the
largest fragments of the partitions. This evolution, also observed 
in~\cite{Wad04}, pleads for the kinetic energy essentially originating 
from Coulomb (and possibly radial expansion) effects rather than from 
thermal motion; in this latter case indeed, the average energy would be 
constant whatever the fragment charge or mass. 
Conversely with a transverse energy selection, the kinetic
energy continuously increases with Z at 50~\AM{}. 
More information was derived in~\cite{I57-Tab05} from the average kinetic 
of fragments as a function of their charge, for events sorted according 
to the fragment multiplicity and the rank of the fragment in the event. 
For each of the two above-mentioned systems, the experimental pattern 
was found independent of the fragment multiplicity, indicating that 
all emitting sources have similar charges. The largest fragment displays 
a specific behaviour which will be discussed later. 
\begin{figure}
\includegraphics[width=0.8\textwidth]{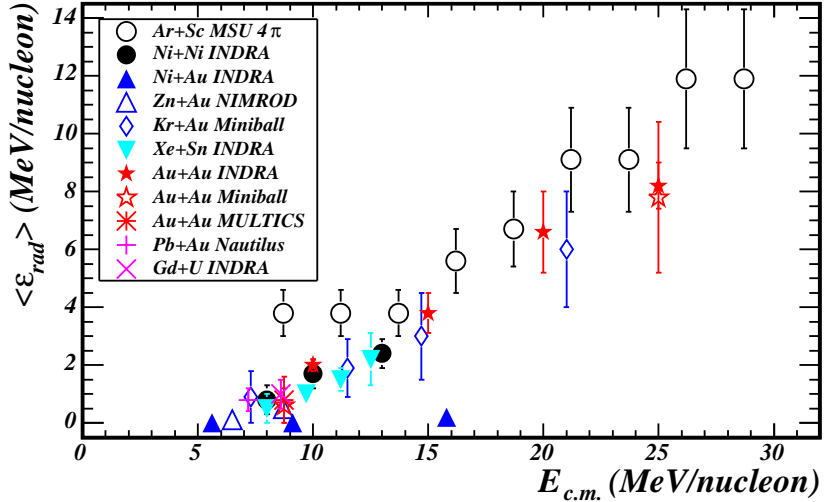}
\caption{Systematics of radial expansion energy per nucleon as a function
of the available energy for central collisions (see text). Data are taken 
from: Ar+Sc~\cite{Pak96}, Zn+Au~\cite{Wad04}, Kr+Au, Au+Au 
MSU~\cite{Wil97}, Au+Au MULTICS~\cite{MDA96}, Pb+Au Nautilus~\cite{Dur96} 
and INDRA (see text).}\label{fig4.6}
\end{figure}
The kinetic energy, for a given charge, increases with the collision 
energy, (see fig.\ref{fig4.5} and fig.39 in~\cite{Wad04}).
If the mass and the volume of the source (i.e. Coulomb effects) remain 
constant, then the extra energy has to come from an expansion which 
increases with the incident energy. \\
For a long time dynamical models have predicted, for central collisions 
between heavy nuclei, the occurrence of a compression phase followed by 
an expansion of the system~\cite{BBro96,ReiR97}; the total collective
energy (thermal plus compressional effects) can be derived at
different times during the collision and at different distances from the
centre of the expanding system. This last point gives some information 
about the evolution of radial velocity flow as a function of the 
radial distance
$\bi{r}$. As for the "Big Bang" a self similar expansion
(collective velocity proportional to $\bi{r}$) is observed up to
($\sim$ 80-100 fm/$c$ after the beginning of the collisions, when the 
system is still homogeneous (pre-fragments begin to appear at that time).
With such a prescription the radial expansion velocity is extracted 
from experiments using: 
$ \bi{v}_{rad} (\bi{r}) = \left( \bi{r}/R \right) v_0$,
where $R$ is the rms of pre-fragment distances to centre and $v_0$ 
the flow velocity. The radial expansion 
energy $E(R)$ is related to the average expansion energy through:
$ \langle E_{rad} \rangle = 3 E(R)/5 $;
$v_0$ and $E(R)$ are used as reference values for comparisons. Note that 
a Coulomb repulsion when fragments are emitted from a spherical volume of 
uniform density also gives a collective velocity field proportional to
$\bi{r}$. Consequently the two effects can not be distinguished and radial 
expansions are extracted from experiments with the help of models as SMM 
or event generators as SIMON, assuming given volumes at 
freeze-out~\cite{T18ADN98,I29-Fra01}. 
When dealing with deformed sources, a different flow profile was found,  
$ \bi{v}_{rad} (\bi{r}) = \left( \bi{r}/R \right)^{\alpha} v_0$,
with $\alpha$ varying between 1 and 2~\cite{I47-Lef04}.
Expansion energy is not a generic feature of statistical multifragmentation
models, the implicit assumption is that partitions result only from the
thermal part of the energy and are not influenced by the flow component.
This hypothesis was shown to be correct in the framework of the lattice gas
model~\cite{Das04,Gul04} even when the flow energy amounts to 50\% of the
available energy.  

A systematics of radial expansion energies derived for central collisions 
and various systems in the Fermi energy domain is reported in
figure~\ref{fig4.6}. In most cases, it was derived from comparison with a
statistical model. In~\cite{Wad04} it was obtained by subtracting the
kinetic energies obtained at low incident energy, as representative of the
Coulomb component, from those measured at higher energy; for compact
sources, the statistical model implies a spherical volume, while for other
selections the authors used SMM with a deformed envelope, or consider only
the energy at 90° in the c.m. For the INDRA data quoted 
in fig.~\ref{fig4.6}
several determinations were made for the same systems. Average values are
reported in the
figure~\cite{I29-Fra01,T16Sal97,T25NLN99,T29Lav01,I47-Lef04,T39Mou04}.
Despite the differences in the methods of determination of the radial 
energy, a general trend emerges, namely the onset of expansion energy 
around $\sim$ 5~\AM{} followed by an increase  with the c.m. available 
energy. Values remain small ($\leq$1~\AM{}) up to available energies 
around 8-9~\AM{}. No clear size effect can be inferred from the available 
data, because of large error bars, and the difficulty to disentangle
Coulomb and expansion energies.
There may be an indication that there is less expansion when the 
entrance channel is very asymmetric, supported by the Ni+Au and Zn+Au 
data, but not by those on Kr+Au. Note also that radial expansion 
energies deduced from the BOB dynamical simulations for the Xe+Sn
system at 32~\AM{} and for the Gd+U system at 36~\AM{}, both studied 
with INDRA, are in good agreement with this systematics~\cite{I29-Fra01}.  

It is much more hazardous to derive fragment kinetic energies, and
expansion energies, from quasi-projectiles, as in this case the 
reference frame is built from the fragment kinetic properties. 
The indications found in the literature are however that there is no 
expansion for QPs as long as their \emph{excitation} energy, 
$\varepsilon^*$, does not reach
$\sim$5~\AM{}; for light Ar QPs, the expansion energy is reported 
to increase from 0.2 to 0.7-1.1~\AM{} for $\varepsilon^*$ varying between
6 and 8~\AM{}~\cite{Jeo96}, while for heavy Au-like nuclei an expansion 
of 0.8~\AM{} was found at $\varepsilon^*$~=~6~\AM{}~\cite{MDA99}. 
A similar small variation versus the excitation energy of the source 
was observed in hadron induced reactions on Au nuclei, with an 
expansion increasing from 0.2 to 0.6~\AM{}
when $\varepsilon^*$ varies between 5.5 and 8~\AM{}~\cite{Bea00}. 
In these two cases, the extra energy can be attributed to thermal 
pressure only, while for central collisions between heavy ions 
a compression phase occurs.
The drawback on the reference frame for QP's can be overcome by using
relative velocities between fragments, in order to compare values from
the different kinds of multifragmenting sources~\cite{T41Bon06}.

\subsection{The specific role of the largest fragment}

Several studies point out the specific properties of the largest fragment 
in each partition, be they static or dynamic. In the MMMC statistical 
model it was shown that the size of the largest fragment strongly decreases
with increasing thermal excitation energy of the source but is
independent of the source total charge (when varied between 56 and 90);
it may thus be used to estimate this energy~\cite{T12LeFe97}. 
 The largest fragment size also decreases when the volume occupied 
by the source grows larger, as indicated in~\cite{Des98},
while those of the fragments of higher rank are roughly constant. 
It must be noted that such a variation of the ranked fragments was 
observed when plotting the fragment size as a function of the fragment
multiplicity~\cite{I57-Tab05}. \\
From an experimental point of view, the independence of the largest 
fragment charge with respect to the system size follows from that of the
charge distribution~\cite{I12-Riv98} and was indeed observed for the
32~\AM{} Xe+Sn and 36~\AM{} Gd+U compact single sources isolated among 
central collisions. Such a study can be extended to the central collisions 
studied with INDRA already used in the previous subsections (transverse 
energy selection).
\begin{figure}
\noindent \begin{minipage}[t]{0.45\textwidth}
\includegraphics[width=1.1\textwidth]{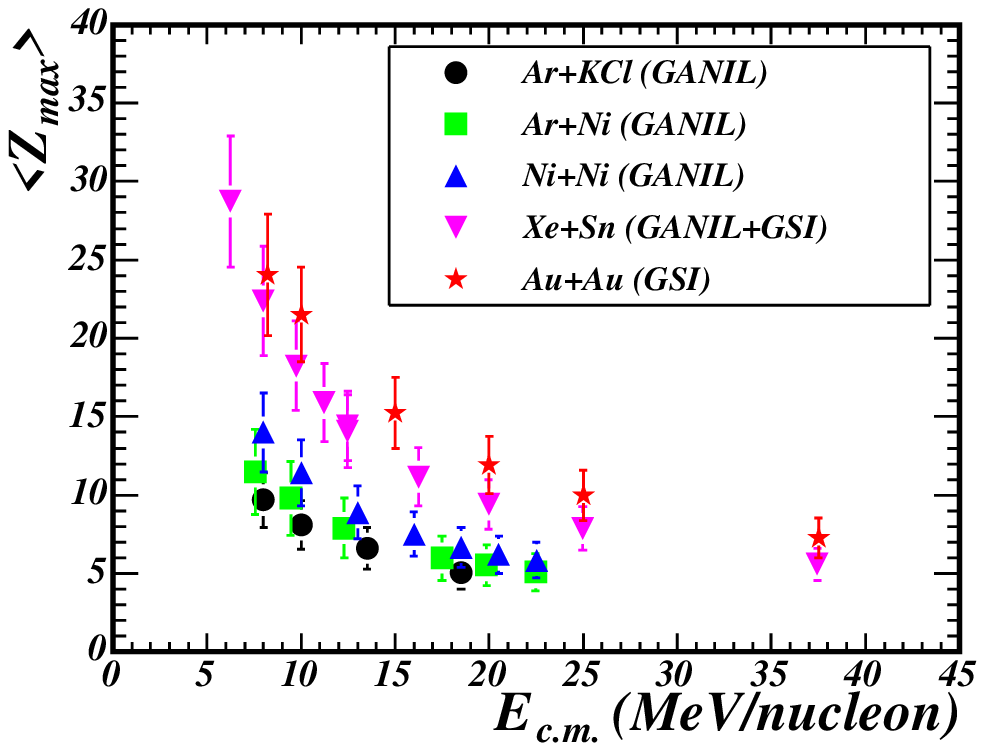}
\caption{Average charge of the largest fragment versus the available energy
for transverse energy selected central collisions measured with 
INDRA.}\label{zmax_e}
\end{minipage}%
\hspace*{0.05\textwidth}
\begin{minipage}[t]{0.5\textwidth}
\includegraphics[width=\textwidth]{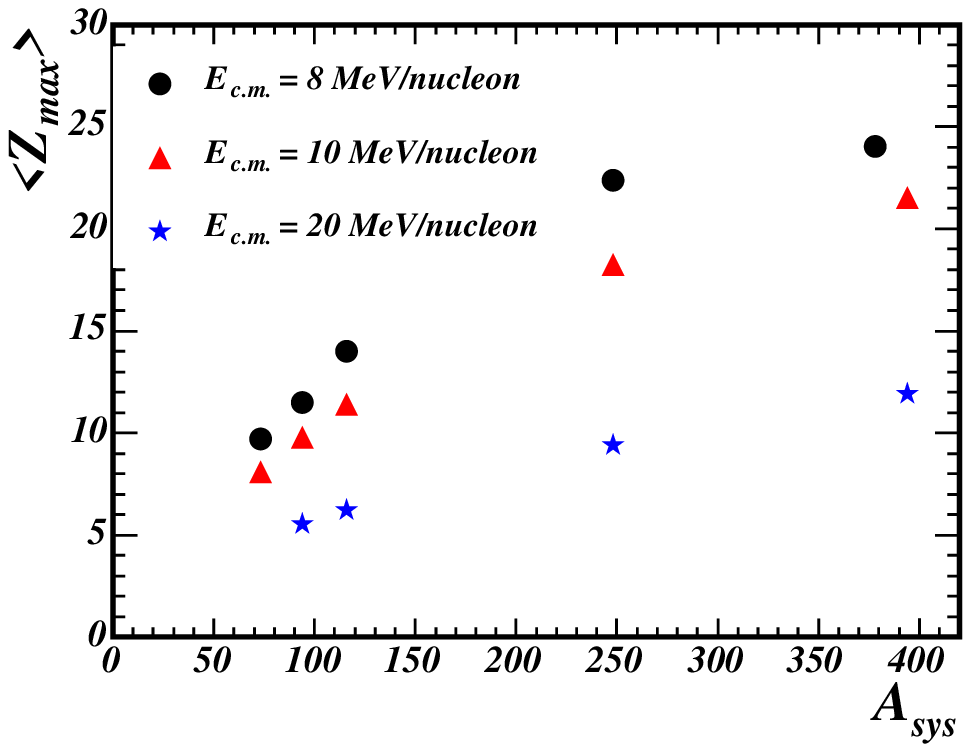}
\caption{Average charge of the largest fragment versus the total system mass
for three available energies and central collisions.}\label{zmax_a}
\end{minipage}%
\end{figure}
The average charge of the largest fragment, for a given system, first 
strongly decreases with increasing available energy and then tends to level
off, due to the fixed lowest charge value Z=3 (fig.~\ref{zmax_e}). 
Following models, this indicates that, as expected, the thermal energy 
increases with the available energy.  For better evidencing the evolution 
with the system mass, several cuts at given available energies are 
performed in fig.~\ref{zmax_a}; one observes that the largest fragment 
charge linearly rises with the system mass up to mass 180-190 
above which a much smoother variation is observed (1-2 units for 100 masses).
 Therefore the independence of the   
$Z_{max}$ distributions on the system only holds for heavy enough systems, 
and starting at higher mass (or charge) than in the MMMC calculation.

$Z_{max}$ also presents some specific dynamical properties. As shown 
in~\cite{I9-Mar97,I57-Tab05} for selected compact multifragmenting sources, 
its average kinetic energy is smaller than that of other fragments with the
same charge. The effect was observed whatever the fragment multiplicity for
Xe+Sn between 32 and 50~\AM{} and for Gd+U at 36~\AM{}. The
fragment-fragment correlation functions are also different when one
of the two fragments is $Z_{max}$ (see section~\ref{FO}).
This observation was
connected to the event topology in multifragmentation, the heavier 
fragments being systematically closer to the centre of mass than the 
others. The BOB stochastic mean field simulations, which reproduce a 
large number of the static and dynamical properties of multifragmentation,
confirm this interpretation, as shown later in fig.~\ref{posfrag}.
It will be shown in the next sections that $Z_{max}$ also displays scaling
properties or bimodal distributions, indicating that it may be considered 
as the order parameter of a phase transition in nuclei.

\subsection{Summary on size effects}
 What has been observed about the influence of the size of the 
 fragmenting system on static variables? Firstly the maximum fragment 
 (Z$\geq$3)  multiplicity is proportional to the system mass, for central 
 collisions as well as for QP's. In central collisions, this maximum 
 occurs at an energy increasing with $A_{sys}$ up to 
 $A_{sys} \sim$ 100-150; the scaling of $M_f$ with $Z_{bound}$ is verified 
 only for
 very heavy systems, and not for a light system with $Z_{sys}$=56; in 
 that case it can be stressed that a very large fraction of $Z_{bound}$
 is exhausted by He nuclei (more than 40\% in the explored range). 
 The charge of the largest fragment, for a given available energy,
 increases with $A_{sys}$ up to $A_{sys} \sim$ 190. Thus there is not 
 a single system mass which would be a milestone for all the variables.
 Finally, because of its special properties,  the largest fragment
 must be distinguished from the others in the partition; being sure of the
charge completeness of the detected events is mandatory in that aim.

\section{Theoretical descriptions of nuclear fragmentation and comparison to
data}\label{models}

Among the existing models some are related to statistical descriptions 
based on multi-body phase space 
calculations~\cite{Mek78,Fai83,Koo87,Gro90,Lee92,%
Hah88,Hahn88,Kon94,Bon95,Gul97,Rad97,Radu00}
whereas others describe the dynamic evolution of systems resulting
from collisions between nuclei via molecular
dynamics~\cite{Pei89,Aic91,Luk93,Ono93,Fel90,Ono96,Sug99,I16-Neb99,Pra95}
or stochastic mean
field approaches~\cite{Ayi88,Ayi90,Ran90,Cho91,Rei92,Rein92,Reinh92,%
Cho94,Gua96,Gua97,Cho96,Mat00}.
The first approach uses the techniques of equilibrium statistical 
mechanics (counting of microstates) with a freeze-out scenario and has 
to do with a thermodynamical description of the phase transition for 
finite nuclear systems. The second, in
principle more ambitious, completely describes the time evolution of the
collisions and thus helps in learning about nuclear matter (stiffness of 
the effective interaction and in-medium NN cross-sections), its phase 
diagram, finite size effects and the dynamics of the phase transition.
Therefore it is highly instructive to compare results of the
two types of models to experimental data. 
 
\subsection{Statistical ensembles and models}

Three different ensembles  are used in statistical models; 
rigorously only the microcanonical ensemble is
adapted to describe isolated excited finite nuclei.\\
The grandcanonical or macrocanonical ensemble corresponds to the rougher
description where the system can exchange particles as well as energy 
with a reservoir. In this ensemble the temperature and the chemical
potential are fixed variables. The total energy but also the total number 
of nucleons and the total charge fluctuate from channel to channel. 
Constraints are only  on the average mass and charge of the system. 
This ensemble is generally used for infinite systems and in relativistic 
quantum systems where
particles are created and destroyed. However it is a good 
approximation for hot nuclei when
one is not interested in an event-by-event analysis and only wants to
calculate mean values at very high excitation energies ($\geq$6-7~\AM{})
where the number
of particles associated to deexcitation is 
large~\cite{Hah88,Hahn88,Kon94,Bon95,Gul97}
(see subsection~\ref{sec:vapo}).
A second ensemble, the canonical ensemble, is used to describe a system 
with a fixed number of particles in contact with a heat reservoir at fixed
temperature~\cite{Lee92,Lee97,Par00}. Here the total energy fluctuates 
from partition to partition and only the mean value of the total energy 
is fixed. 
Considering hot nuclear systems and using the Fermi gas model, one 
can estimate the standard deviation of the excitation energy distribution,
$\sigma$, as a function of the temperature T and of the number of 
nucleons A.
It is given by the relation $\sigma = 4E^*/ \sqrt{AT}$ 
which shows that the canonical ensemble becomes a reasonable 
approximation for
A$\gtrsim$200 and temperature T$\gtrsim$6 MeV.
Within the formalisms of both these ensembles
many studies have been performed to derive mean properties
and  to discuss phase
transitions in terms of intensive variables such as temperature. The 
third ensemble, the microcanonical one, is the most relevant
for studying isolated systems like nuclei. It is used to describe a 
system which has 
fixed total energy and  particle number~\cite{Rad97,Gro90,Bon95,Koo87}.
In this ensemble the temperature is no longer a natural concept and a
microcanonical temperature can be introduced through the thermodynamic
relation: $T_{micro}^{-1} = \partial S / \partial E$.
Such an ensemble is fully appropriate if one wants to study, for example,
partial energy fluctuations (see section~\ref{SignPT})
and/or to perform analyses on an event-by-event basis. 
Results have to be discussed as a mixing of microcanonical ensembles
in order to be compared to those of canonical ensembles.
Numerical realizations are possible after elaborating specific algorithms
based on the Monte Carlo method. Finally one can conclude about the choice 
of the different ensembles by saying that the excitation energy domain,
the size of the system,
the pertinent observable to study  and the event
sorting chosen impose (or permit with some approximation
the statistical ensemble to be used. For comparison with data
additional  constraints (volume, pressure, average volume\ldots ) 
are added to these ensembles~\cite{Das01,Cho00}.

\subsection{Statistical descriptions of multifragmentation}

\begin{figure}[htb]
\begin{center}
\includegraphics[scale=0.9]{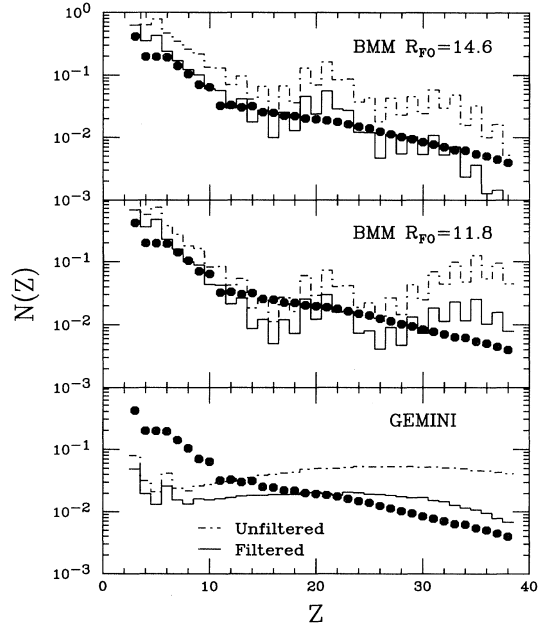}
\end{center}
\caption{Experimental charge distribution (full points) in central events 
compared with unfiltered (dot-dashed curves) and filtered (solid curves)
predictions by the BMM model with two freeze-out radii (top and central
panels, see text) and by GEMINI (from~\protect\cite{Bow95}).} \label{PRC52}
\end{figure}

\begin{figure}[htb]
\includegraphics[width=0.45\textwidth]{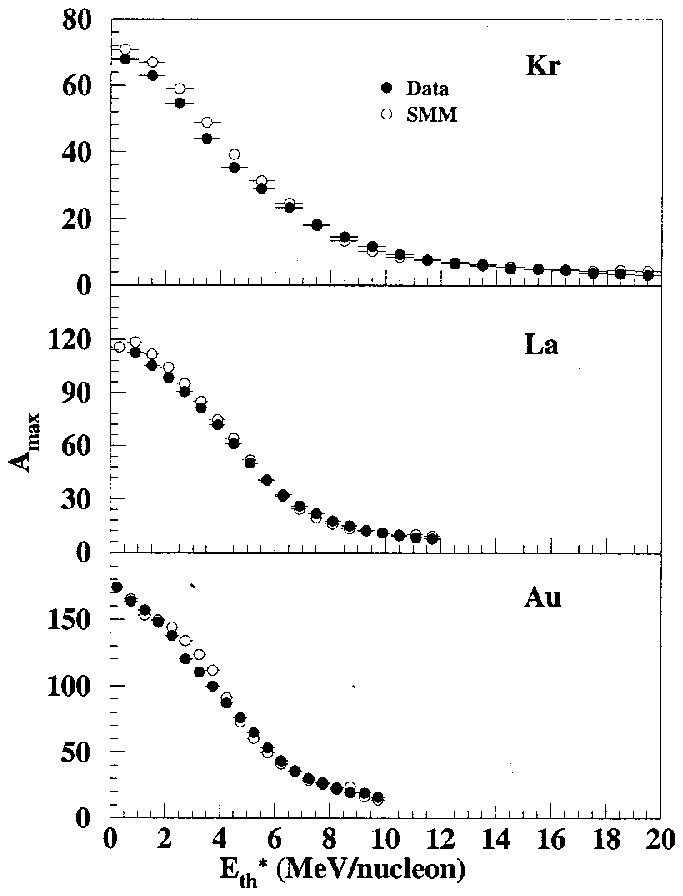}
\hspace*{0.1\textwidth}
\includegraphics[width=0.45\textwidth] {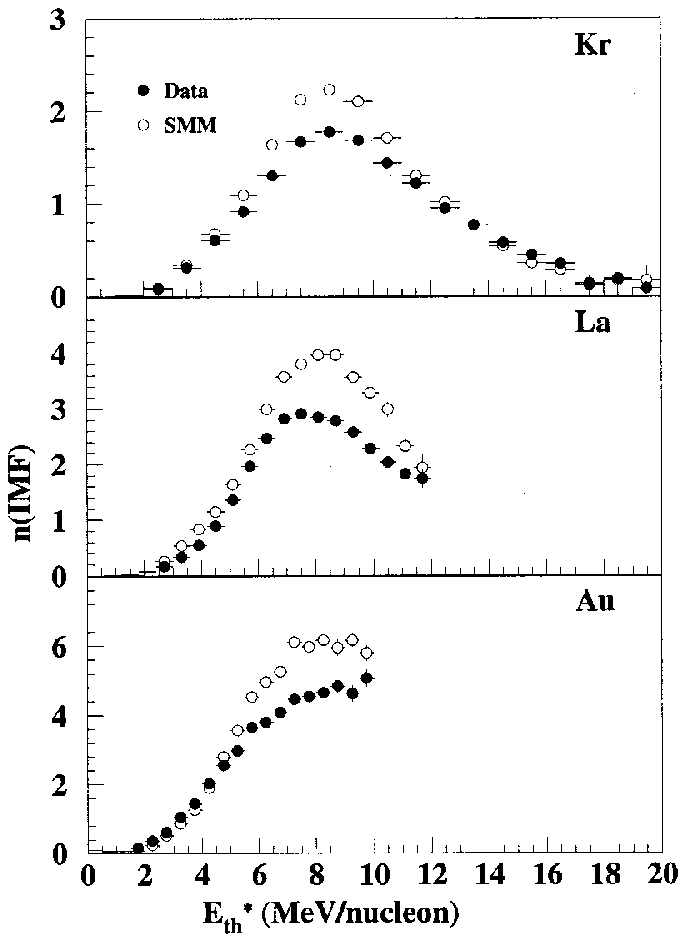}
\caption{Left panel: size of the largest fragment as a function of thermal
excitation energy, $E^*_{th}$ for multifragmentation of 1 GeV/nucleon Kr, 
La and Au (data) and SMM. Right panel: idem for average fragment 
multiplicities (from~\protect\cite{Sri02}).} \label{PRC65}
\end{figure}

\begin{figure}[htb]
\begin{minipage}[t]{0.45\textwidth}
\includegraphics[width=\textwidth]{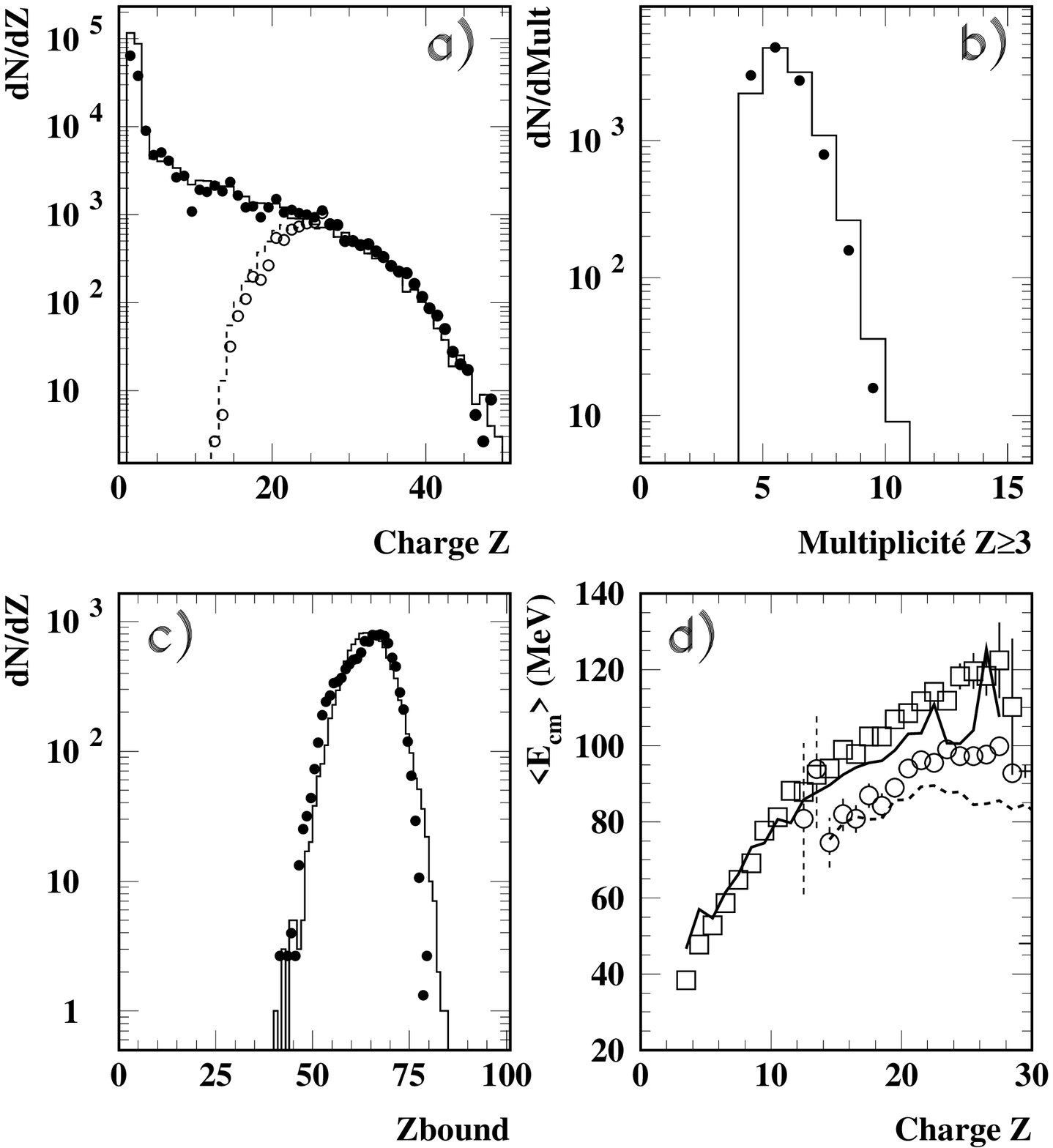}
\caption{Comparison of experimental data (Xe+Sn compact single sources
produced in central collisions-32~\AM{})
with SMM simulations. The lines are for data and symbols for SMM
(all fragments, except open circles and dashed lines which refer to the
largest fragment of each partition). Z$_{bound}$ represents the sum of the
charges of all fragments (from~\protect\cite{T25NLN99}).} \label{smmexp}
\end{minipage}%
\hspace*{0.05\textwidth}
\begin{minipage}[t]{0.45\textwidth}
\includegraphics[width=1.1\textwidth]{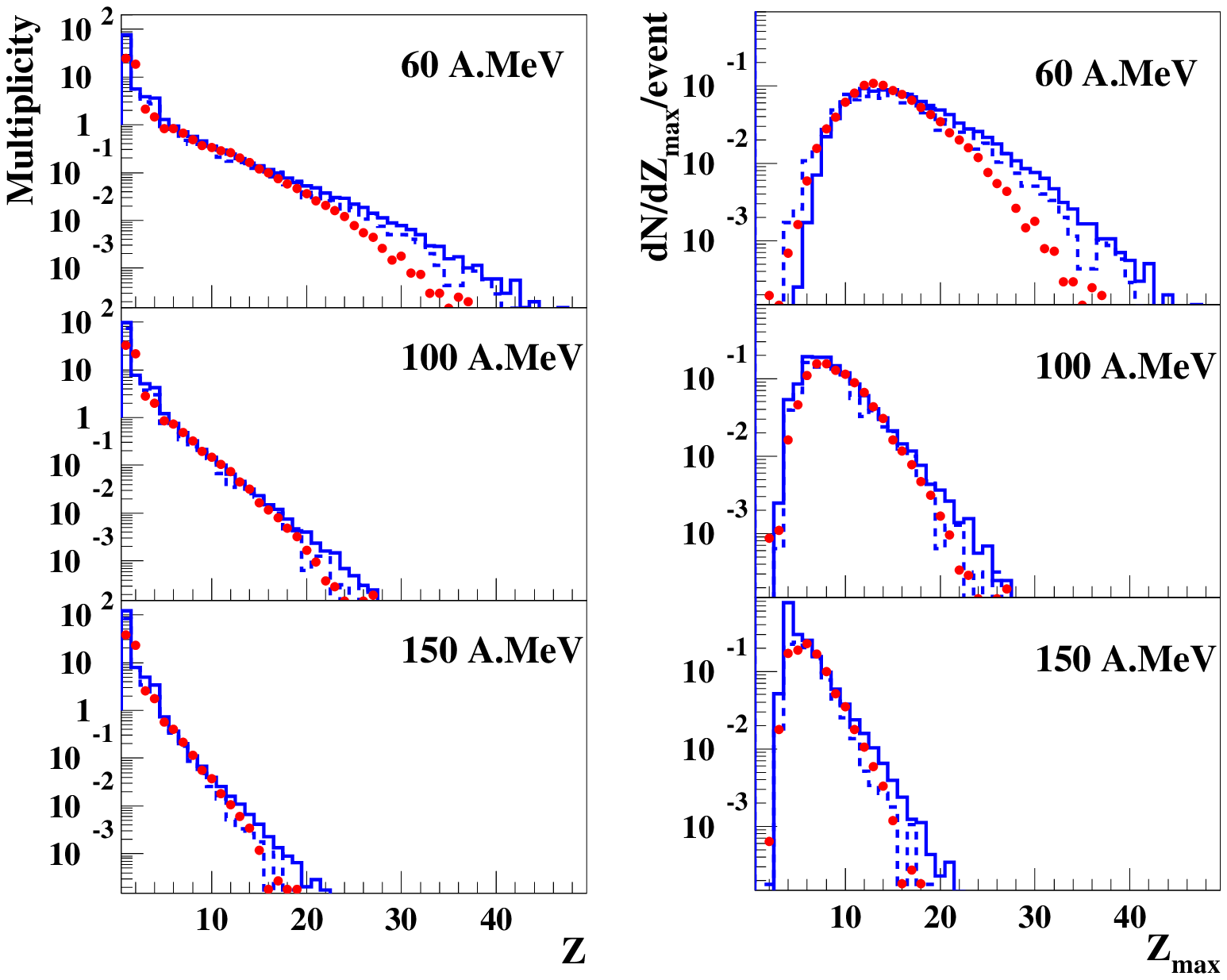}
\caption{Fragment (left panel) and heaviest fragment (right panel)
charge distributions for central $^{197}Au$+$^{197}Au$ collisions at
different incident
energies. Points refer to data and full lines to fragment from QMD events.
Fragments from QMD events filtered by the experimental device correspond 
to dashed lines (from~\protect\cite{I62-Zbi07}).} \label{qmdexp}
\end{minipage}
\end{figure}

A statistical theory of multifragmentation is supposed to predict 
partition probabilities at statistical equilibrium. Thus the
weight of a given break-up
channel $f$, i.e. the number of microstates leading to this
partition, is determined by its entropy,
$\Delta\Gamma_f$=exp$S_f$, within the microcanonical
framework. Statistical model event generators have been developed for
comparisons with experiments and in such models the set of fragments
corresponding to a given
partition is considered to be distributed randomly in a freeze-out volume
equal to 3-10 times the corresponding volume at normal density; the
 freeze-out stage is assumed to follow a
compression-expansion or/and a thermal expansion~\cite{Fri90} phase.
These fragments interact via Coulomb forces and are endowed with some 
internal excitation energies and initial
thermal velocities. Radial expansion velocities, fully decoupled from
thermal properties, are also added in some models. The subsequent
 evolution including sequential decays of primary fragments  is performed,
 preserving in some models space-time correlations. Finally generated events
are filtered to account for the experimental device.
 Detailed presentations of models can be found in 
 references~\cite{Gro90,Bon95,Rad97,Radu00,Col00}. Within such
an approach the input parameters: mass and charge of the
system at break-up density, its excitation energy, its volume at 
freeze-out and the eventual added radial expansion have to be backtraced 
to experimental
data, estimated from dynamical simulations or derived from data related to
properties of systems at break-up. Statistical models are very useful to 
compare data sets and to produce comparisons or systematics.
The following examples illustrate their different utilizations.\\
In fig.~\ref{PRC52} the fragment charge distribution measured for central
$^{129}Xe$+$^{nat}Cu$ collisions at 30~\AM{} incident
energy~\cite{Bow95} is
compared to the Berlin Multifragmentation Model (also called 
MMMC-Microcanonical
Metropolis Monte Carlo)~\cite{Gro90} and to the sequential statistical model
GEMINI~\cite{Char88}. The BMM calculations give good qualitative agreement
with the experimental charge distribution over the entire Z range measured.
Input parameters of the source (A=177, Z=76 and 750 MeV excitation
energy) were estimated from dynamical simulations at
a freeze-out time corresponding to a volume of 5.6 $V_0$ ($R_{FO}$=11.8~fm
for the sphere including all the products at freeze-out). Results
with a larger freeze-out volume (10.7 $V_0$) which show a better overall
agreement are also shown. Conversely, the GEMINI calculation  
underpredicts the yields of low Z  by an order of
magnitude and predicts a nearly flat charge distribution.
A comparison with the most popular model, namely the Statistical
Multifragmentation Model SMM~\cite{Bon95}, is shown
in fig.~\ref{PRC65} for the largest fragment and the fragment multiplicity
as a function of the thermal excitation energy of the source. 
It concerns multifragmentation of 1 GeV/nucleon Kr,
La and Au projectiles produced in collisions with a carbon target,
studied by the EOS collaboration~\cite{Sch01,Sri02}.
Inputs for SMM (thermal excitation energy, mass, charge and freeze-out volume
of the source) were derived from data including p-p correlations for
freeze-out volume and the ISABEL cascade
model for estimate of preequilibrium neutrons, which were not detected.
The ranges of mass and charge of the SMM sources which cover the
explored thermal excitation energy domains are A=80-30 and Z=34-16 for Kr,
A=130-80 and Z=54-26 for La and A=190-120 and Z=76-48 for Au.
The standard value of SMM (16 MeV) was used for the inverse level density
parameter and derived freeze-out volumes correspond to three times the
volume of Kr, La and Au nuclei at normal density. In SMM the definition of
the freeze-out volume differs from that of BMM and includes only the centre
of all the fragments. Very good agreements
are observed between data and SMM for the
largest fragment over the whole range of thermal excitation energy
whereas fragment multiplicities are overestimated by about
20\% in the excitation energy range 5-10~\AM{}.
Finally fig.~\ref{smmexp} shows
different static and dynamic fragment observable distributions measured for
central Xe+Sn collisions at 32~\AM{} incident energy and compared again with
SMM. 
To get this agreement the input parameters of the
source are the following: A=202, Z=85 as compared to A=248 and Z=104 for the
total system, freeze-out volume 3$V_0$, 
partitions fixed at thermal excitation energy = 5~\AM{} and added
radial expansion energy of 0.6~\AM{}.
As we have shown with those examples,
but it is a general trend, the observed properties of fragments are
compatible with the hypothesis of
sources in thermal equilibrium which undergo multifragmentation.

\subsection{Dynamical descriptions of multifragmentation}

Beside statistical descriptions, there are microscopic frameworks that
directly treat the dynamics of colliding systems such as 
the family of semi-classical simulations based on
the nuclear Boltzmann equation (the Vlasov-Uehling-Uhlenbeck (VUU), 
Landau-Vlasov (LV),
Boltzmann-Uehling-Uhlenbeck (BUU) or Boltzmann-Nordheim-Vlasov (BNV)
codes~\cite{Kru85,Gre87,Ber88,Bon94}),
classical molecular dynamics (CMD)~\cite{Pra95,Str99,Cus02,Che04}, 
quantum molecular 
dynamics (QMD)~\cite{Pei89,Aic91,Luk93}, fermionic molecular dynamics
(FMD)~\cite{Fel90}, antisymmetrized molecular dynamics 
(AMD)~\cite{Ono93,Ono96,Sug99} and stochastic mean field approaches 
related to simulations of the Boltzmann-Langevin equation
~\cite{Ayi88,Ayi90,Ran90,Rei92,Rein92,Reinh92}.
Boltzmann type simulations
follow the time evolution of the one body
density. Neglecting higher than binary correlations, they ignore
fluctuations about the main trajectory of the
system (deterministic description), which becomes a severe
drawback if one wants to describe processes involving instabilities,
bifurcations or chaos expected to occur during the multifragmentation 
process. Such approaches are only appropriate during
the first stages of nuclear collisions, when the system is hot and 
possibly compressed and then expands to reach a uniform low density.
They become inadequate to correctly treat the fragment formation,
and for the description of multifragmentation it is
essential to include fluctuations. This is done in quantum molecular
dynamics methods and in stochastic mean field approaches.

\subsubsection{Quantum molecular dynamics: QMD and AMD simulations}
\begin{figure}[htb]
\begin{minipage}[t]{0.5\textwidth}
\includegraphics[width=\textwidth]{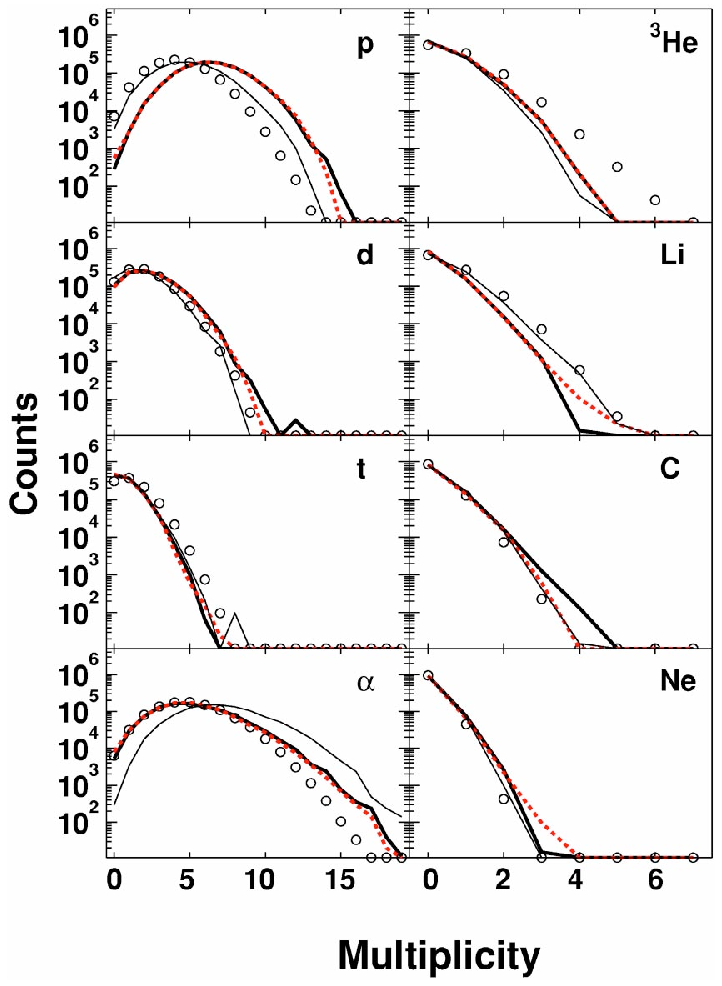}
\caption{Multiplicity distributions of selected particles and fragments
produced in central $^{64}$Zn+$^{92}$Mo collisions at 47~\AM{} incident 
energy. Experimental results are shown by circles and calculated 
results (AMD-V) correspond to different lines (see text). All calculated 
results have been treated with the experimental filter and all 
distributions are normalized to one million events in total 
(from~\protect\cite{Wad04}).} \label{amdexp}
\end{minipage}%
\hspace*{0.03\textwidth}
\begin{minipage}[t]{0.47\textwidth}
\includegraphics[width=\textwidth]{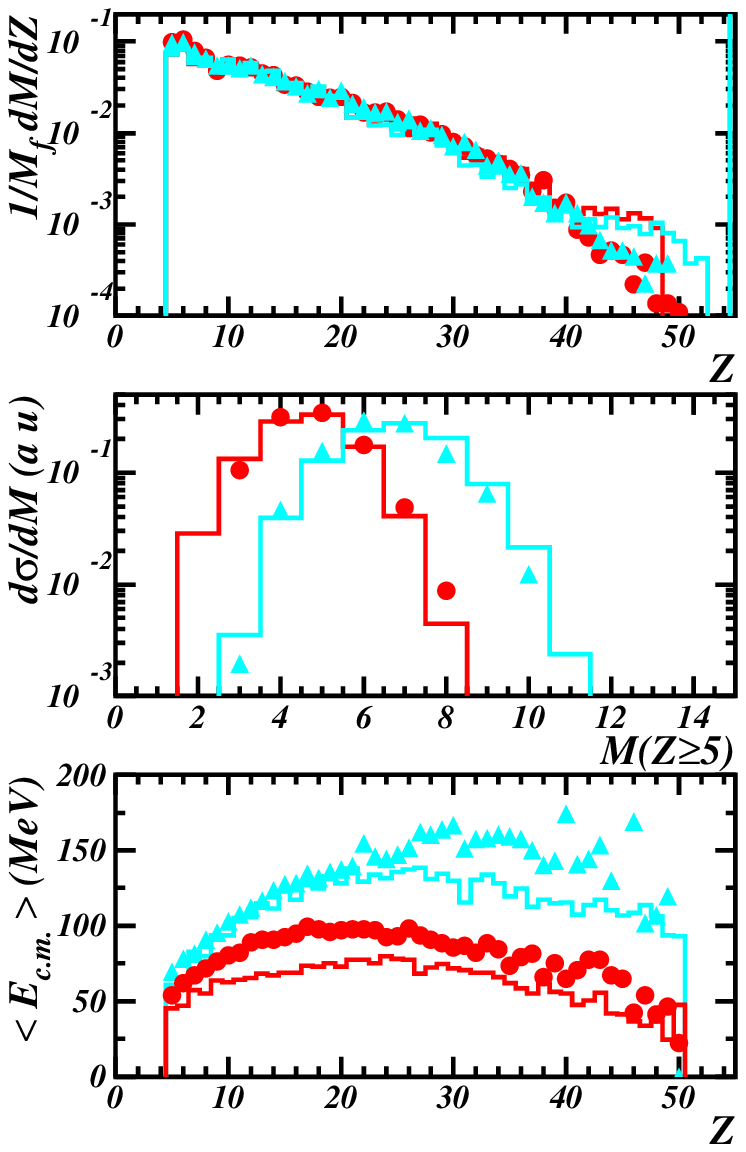}
\caption{Comparison of experimental data (central collisions: 
Gd+U-36~\AM{} and Xe+Sn-32~\AM{}) with BOB simulations for charge 
and multiplicity 
distributions of fragments (top and middle panels) and for their 
average kinetic energies (bottom panel). The symbols are for data and 
the lines for BOB simulations. Light grey lines and triangles stand 
for Gd+U and black lines and circles for Xe+Sn 
(adapted from~\protect\cite{I29-Fra01}).}
\label{databob}
\end{minipage}
\end{figure}

QMD is essentially a quantal extension of the molecular dynamics approach
widely used in chemistry and astrophysics. Starting from the n-body
Schrödinger equation, the time evolution equation for the Wigner 
transform of the n-body density matrix is derived. Several approximations 
are made. 
QMD employs a product state of single-particle states where only the mean
positions and momenta are time-dependent. The width is fixed and is
the same for all wave packets. The resulting equations of motion are 
classical. Also the interpretation of mean position and momenta is purely
classical and the particles are considered distinguishable; this simplifies
the collision term which acts as a random force. 
All QMD versions use a collision term with Pauli blocking 
in addition to the classical dynamics.
Some versions consider spin and isospin and others do not distinguish
between protons and neutrons (all nucleons carry an average charge).
Finally one has to stress that for most of the QMD versions a statistical
decay code must be coupled to describe the long time dynamics. 
However for the
code of ref.~\cite{Aic91} there is no need to supplement the QMD
calculations by an additional evaporation model~\cite{Mul93}. This code 
was very recently used in a rather complete comparison with
data measured for central $^{197}$Au+$^{197}$Au collisions over the incident
energy range 60-150~\AM{}~\cite{I62-Zbi07}. Using a soft equation of state 
(incompressibility K$_{\infty}$= 200 MeV),
static properties like fragment multiplicity and charge distributions
are rather well reproduced, particularly for the higher incident energies
where QMD codes are certainly better adapted.
Figure~\ref{qmdexp} illustrates the comparison for fragment and heaviest
fragment charge distributions.\\
An antisymmetrized version of
molecular dynamics (AMD) was constructed by incorporating two-nucleon
collision process as the residual interaction into the fermionic molecular
dynamics (FMD). AMD describes the system with a Slater determinant of
Gaussian wave packets and therefore can describe quantum-mechanical features.
However, in the dynamics of nuclear reactions, there may be other phenomena
caused by the wave packet tail that are completely lost in AMD due to the
restriction of the single-particle states. So an improvement was realized
(called AMD-V) with the stochastic incorporation of the diffusion and the
deformation of wave packets which is calculated by the Vlasov equation
without any restriction on the one-body distribution~\cite{Ono96}. More
recently the quantum branching process due to the wave packet diffusion
effect was treated as a random term in a Langevin-type equation of motion
whose numerical treatment is much easier. Moreover a new approximation 
formula was also introduced in order to evaluate the Hamiltonian in 
the equation of motion with much less computation time than the exact 
calculation, so that systems like Au+Au are now treatable~\cite{Ono99}. 
As for QMD the stiffness of the effective interaction and the in-medium 
NN cross-section are both important ingredients for determining the 
degree of agreement with experimental data. In order to test the 
sensitivity of the ingredients, a detailed study of reaction dynamics 
and multifragmentation was done in ref~\cite{Wad04} by comparing 
AMD-V calculations with data from heavy-ion reactions around the Fermi 
energy. Figure~\ref{amdexp} presents 
multiplicity distributions of selected particles and fragments
produced in central $^{64}$Zn+$^{92}$Mo collisions at 47~\AM{} incident 
energy. Thin solid, dashed and thick solid lines indicate the results of
Soft+$NN_{emp}$, Stiff+$NN_{emp}$ and Stiff+$NN_{LM}$ respectively. As
previously observed with QMD a
better global description of data is obtained with a soft equation of state
(incompressibility K$_{\infty}$= 228 MeV). We refer
the reader to ref.~\cite{Wad04} for more information on in-medium NN
cross-section.
\subsubsection{Stochastic Mean Field approach: Brownian One-body Dynamics 
(BOB)}\label{BOB}
In many domains of physics a diffusive behaviour is described by transport
theories which were originally developed for Brownian motion.
The effects of the disregarded degrees of freedom are simulated by a random
term in the dynamics of the retained variables. That idea is the starting
point of the so-called Boltzmann-Langevin equation (BLE):
$\partial f / \partial t = \{h[f],f\} +  \overline{I}[f] + \delta I[f] $ 
which was introduced for heavy-ion collisions in 
references~\cite{Ayi88,Ayi90,Ran90}.
$f$ is the one-body phase space density. The first term on the r.h.s.
produces the collisionless propagation of $f$ in the self-consistent 
one-body field described by the effective Hamiltonian. The second term,
called collision term, represents the average effect of the
residual Pauli-suppressed two-body collisions; this is the term included 
in LV, BUU and BNV simulations. The third term is the Langevin term which
accounts for the fluctuating part of the two-body collisions.
Exact numerical solutions of the BLE are very difficult to obtain and
have only been calculated for schematic cases in one or two
dimensions~\cite{Cho91}. Therefore various approximate treatments of the BLE
have been developed. The basic idea of BOB~\cite{Cho94} is to replace the
fluctuating term  by
$\delta \tilde{I} [f] = - \delta \bi{F} [f] . \partial f / \partial \bi{p}$
where $\delta \bi{F} (\bi{r},t)$  is the associated Brownian force
($<\delta \bi{F}>=0$). Since the
resulting Brownian one-body dynamics mimics the BL evolution, the 
stochastic force is assumed to be local in space and time. The strength 
of the force is adjusted to reproduce the growth of the most unstable 
modes for infinite nuclear matter in the spinodal region (see 
section~\ref{FinitePT}). Quantal fluctuations connected with collisional 
memory effects are also taken into account as calculated in~\cite{Ayi94}.\\
An extensive comparison data-BOB was made for two very heavy fused 
systems produced in Xe+Sn and Gd+U  central collisions which undergo 
multifragmentation with about the same excitation energy 
($\sim$ 7~\AM{})~\cite{I29-Fra01,I40-Tab03,I57-Tab05}. Stochastic 
mean-field simulations were performed for head-on collisions 
with a self-consistent mean field potential chosen to give a
soft equation of state (K$_{\infty}$= 200 MeV). The
finite range of the nuclear interaction was taken into account using a
convolution with a Gaussian function with a width of 0.9 fm.
A term proportional to $\Delta \rho$  in the mean-field
potential was added; it allows to well reproduce the surface 
energy of ground-state
nuclei, which is essential in order to correctly describe the
expansion dynamics of the fused system.  In the collision term a constant
NN cross-section value of 41 mb, without in-medium, energy, isospin or
angle dependence was used.
As a second step the spatial configuration of the primary fragments,
with all their characteristics as given by BOB, was taken as input in a
statistical code to follow the fragment deexcitation while preserving
space-time correlations. Finally the events were filtered
to account for the experimental device. 
These simulations well reproduce the observed charge and multiplicity
distributions of fragments (see fig.~\ref{databob}).
Particularly the independence of the charge distribution against the mass 
of the system experimentally observed was recovered~\cite{I12-Riv98}. 
More detailed comparisons of the charge distributions of the three 
heaviest fragments also show a good agreement~\cite{I29-Fra01}.
Kinetic properties of fragments are rather well reproduced
 for the Gd+U system, whereas 
for Xe+Sn the calculated energies fall $\sim$ 20\% below the measured 
values. We also refer the reader to 
sections~\ref{FO} and~\ref{FinitePT} for further comparisons.

\subsubsection{Concluding remarks on dynamical simulations}
To conclude on dynamical descriptions of multifragmentation one 
can make a few general comments. Better 
agreements between data and calculations are
generally observed with a soft equation of state
(incompressibility K$_{\infty}\sim$ 200-230 MeV). One can however
not trust the in-medium NN cross-sections extracted as the
self-consistency between the mean field potential and the two-body
collision term is not fulfilled in models. QMD and AMD
calculations lead to too much transparency at low incident energies
(typically below 50-100~\AM{}). For AMD-V and BOB calculations
at incident energies around 35~\AM{} a maximum density of 1.2-1.4 $\rho_0$
is observed at 30-40~fm/$c$ after the beginning of central heavy-ion collisions 
and the normal density is recovered around 70 fm/c. Thermal equilibrium 
times  are found in the range 100-140 fm/c after the beginning 
of collisions, well before
freeze-out configurations (200-300 fm/$c$). Primary fragments exhibit
an equal or almost equal excitation energy per nucleon of 3-4~MeV.
The mechanism of fragment production differs depending on model type.
 In molecular dynamics models fragments are
preformed  at early stages close to the normal nuclear density whereas
in stochastic mean field calculations, as BOB, fragment formation is
linked to the spinodal decomposition mechanism: mononuclear systems at
low density ($\sim$0.4$\rho_0$) formed at around 100~fm/$c$ develop 
density fluctuations during about 100~fm/$c$ to
form fragments. More constrained observables related to the formation 
of fragments by spinodal instabilities will be discussed in 
section~\ref{FinitePT}.\\
Last point, dynamical calculations exhibit radial collective energies
for fragments with average values in the range 0.1-2.0~\AM{} for 
heavy-ion collisions in the Fermi energy domain which fairly agree 
with values derived from experiments (see subsection~\ref{EneF}).

\subsection{The link between dynamical and statistical descriptions}

While both descriptions show reasonable agreement with data in 
reproducing average static and kinematic properties of fragments,
sharp conclusions on  multifragmentation scenarii can not yet be
derived. However from dynamical descriptions like  AMD-V and BOB we 
learnt that, in average, thermal equilibrium was reached before the 
breaking stage, which is also the hypothesis of statistical approaches. 
A tentative to derive more precisely at what time, during the reaction, 
the statistical description takes place was done in 
refs.~\cite{T28Tab00,I57-Tab05}: in BOB simulations of Xe+Sn central 
collisions at 32~\AM{} incident energy, the volume of the system was 
calculated every 20~fm/$c$ from 100 to 250~fm/$c$
to be compared with the freeze-out volume of 3$V_0$ as input in the SMM
code~\cite{T28Tab00,I57-Tab05}.
The result is the following: at 200~fm/$c$, during the formation of
fragments, a freeze-out volume of $\sim$3$V_0$ (calculated à la SMM)
is reached and the characteristics in size
of the source are A=190 and Z=80, in good agreement, within 10\%,
with SMM inputs. Very recently a remarkable result was also
obtained by checking the consistency of predictions between the 
dynamical evolution of Xe+Sn central collisions at 32~\AM{} via a 
stochastic mean field approach and a microcanonical multifragmentation 
model (MMM)~\cite{Rad06}: a statistically equilibrated stage was 
identified at $\sim$140~fm/$c$ just intermediate between the beginning 
of the spinodal decomposition and the freeze-out configuration. 
In both cases  a rather coherent link between
dynamical descriptions, like stochastic mean field approaches, and 
statistical descriptions is derived.

\section{Calorimetry and thermometry}\label{CaloTh}

The knowledge of the excitation energy and temperature, as well as of 
its  numbers of neutrons and protons (mass and charge) is mandatory
for any description of a system in terms of thermodynamical variables.
Determining these quantities remains the most challenging task for the
groups involved in the study of multifragmentation. \\
As previously mentioned, in nuclear collisions the formed
multifragmenting systems are accompanied by a quite abundant preequilibrium
emission. These early emitted products should not be included in the
calculation of the mass and of the excitation energy of the fragmenting 
system; it is unfortunately  difficult to unambiguously attribute the
observed final products to preequilibrium or to the fragmenting system.
This is particularly true for neutrons - when measured - and for light 
charged products, H and He and to a lesser extent Li and Be isotopes.
Conversely the properties of the heavier fragments  
indicate that they do originate from the fragmenting system.
Preequilibrium is removed either with the help of models, or 
through angular and energetic properties of the observed 
products~\cite{I29-Fra01,T32Hud01}.

\subsection{Calorimetry} \label{Calo}

All procedures for obtaining the excitation energy of a fragmenting 
source, observed with a 4$\pi$ array, are based on the determination 
of its velocity. For central 
collisions the reaction centre of mass velocity is most often chosen
whereas the quasi-projectile velocities are identified to  either 
that of the biggest fragment, or that of the subsystem containing all 
the fragments (Z$\geq$3 or 5), forward emitted in the centre of mass.
The excitation energy, $E^*$, of the source is then calculated event 
by event with the relation
$E^* = \sum_{M_{cp}} E_{cp} + \sum_{M_n} E_{n} - Q$.
$E_{cp}$ and $E_{n}$ are respectively the kinetic energies of charged
products and neutrons belonging to the source, Q is the mass difference
between the source and all final products.  
Energies are expressed in the source reference frame. $M_{cp}$ is in most
cases the detected multiplicity of charged products. The energy removed
by gamma rays is small and most often neglected in the calculation.
The decision of including or not one observed charged product in the
source differs with the experimental apparatus and the type of collision
under study. 
\begin{itemize}
\item[CC] for central heavy-ion collisions, all fragments with Z$\geq$3
(or 5) are attributed to the source. Preequilibrium in that case is
mostly forward/backward emitted, and indeed the angular distributions of the
light products appear isotropic between 60 and 120$^o$. The charge,
mass and energy
contributions of these particles are doubled for the calculation of the
characteristics of the source (i). Another possibility, to account for the
detector inefficiency, is to calculate the charge, mass and energy of the
anisotropic part, and to remove it from those of the composite system (ii).
\item[QP] for quasi-projectiles the most important contamination comes from 
mid-rapidity products and several techniques are used for the QP
reconstruction. i) All fragments forward emitted in the reaction 
c.m. system are attributed to the QP. Variants consist either
in putting a low velocity cut for the lighter fragments~\cite{Lle93,Pla01},
or in keeping only events with a compact fragment configuration in velocity
space~\cite{I61-Pic06,T41Bon06}. Then twice the light 
elements in the QP forward hemisphere are added.
ii) Fragments are treated as above, but particles are attributed 
a probability to come from the QP emission, either using a 3-source 
fit~\cite{Ma05}, or by taking a well characterized subspace as 
reference~\cite{H5Vie06,Pia06}.
The velocity of the QP is then recalculated by including all its components.
\item[hIC] finally in hadron-induced collisions, products emitted from the
source are chosen from energetic considerations, by excluding those 
with an energy per nucleon above a given threshold either 
fixed~\cite{Hau96} or varying with Z~\cite{Vio06}.
\end{itemize}
All those procedures assume forward-backward 
symmetry of  particle emission in the source frame. For QP's the
symmetry of the source emission may be questionable when highly excited 
QP's and QT's start emitting right after their
separation~\cite{Jan05,Hud05}: the close proximity of the partner
deforms phase space and emission is favoured between the QP and QT.
This possible effect is generally ignored.\\
Once counted the charged products, the charge of the source is known. 
A first uncertainty is introduced in calculating the associated mass, as 
that of heavy fragments is not measured. A single mass is attributed 
to all nuclei with a given atomic number, either that of the most stable 
species, or that derived from formulae existing in the literature 
(EPAX~\cite{Sum00} or EAL~\cite{Cha98}). 
At that point neutrons must be included. Except in
experiments using a neutron ball, neither their multiplicity not their
energy is known. The neutron number is thus calculated by assuming that the
source has the same N/Z ratio as the total system (central or hadron-induced 
collisions) or as the projectile. The average neutron energy is taken equal
to the average proton energy over the event sample, removing some Coulomb 
barrier. Note that with neutron balls only the neutron multiplicity is 
measured, at the price of a poor geometrical coverage for charged 
products. In that case corrections accounting for the undetected particles
and neutrons are made~\cite{Gal05}. \\
\begin{figure}
\begin{minipage}[t]{0.60\textwidth}
\includegraphics[width=\textwidth]{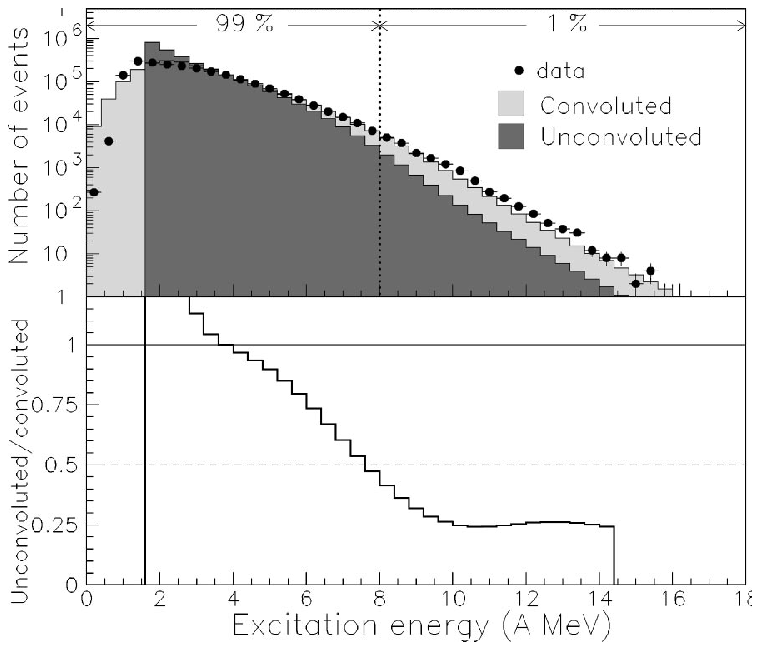}
\caption{Top : Convoluted, unconvoluted and experimental excitation 
energy distributions obtained in $\pi$+Au reactions. Bottom :
Ratio of unconvoluted-to-convoluted distribution as a function of
the excitation energy per nucleon.  
Adapted from~\cite{Lef01}}\label{fig5.1}
\end{minipage}%
\hspace*{0.02\textwidth}
\begin{minipage}[t]{0.35\textwidth}
\includegraphics[width=\textwidth]{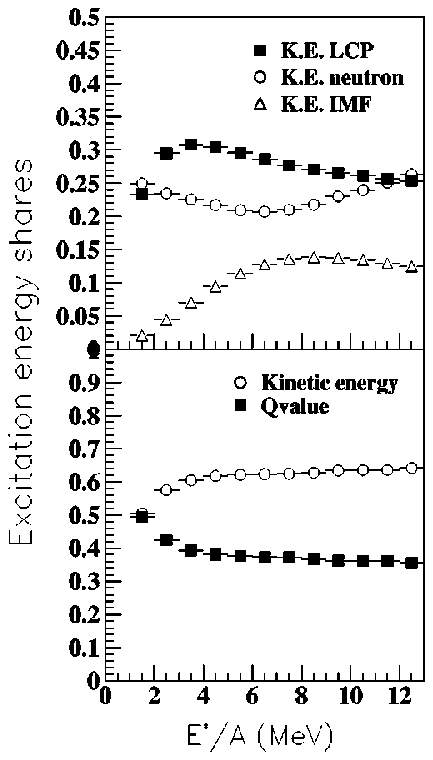}
\caption{Relative share of excitation energy for various components of 
the reconstruction procedure as a function of the excitation energy per 
nucleon. From~\cite{Lef01}}\label{fig5.2}
\end{minipage}%
\end{figure}
In central heavy-ion reactions, the excitation energy domain populated is
narrow : $\sigma_{E^*} \sim $0.7-1.25~\AM{}; the width includes 
experimental effects (detector efficiency and resolution), calculation 
assumptions and physical effects (pre-equilibrium).
Conversely, in hadron-induced reactions as well as in quasi-projectile
studies, a broad domain of excitation energy is populated, proportionally 
to the partial cross section, function of the impact parameter. However, 
due to on-line trigger effects, very low energies are poorly sampled, 
by particular events matching the trigger requirement; 
indeed neutron
emission - not detected - is dominant in this region. 
At the other end of the distribution, the very high energies 
probably result from significant fluctuations. In
all cases the reliable domain extends from about 2 to 8~\AM{}:
for example in fig.~\ref{fig5.1} obtained in $\pi$+Au reactions, the
excitation energy distribution is unconvoluted assuming Gaussian 
fluctuations. More than half of the 1\% of events above the vertical 
dotted line have an energy overestimated by 1-2~\AM{}~\cite{Lef01}.\\
How reliable are the energies so obtained ? 
Because of compensation of the errors on the mass and on the 
energy, the energy per nucleon is a more robust experimental
observable than the energy.
By comparing values obtained by different methods for quasi-projectiles,
differences on $E^*/A$ smaller than 10\% were 
found~\cite{I27-Ste01,T41Bon06}. From
simulations with the HIPSE event generator, the reconstructed
values were found to differ from the true values by less than 
10\%, except for very peripheral collisions where the
discrepancies are much larger~\cite{H5Vie06}. In central collisions, 
excitation energies slightly smaller than the available
energies are generally found, which is what can reasonably be expected.
It was verified in the INDRA Xe+Sn data that procedures CCi and CCii give 
the same results when a high
degree of completeness is required for CCi, e.g. 90\% of the system charge. 
For lesser completeness (80\%), the difference between both types of
calculation increases with the incident energy, reaching 1~\AM{} (10\%) 
at 50~\AM{}.
The main source of uncertainty in the calculation of $E^*$ comes from the
neutron terms. However, compensation occurs in the calorimetry equation
between the kinetic and the mass balance terms; indeed the weight of 
these two terms is similar for quasi-projectiles and in hadron-induced 
reactions 
($Q/E^* \sim$30-36\%~\cite{Lef01,T41Bon06} - see fig.~\ref{fig5.2}); in 
central collisions the Q term accounts only for $\sim$20\% of the 
excitation energy.

\subsection{Temperature measurements} \label{Temp}
Two recent reviews extensively describe the methods used for temperature 
measurements~\cite[and references therein]{DasG01,Kel06}. A brief summary
will be given here. The concept of temperature for a nucleus, which is a 
microscopic, isolated Fermionic charged system, is not \emph{a priori}
obvious. 
According to statistical mechanics, temperature can be defined
as $T^{-1}=\partial S(E,N,V)/\partial E$, where $S$, $E$, $V$, $N$, 
are the entropy, energy, volume and (fixed) number of particles of the 
system. This definition is valid only if the system is 
in statistical equilibrium and its density of
states is known. For compound nuclei at low excitation energy, both 
conditions are fulfilled, which might not be the case at higher
energies, and for fragments resulting from multifragmentation. Moreover
no probe can be used to measure the temperature of these small systems,
it has to be derived from the properties of particles that they emit
during their cooling phase. The abovementioned issues of   
identifying an equilibrated emitting source by distinguishing 
pre-equilibrium particles from those emitted by that source is
therefore common to all the methods.
Three families of methods are used to "measure" temperatures.   
\begin{enumerate} 
\item \textit{Kinetic approaches.}
Historically, temperatures of compound nuclei were derived from the 
slopes of the kinetic energy spectra of the emitted neutrons or 
charged particles that they evaporate, as the spectra can be fitted with 
Maxwell-Boltzmann distributions. At higher energies, when long 
chains of particles are emitted, the obtained result is an average over
the deexcitation chain, and may differ from one particle to another,
depending on the emission sequence. To retrieve the initial 
temperature, it was proposed to subtract from the spectra those 
of particles coming from the same nucleus formed at lower
excitation energies~\cite{Gon90}. For multifragmenting systems, the 
slopes of light product spectra lead to very high
"temperatures", and do not probably reflect only the thermal 
properties of the system, but also the collective energies coming 
from the dynamics of the nuclear collision. It was recently proposed 
to derive the temperature of the fragmenting system from the slope of
the thermal hard photons~\cite{Ort06}, which have the advantage of 
being insensitive to Coulomb field and final-state effects: temperatures
close to 7~MeV are for example obtained for central Xe+Sn collisions at
50~\AM{}. 
\item \textit{Population of excited states.}
The underlying idea for this method is that the population of the excited
states of a system in statistical equilibrium is given by the temperature 
of the system and the energy spacing, $\Delta E$, between the levels. This
definition in itself bears the limits of the method:  when the temperature
is higher than $\Delta E$, the ratio between the population of two states
saturates. Considering particle-unbound states is thus interesting as it
allows to measure higher temperatures, and the population ratio should
in that case be less influenced by secondary decays.
\item \textit{Double ratios of isotopic yields.}
This method uses the yields of different light isotopes produced by the
system. It was developed in the grandcanonical approach, and is valid for
systems at densities low enough to make fragment nuclear interaction 
negligible,
thus the composition of the system is frozen~\cite{Alb85}. The basic
assumption is that free nucleons and fragments are in thermal equilibrium
within an interaction volume V. The density of an isotope reads:
$\rho(A,Z)=N(A,Z)/V =A^{3/2} \omega(A,Z) \lambda_{T_N}^{-3} \exp (\mu(A,Z)/T)$,
where $\omega$ is the internal partition function of particle($A,Z$), $\mu$
its chemical potential and $\lambda$ the thermal nucleon wave length.
The condition of chemical equilibrium allows to define the chemical
potential of a species in terms of those of free neutrons and protons and of
its binding energy. Using two sets of two nuclei differing only by one
nucleon, the temperature is derived from the double yield ratio, the binding
energy differences and the partition functions only, the other terms 
disappear. The results depend on the validity of the assumptions: is the
grandcanonical ensemble relevant in the studied case? Is the
system really in thermal and chemical equilibrium? The considered particles
should be present at freeze-out, and not  produced by secondary decays.
Another problem lies in the calculation of the binding energies which might
depend on density and temperature. Different corrections were proposed to
account for finite-size effects~\cite{Radu99} or secondary
decays~\cite{Tsa97,Rad00}.
\item  \textit{Isospin thermometer approach.}
At relativistic energies, it was proposed to derive the temperature 
at freeze-out from the isotopic distributions of the final
residues~\cite{Sch02}. The assumptions are that the fragments at 
freeze-out have the same N/Z ratio as
the projectile; their thermal energy is dissipated through an evaporation
cascade, which is reconstructed with the help of evaporation codes. The
results thus entirely rely on the reliability of these codes. 
\end{enumerate}

A recent study on the reliability of temperature measurements for Xe QPs
from Xe+Sn collisions at various energies shows that, whatever the
method used, temperatures are determined at best within 
10-20\%~\cite{H5Vie06}.

\subsection{Caloric curves} \label{Calcur}
\begin{figure}[htb]
\begin{center}
\includegraphics[width=8cm]{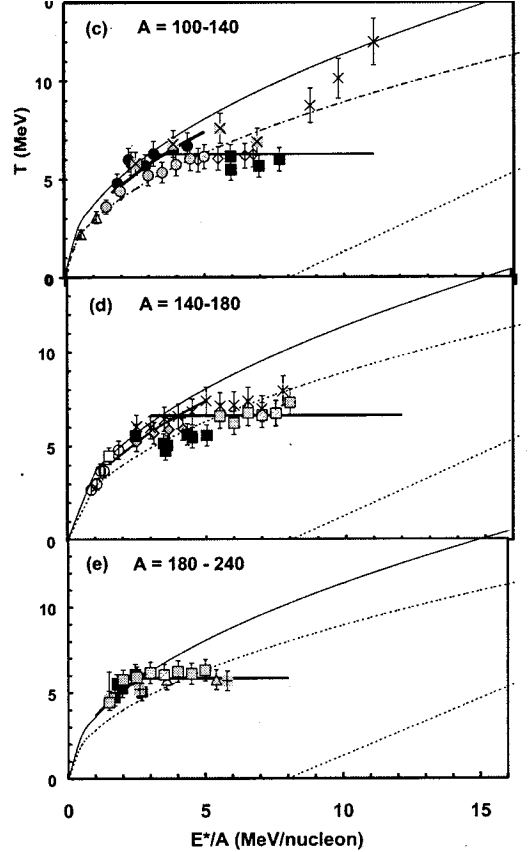}
\end{center}
\caption{Caloric curves for three selected regions of mass. The initial 
temperatures
are plotted vs the thermal excitation energies of the systems.
(Adapted from~\protect\cite{Nato02}). \label{calocurve}}
\end{figure}
The combined and independent determinations of the thermal excitation 
energy and of the temperature of a system allow to draw the caloric 
curve, which gives the relation between these variables. 
The first example of caloric curve was given in~\cite{Poc95}
and taken as the evidence of the occurrence of a liquid-gas type phase
transition. Many  data followed, which were compiled in~\cite{Nato02},
from which is extracted fig.~\ref{calocurve}. 
Care was taken to recover 
the initial temperature of the systems by applying corrections, as mentioned
in sect.~\ref{Temp}, to account for the deexcitation cascade, or with the 
help of a statistical model. The published excitation energies were
considered as pure thermal energy, except in the case of a heavy ion 
reaction for which the thermal part was evaluated with the SMM model.
The general
trend of these curves, classified in different mass zones, is first the
well-known rising part attributed to the Fermi gas ($E^*=aT^2$, with $a$ 
the level density parameter), followed by a plateau. The temperature at 
the plateau, as well as the energy at which it is first reached, were 
seen to decrease for larger masses. This reminded of the evolution of the
predicted limiting temperatures caused by Coulomb
instabilities~\cite{Bon84,Bon86}. More recently, caloric curves were 
published which use the temperatures obtained from hard thermal photon
spectra~\cite{Ent02,Ort06}. The results do not clearly exhibit a plateau, 
but above 3-4~\AM{} of excitation the caloric curves fall below that 
expected for the Fermi gas. Finally a theoretical study in the 
microcanonical framework of the MMM model calibrated the different 
isotopic temperatures against the true microcanonical one~\cite{Radu99}: 
the authors 
found a universal relation for all masses, and used it to re-evaluate 
some published caloric curves. The corrected curves all exhibit three 
parts, the Fermi gas one, a more or less broad plateau followed by a
linearly rising part attributed to a ``classical'' gas region. 
Further discussion on
caloric curves will be presented in sect.~\ref{FinitePT} in relation 
with phase transitions.

\section{Freeze-out properties} \label{FO}

The concept of freeze-out was first introduced as a starting point in 
statistical models. It can be defined as a configuration for which mutual 
nuclear interactions between primary products of a multifragmenting 
source become negligible. Moreover, for the sake of simplicity, those 
products are supposed to be spherical and having eventually recovered 
normal density. It was shown recently, using dynamical simulations as BOB, 
that such a simple geometrical picture can be relevant on the
event by event basis~\cite{Par05}. \emph{The freeze-out configuration
characterizes the new physical state produced by the excited finite system}. 
In particular the freeze-out volume appears as a key quantity and its 
knowledge is particularly important in the extraction of thermostatistics 
observables as the microcanonical heat capacity (see 
section~\ref{FinitePT}). Relative velocity correlations including 
fragments and particles as well as different types of simulations
compared to data have been used to progress with the difficult task
of determining the freeze-out volume.

\subsection{Fragment velocity correlations and event topology at 
freeze-out} \label{Corv}

Fragment-fragment relative velocity correlations functions were widely
used in the 90's to derive information on emission time scales and the 
disassembling source volume. Most of the papers published on the subject 
make use of the
formalism developed in~\cite{KimY92}, where arguments are presented
which justify a classical treatment of such correlations. The importance 
of Coulomb effects and momentum conservation laws is underlined. In this
picture, it is suggested to mix correlations obtained with fragments of
different sizes by replacing the relative velocity, $v_{rel}$, between 
two fragments of charge $Z_i$ and $Z_j$ by a reduced velocity, 
$v_{red} = v_{rel}/ \sqrt{Z_i+Z_j}$; in this formula the mass of 
the fragments is supposed to be twice their charge, meaning that it is 
more appropriate for light fragments. The uncorrelated yield necessary 
to build
the correlation function was generally obtained from event mixing between
events of the same class~\cite{Zaj84,Lis91}. As for particle-particle
correlations, one obtains information on the space-time extent of the
emitting source. A majority of the results deal with trajectory
calculations, assuming an exponential probability distribution of the
time delay  between emitted fragments, $P(t) \propto \exp{-t/\tau}$. 
Excitation functions show that, for central collisions, the space-time
extent decreases when the bombarding energy is raised: times around
300-500~fm/$c$ are reported around 30~\AM{}, while they are between 50 
and 100~fm/$c$ above 60~\AM{}~\cite{Fox93,Bau93}.
For multifragmenting quasi-projectiles produced in semi-peripheral 
collisions, fragment emission times of 200-500~fm/$c$ are reported 
in~\cite{Bow93}, for the Cu+Au reaction at 50~\AM{}, independently of 
the assumed volume of the QP source; for Ni QP
from the Ni+Au reaction at 34.5~\AM{}, $\tau$ decreases from 550 to
75~fm/$c$ when the excitation energy increases from 2 to 7~\AM{} and 
remains constant beyond~\cite{Zhi00}. Thus on
average, QP fragment emission times are comparable to those of central
collisions, as appears in fig.~\ref{figdtf}, where the emission times 
are plotted versus the source excitation energy.
Finally velocity correlations were also studied in hadron-nucleus 
reactions as a function of the excitation energy deposited in the 
nucleus~\cite{Bea00}. Emission times were shown to decrease
from $\tau$=500~fm/$c$ at E$*$/A=2~MeV to $\tau \sim$20-50~fm/$c$ for
E$*$/A$\geq$5~MeV. The time range quoted at high excitation accounts for
corresponding associated volumes 6-4V$_0$. Figure~\ref{figdtf} shows
that the times obtained for these reactions are systematically shorter 
than those measured in nucleus-nucleus collisions, by a factor of about 5.
The authors of~\cite{Bea00} tentatively explain this observation by a 
better source selection in hadron-induced reactions. This would act on 
the excitation energy axis only (through the energy range of included lcp, 
the collective energies).
Altogether, these results testify for an increased thermal energy 
deposition in collisions when raising the excitation energy. \\
In a few cases,  fragment-fragment velocity correlation functions were
compared to a full calculation coupling a dynamical simulation of the
collision to an after-burner. Phase-space coordinates of the fragments
produced in the first phase are injected in the second one.
Two examples will be given here, one coupling QMD and SMM for central
100~\AM{} Fe+Au collisions~\cite{San95}, and the other using the stochastic 
mean field BOB simulation followed by the SIMON deexcitation for central
32~\AM{} Xe+Sn and 36~\AM{} Gd+U reactions~\cite{I57-Tab05}.
\begin{figure}
\includegraphics[scale=0.9]{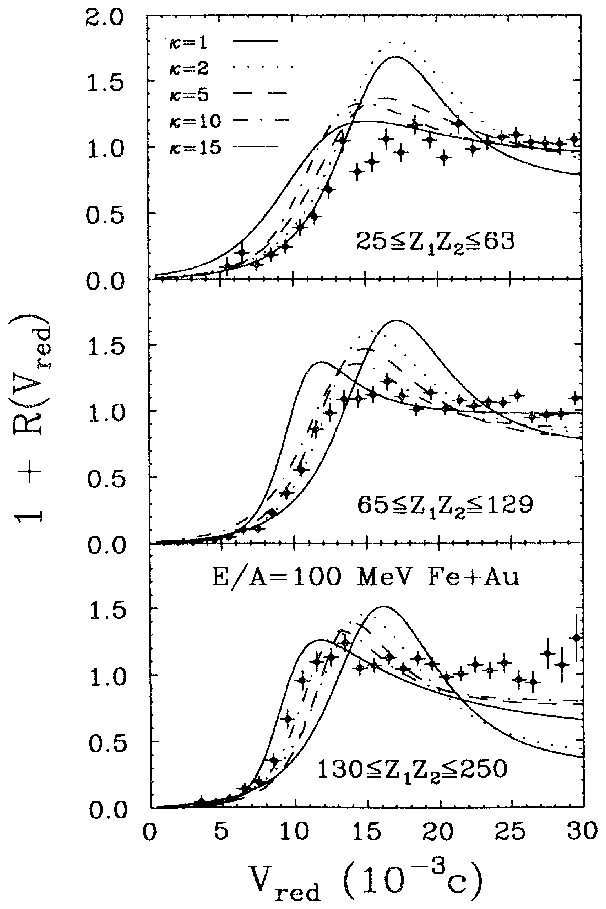}
\includegraphics[scale=0.65]{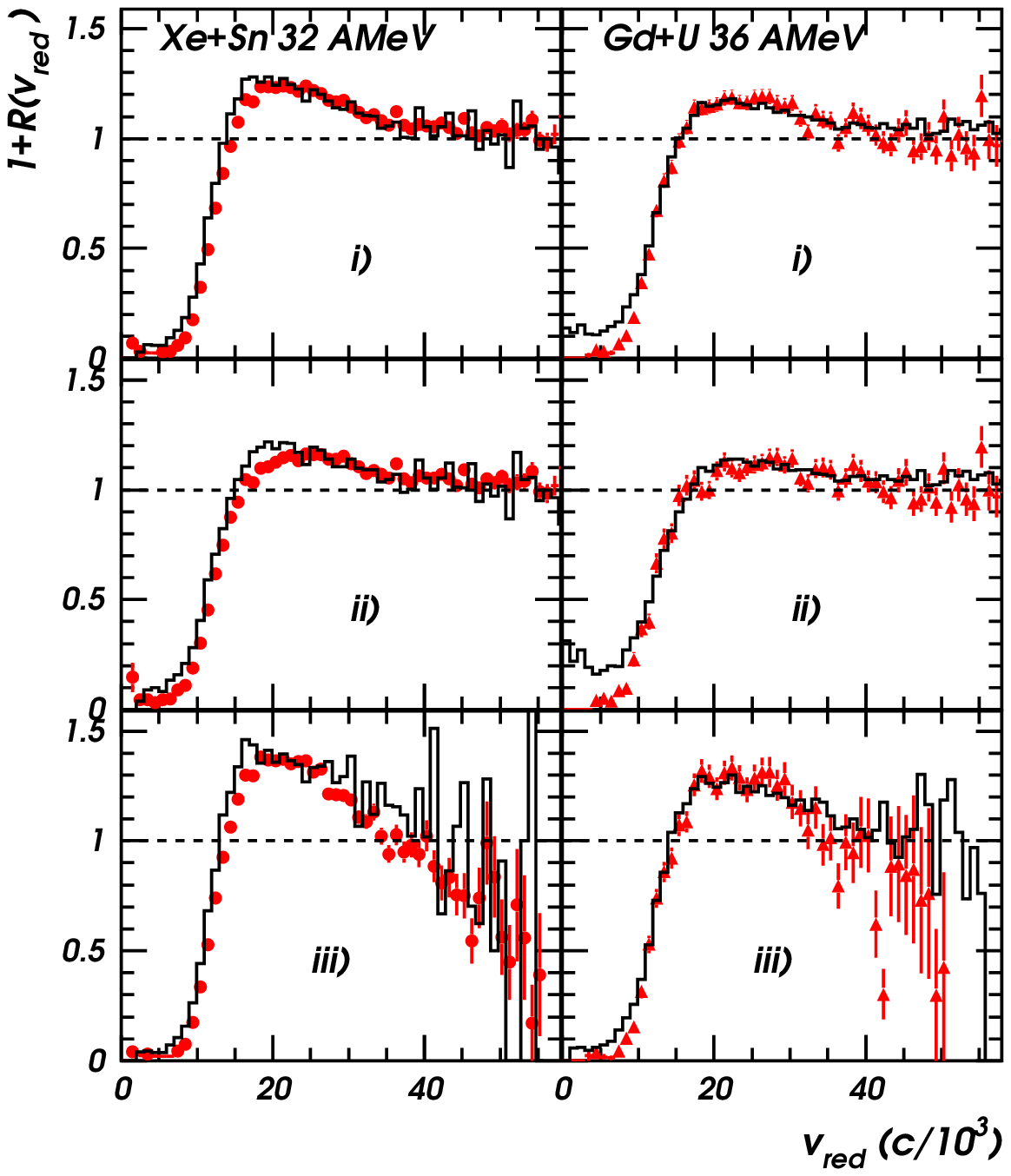}
\caption{Experimental reduced velocity correlation functions (symbols) for
central collisions in the reactions Fe+Au (left~\cite{San95}) and Xe+Sn and
Gd+U(right~\cite{I57-Tab05}) for different charges of the fragment pairs. 
The lines represent calculations from QMD+SMM and different freeze-out
volumes in the left panel, and from BOB in the right panel. See text for
details}\label{fig_corv}
\end{figure}
Provided that the models accounts sufficiently well for more simple
fragment properties, which was verified in both cases, a good reproduction
of the correlation functions (filtered by the experimental apparatus) 
allows to trust the freeze-out volume and the
multifragmentation time given by the dynamics. For the Fe+Au reaction,
thermal multifragmentation of the quasi-target was shown to be the dominant
mechanism. The freeze-out volume is given by the SMM parameter $\kappa$. 
The overall agreement between data and calculations is not very 
good (fig.~\ref{fig_corv} left),
nevertheless the rising part after the Coulomb hole allows to derive
freeze-out volumes which increase between 2 and 10V$_0$ (or alternatively
shorter emission times) with the charge of the fragment pair. 
Times shorter than 500~fm/$c$ are indicated. That small fragments are 
emitted from a hotter and more dense source was also suggested 
in~\cite{Wan99} from large angle correlation results. In the case of
reactions between heavier ions near the Fermi energy, where the formation of
compact single sources was evidenced, the BOB+SIMON model well reproduces
the three types of pair selections, with no cuts on the fragment
energy/velocity: \\
i) all fragments considered ($Z_{\mathrm{i,j}} \geq 5$); \\
ii) intermediate mass fragments (IMF) 
$5 \leq Z_{\mathrm{i,j}} \leq 20$; \\
iii) reduced velocity correlation of the heaviest fragment
$Z_{max}$ with each of the others in the event $Z_{\mathrm i} < Z_{max}$. \\
Note that Li and Be are not considered in that case. The almost perfect
matching of  the calculated correlation functions with the data, particularly
for Gd+U (fig.~\ref{fig_corv} right), led the authors of ref~\cite{I57-Tab05}
to extract average freeze-out volumes (with a definition close to the SMM one)
around 4V$_0$ for Xe+Sn and 8V$_0$ 
for Gd+U. The freeze-out instant was defined as that when the average
fragment multiplicity becomes frozen~\cite{Par05}, namely 200 (240)~fm/$c$ 
after the incident partners come to contact for Xe+Sn (Gd+U).
The freeze-out topologies found in the simulations imply that the heavier
fragments are located closer to the center, as shown in fig.~\ref{posfrag}.
\begin{figure}
\includegraphics{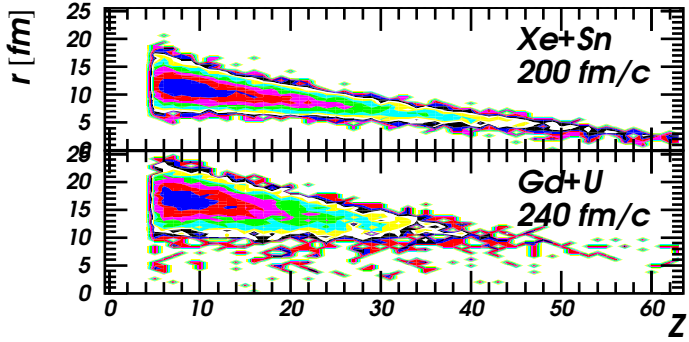}
\caption{Fragment positions as a function of the atomic number at the
freeze-out instant for 32~\AM{} $^{129}$Xe+$^{119}$Sn  - upper panel, 
and 36~\AM{} $^{155}$Gd+$^{238}$U  - lower panel. The contour scale is
logarithmic. Adapted from~\cite{I57-Tab05}} \label{posfrag}
\end{figure}
Note that in $^3$He induced reactions it was conversely found that the 
heavier fragment should be randomly positioned in a dilute source in
order to well account for the experimental correlation functions with 
a trajectory calculation\cite{Wan99}. This observation may sign a 
difference in the fragmentation process whether it is thermally driven 
as it seems for QP and hadron induced reactions or it follows a 
compression phase as in very heavy-ion central collisions. In that case 
self similar expansion should favour coalescence of primary fragments 
close to the centre of mass.

\subsection{Fragment-particle correlations and fragment excitation at
freeze-out} \label{Corfp}
\begin{figure}
\includegraphics[scale=0.65]{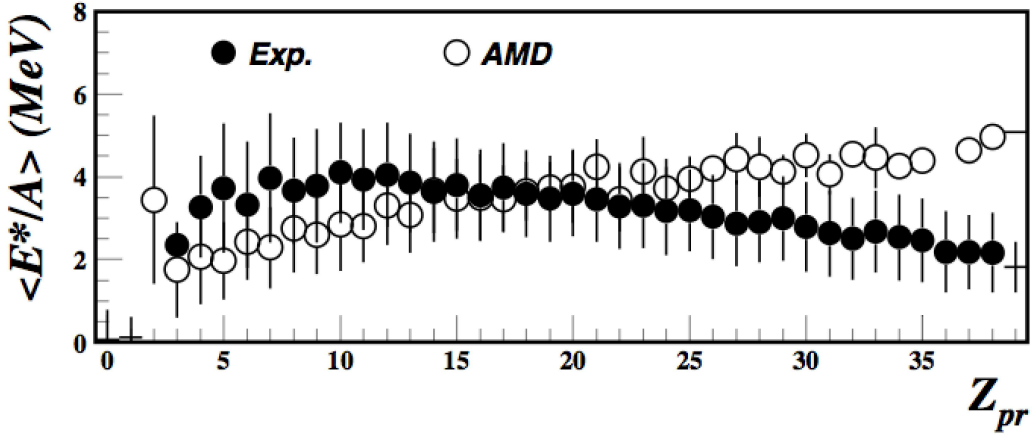}
\caption{Average excitation energy of primary fragments as a function 
of the atomic number
for 50~\AM{} Xe+Sn central collisions. Full points refer to experimental
evaluations using fragment-particle correlations and open points to AMD
simulations. From~\cite{T32Hud01}} \label{Exfrag}
\end{figure}

Excitation energy measurements of primary fragments which are present 
at the freeze-out
stage are also of large interest. They can provide information on the
degree of equilibration at freeze-out and put strong constraints on
freeze-out characteristics through a microcanonical description of
multifragmentation events. Such measurements are derived from
fragment-particle correlations. By using the correlation technique for the
relative velocity between light charged particles and fragments, one can
extract the multiplicities and average kinetic energies of particles 
emitted by fragments with a given final charge, and then reconstruct 
the sizes  and excitation energies of the
primary fragments. This technique, first proposed in~\cite{I11-Mar98} for
Xn+Sn central collisions at 50~\AM{} incident energy, was  applied from 
32 to 50~\AM{}~\cite{I39-Hud03} and for Kr+Nb central collisions at
45~\AM{}~\cite{Sta01}. For each bombarding energy, a constant value of 
the mean excitation energy per nucleon has been found over a primary
fragment charge range 5-20, which strongly suggests that fragments are 
on average in thermal equilibrium at freeze-out. This average excitation 
energy, equal to 2.2~\AM{} at the lower incident energy, 
saturates around 2.5-3.5~\AM{} 
for beam energies 39~\AM{} and above. Note that dynamical simulations
(AMD, BOB) performed for Xe+Sn central collisions at similar incident
energies~\cite{I29-Fra01,T32Hud01} predict average values close to 
3~\AM{} for fragment excitation at freeze-out. However a detailed 
comparison between data and AMD simulations, displayed in 
fig.~\ref{Exfrag}, shows different evolutions with fragment charge; 
the increase of the fragment excitation energy per nucleon with its 
charge  is a general trend of AMD
results in the Fermi energy region (see also~\cite{Wad04}). 
Going more into the details of correlation results, one learns that
the corresponding secondary evaporated light
charged particles represent less than 40\% of all produced charged particles
and decreases down to 20\% for 45-50~\AM{} incident energies. Finally
experimental data are better reproduced with secondary 
deexcitation simulations assuming that the fragment N/Z ratio at 
freeze-out is the same as that for the combined projectile-target system.

\subsection{Break-up densities and freeze-out volumes} \label{Fov}
From an experimental point of view the location of nuclear
multifragmentation data in the phase diagram requires accurate 
independent measurements of temperature and density at the break-up 
stage. While the problem of temperature has been solved with acceptable 
accuracy using He-Li isotope ratios up to 5-6~MeV ~\cite{I46-Bor02}, 
no accurate enough method is available to determine the spatial 
extension of mononuclear systems which undergo multifragmentation.  
Therefore estimates of break-up densities have been obtained using 
various approaches. For instance,
break-up densities for spectator fragmentation in Au+Au collisions at 
1000~\AM{} were estimated by using selected particle-particle 
correlations (particles from secondary decays are excluded by imposing 
an energy threshold)~\cite{Frit99}. Assuming a zero lifetime,
the volumes of spectator sources were extracted 
and densities calculated by dividing the number of spectator 
constituents by the source volume. The estimated average values slowly 
decrease from about 0.3 to 0.2~$\rho_0$ when excitation energies of 
spectators move from 4 to 10~\AM{}. Caloric curves were  analyzed 
within the framework of an expanding Fermi gas hypothesis 
to extract estimates of break-up densities~\cite{Natow02}.
In this approach the observed flattening of the caloric curves reflects 
an increasing expansion with increasing excitation energy (the product
$T$x$\rho^{-2/3}$ is a constant within a isoentropic hypothesis). 
For nuclei of medium to heavy mass, the derived density values vary 
from 0.7-0.6 to 0.4~$\rho_0$ when excitation energies increase from 4 
to 6-8~\AM{}. Starting from the definition of freeze-out given at the 
beginning of that section, one can compare break-up density/volume to 
freeze-out volume obtained from dynamical simulations as BOB. 
Taking the case 
of Xe+Sn central collisions at 32~\AM{} (source excitation energy 
close to 6~\AM{}), break-up density coming from simulations is 
0.41$\rho_0$ or 2.4$V_0$ and the corresponding average freeze-out 
volume $V_{F.O.}^{SMM}$ is 2.8$V_0$ with the definition
of SMM (which corresponds to a sphere including all the centres of primary
fragments) and 6.1$V_0$ for $V_{F.O.}^{sph.}$ a sphere including all
fragments~\cite{I28-Fra01,I57-Tab05}.
As we see $V_{F.O.}^{SMM}$ well compares, at least for that system size, 
to the break-up volume.\\
Average freeze-out volumes have been extracted first
from comparisons with statistical models like SMM as shown for some 
examples in subsection 5.2. 
In general, experimental results can be well reproduced over a large 
range of excitation energy (from 2 to 10, 12~\AM{}), by keeping the same
$V_{F.O.}^{SMM}$ typically 3 or 6$V_0$.
Recently a first attempt  of estimating the freeze-out volume in a 
fully consistent way was done, by employing a simulation 
built event by event from all the available asymptotic experimental
information (charged particle spectra, average and standard deviation of 
fragment velocity spectra and calorimetry)~\cite{I58-Pia05}. 
Dressed excited fragments, which statistically deexcite, and particles 
at freeze-out are described by spheres
at normal density. Four free parameters are used to recover the data: 
the percentage of measured particles which were  evaporated from primary 
fragments, the collective radial energy, a minimum distance between 
the surfaces of products at freeze-out and a
limiting temperature for fragments (see ~\cite{I58-Pia05} for more details).
For Xe+Sn data (32~\AM{}, central collisions) an average
volume $V_{F.O.}^{sph.}$=7.6$\pm$2.0$V_0$ was derived. However
the deduced average excitation energy for primary
fragments is close to 4~\AM{}, significantly higher that the 
value of 2.2~\AM{} evaluated from fragment-particle correlations 
(see previous subsection). Note that the use of the widths 
of fragment velocity spectra in the comparison between data and 
simulation has shown that the introduction of a limiting temperature 
for fragments seems mandatory. The strong sensitivity 
of the freeze-out volume to limiting temperature values was also 
demonstrated in the microcanonical multifragmentation model MMM as far 
as average properties of fragments are used to put constraints on 
volumes~\cite{Radu05}.\\
In conclusion we can say that working hypotheses
and approximations are used to give semi-quantitative information on
average break-up densities or freeze-out volumes. Values remaining 
constant around 0.2-0.3$\rho_0$ or slowly decreasing from 
0.6 to 0.3$\rho_0$  over the excitation energy range 3-12~\AM{} are 
derived. The restoration of freeze-out stage using all the asymptotic 
experimental information appears as promising.

\section{Finite systems and phase transitions}\label{FinitePT}

Phase transitions are universal properties of interacting matter and
traditionally they have been studied in the thermodynamical limit of
infinite systems. A phase transition occurs when a phase becomes unstable 
in given thermodynamical conditions described with intensive variables 
like temperature, pressure \ldots. However in physical situations, as the
one encountered in the present studies concerning isolated finite systems 
like nuclei, the concept of thermodynamical limit can not apply. Extensive
variables like energy and entropy are no more additive due to the 
important role played by the surfaces of particles and fragments
which are produced. The entropy of the surfaces which separate the 
coexisting phases does not scale with the size of the system. The entropy
per particle at equilibrium s=S(E)/N shows a convexity  with a depth
proportional to $N^{-1/3}$ which is suppressed at the thermodynamical limit.
Consequently phase
transitions should be reconsidered from a more general point of view.
An important theoretical effort started a few years ago
to propose and discuss signatures of phase transitions in small
systems~\cite{Gro01,Gro95,Gro97,Gul99,Cho99,Gro00,Bor00,Cho00,Bot00,Jel00,%
Mul01,Cam01,Cho01,Car01,%
Radut01,Radut02,Mor02,Gulm04,Rad03,Cho03,Gul03,Cho04,Mor04,Das05,%
Cam05,Gulm05,Gul05,More05,De06}.
``Small''
systems have been defined in reference~\cite{Gro00} as  systems where the
linear dimension is of the same order of magnitude as
 the characteristic range of the interaction. This is for example 
the case for atomic clusters but also for astrophysical systems,
due to the long range of gravitation. Such systems are also classified as
non-extensive systems.\\
This section is divided into three parts. One is devoted to the possible 
dynamics of phase transition for hot unstable nuclei. In a second part we
shall discuss the thermostatistics involved and
the associated relevant signatures. Finally critical behaviours early
observed in multifragmentation data will be discussed in the context 
of finite systems.

\subsection{Dynamics and spinodal instability}
We have noticed in previous sections  experimental evidence
for a radial extra energy boost (radial expansion energy) associated to
multifragmentation products. It can be attributed either to a dominant 
compression-expansion phase in central nucleus-nucleus
collisions or to thermal pressure for more gentle collisions:
hadron-nucleus or semi-peripheral nucleus-nucleus collisions.
The system might then reach densities and temperatures that
correspond to spinodal instability  and clusterization would ensue as the
system seeks to separate into the corresponding coexisting liquid and gas
phases.\\
In the last fifteen years major theoretical progresses have been realized 
to understand and learn about spinodal decomposition in the nuclear 
context and a review can be found in reference~\cite{Cho04}.
We shall first briefly discuss what are the specificities of spinodal
decomposition as far as infinite nuclear matter is concerned.
\begin{figure}[htb]
\begin{minipage}[t]{0.45\textwidth}
\includegraphics[width=\textwidth]{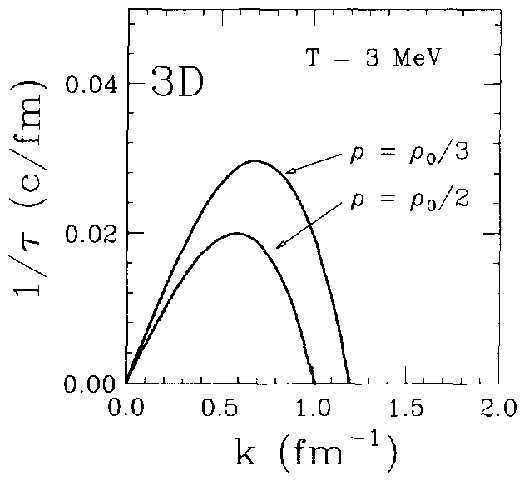}
\caption{Nuclear matter dispersion relation at 3~MeV temperature for
two different densities; $\rho_{0}$ is the normal density.
(from~\protect\cite{Colo97}).} \label{disper-relation}
\end{minipage}%
\hspace*{0.05\textwidth}
\begin{minipage}[t]{0.5\textwidth}
\includegraphics[width=1.1\textwidth]{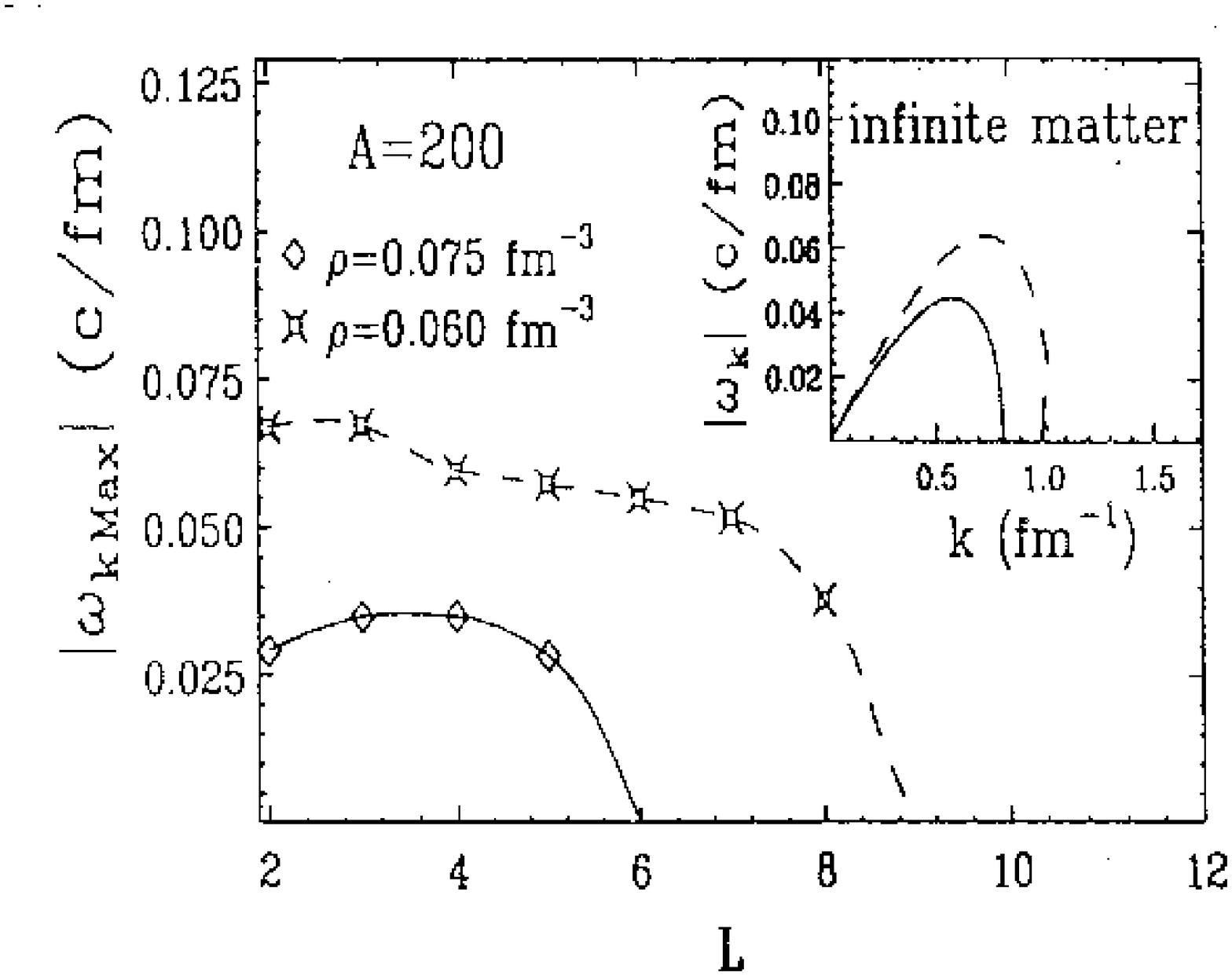}
\caption{Growth rates of the most unstable modes for a spherical source 
with 200 nucleons as a function of the multipolarity $L$ and for
two different central densities.
(from~\protect\cite{Jac96}).} \label{disper-relationfini}
\end{minipage}
\end{figure}
Associated to negative compressibility the mechanically unstable spinodal
region can be investigated by studying the propagation of small density
fluctuations~\cite{Colo97,Ayi95}. By analogy with optics,
the nuclear dispersion relation can be calculated  for different 
conditions of temperature and density
by introducing the Boltzmann-Langevin equation (see subsection~\ref{BOB}).
Within the linear response theory framework, if one expands the solution
of the Boltzmann-Langevin equation as $ f = f_0 + \delta f$, where $f_0$ is
a solution of the Boltzmann equation and $\delta f$ the fluctuating part,
one finds the equation of motion
$\partial \delta f / \partial t = -i\mathcal{M} \delta f + \delta I[f_0]$
at the leading order in $\delta f$. The extended RPA matrix $\mathcal{M}$
represents the combined action of the effective field and of the average
collision term.
In the spinodal region the eigenvalues of the matrix $\mathcal{M}$
become imaginary. Consequently the fluctuations associated with a given
eigenmode agitated by the source term $\delta I$ are exponentially 
amplified or suppressed, depending upon the sign of the imaginary 
part of the frequency. In the case of infinite nuclear matter the 
eigenmodes of the linearized
dynamics are plane waves, characterized by a wave number $k$ and
 an imaginary eigenfrequency, which is the inverse of the instability 
 growth time.
Figure~\ref{disper-relation} presents an example of nuclear dispersion
relation at 3~MeV temperature for two different densities $\rho_{0}/2$ 
and $\rho_{0}/3$. Imaginary RPA frequencies are reported as a function 
of the wawenumber k of the considered perturbation. This dispersion 
relation exhibits a strong maximum at a given wave number followed by 
a cut-off at large k
values. This cut-off reflects the fact that fluctuations with wavelength
smaller than the range of the force can not be amplified. The most 
unstable modes correspond to wavelengths lying around 
$\lambda \approx$~10 fm
and the associated characteristic times are almost identical, 
around 30-50~fm/$c$, depending on density ($\rho_{0}/2$-$\rho_{0}/8)$ 
and temperature (0-9~MeV)~\cite{Colo97,Idi94}.
A direct consequence of the dispersion relation is the 
production of ``primitive'' fragments with size $\lambda/2$ $\approx$ 5~fm 
which correspond to Z $\approx$ 10. However this simple and rather 
academic picture is expected to
be largely blurred by several effects. We do not have a single unstable
mode and consequently the beating of different modes occurs. 
Coalescence effects due to the residual
interaction between fragments before the complete disassembly are also
expected~\cite{Colo97} and finally primary fragments deexcite by 
secondary decay.\\
Does the signal discussed for nuclear matter survive (in final fragment
partitions experimentally measured) if we consider the
 case of a hot expanding nucleus which undergoes multifragmentation? 
 First of all, the system produced by the collision has to stay long 
 enough in the spinodal region ($\approx$3
characteristic time: 100-150~fm/$c$) to allow an important amplification 
of the initial fluctuations. Secondly, the presence of a surface
introduces an explicit breaking of the translational symmetry.
Figure~\ref{disper-relationfini}
shows the growth rates of the most unstable modes for a spherical source
of A=200 with a Fermi
shape profile and for two different central densities~\cite{Jac96}.
The growth rates are nearly the same for different multipolarities $L$
up to a maximum multipolarity $L_{max}$ (see also~\cite{Nor00}).
This result indicates that the unstable
finite system breaks into different channels.
Depending on multipolarity $L$, equal-sized ``primitive''
fragments are expected to be produced with sizes in the range
$A_{F}/2$-$A_{F}/L_{max}$;
$A_{F}$ being the part of the system leading to fragments during the
spinodal decomposition.
One can also note that the Coulomb potential has a very small effect on the
growth rates of unstable collective modes except close to the border of the
spinodal zone where it stabilizes very long wave-length unstable
modes~\cite{JacT96}.
On the other hand, for a finite system, Coulomb
interaction reduces
the freeze-out time and enhances the chance to keep a memory of the dynamical
instabilities; a similar comment can be made as far as collective expansion 
is concerned. Both effects push away the ``primitive'' fragments 
and reduce the time of their mutual interaction. So finally, even if
expected strongly
reduced, the presence of partitions with nearly equal-sized fragments is a
good candidate to sign the role of spinodal instability in
multifragmentation.\\
Following early studies related to nearly equal-sized fragment 
partitions~\cite{Bru92},
ten years ago a method called higher order charge
correlations~\cite{Mor96} was proposed to enlighten any extra production 
of events with specific fragment partitions. The high sensitivity 
of the method makes it particularly appropriate to look for small numbers 
of events, as those
expected to have kept a memory of spinodal decomposition properties.
Thus the charge correlation method allows to examine model independent
signatures that would indicate a preferred decay into a number of 
equal-sized fragments in events from experimental data or from 
simulations. All fragments of one event with fragment multiplicity 
$M_f$ = $M = \sum_Z n_Z$, where $n_Z$ is the number of fragments
with charge $Z$ in the partition,  are taken into account. By means of the
normalized first, $\Zmoy$, and second order, $\sigma_Z^2$,
moments of the fragment charge distribution in the event, 
one may define the higher order charge correlation function as:
$\left. 1+R(\sigma_Z, \Zmoy)=(Y(\sigma_Z, \Zmoy)/Y'(\sigma_Z, \Zmoy)) 
\right|_{M}$. 
Here, the numerator $Y(\sigma_Z, \Zmoy)$ is the yield of events with 
given $\Zmoy$  and $\sigma_Z$ values. The denominator $Y'(\sigma_Z, \Zmoy)$, 
which represents the uncorrelated yield
of pseudo-events, can be built in different ways; details and discussions
can be found in~\cite{I40-Tab03,I41-Cha03,Bor06}. The exact identification
of fragments up to at least Z=20 is mandatory to use that method.
\begin{figure*}[htb]
\includegraphics[width=0.5\textwidth]{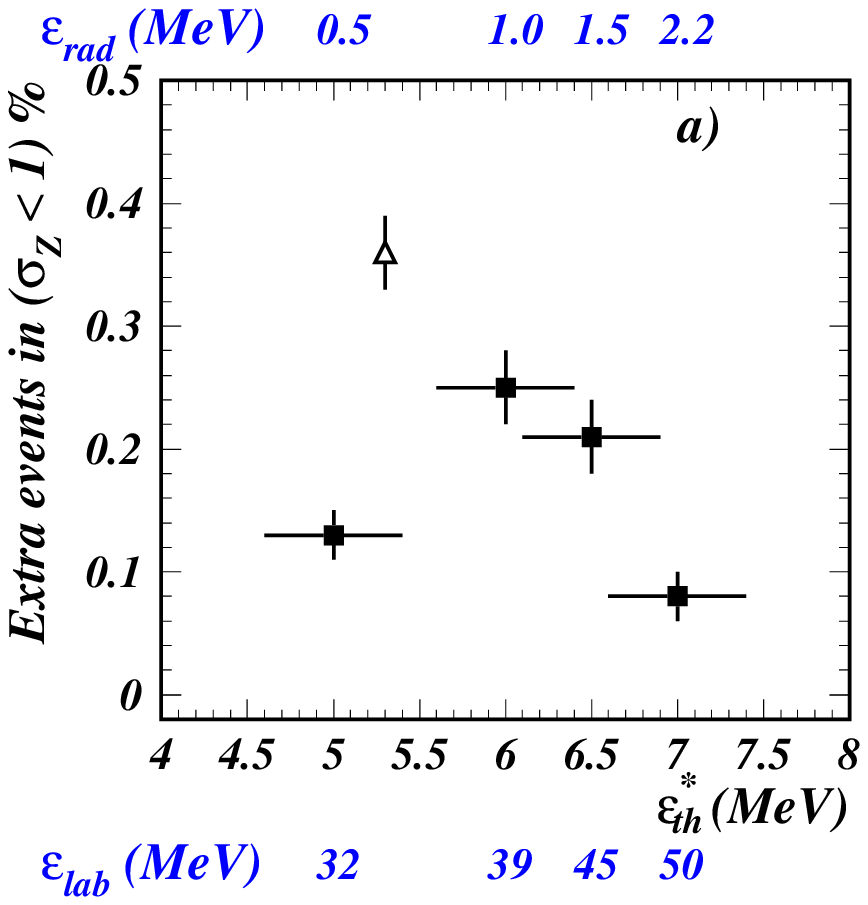}%
\includegraphics[width=0.5\textwidth]{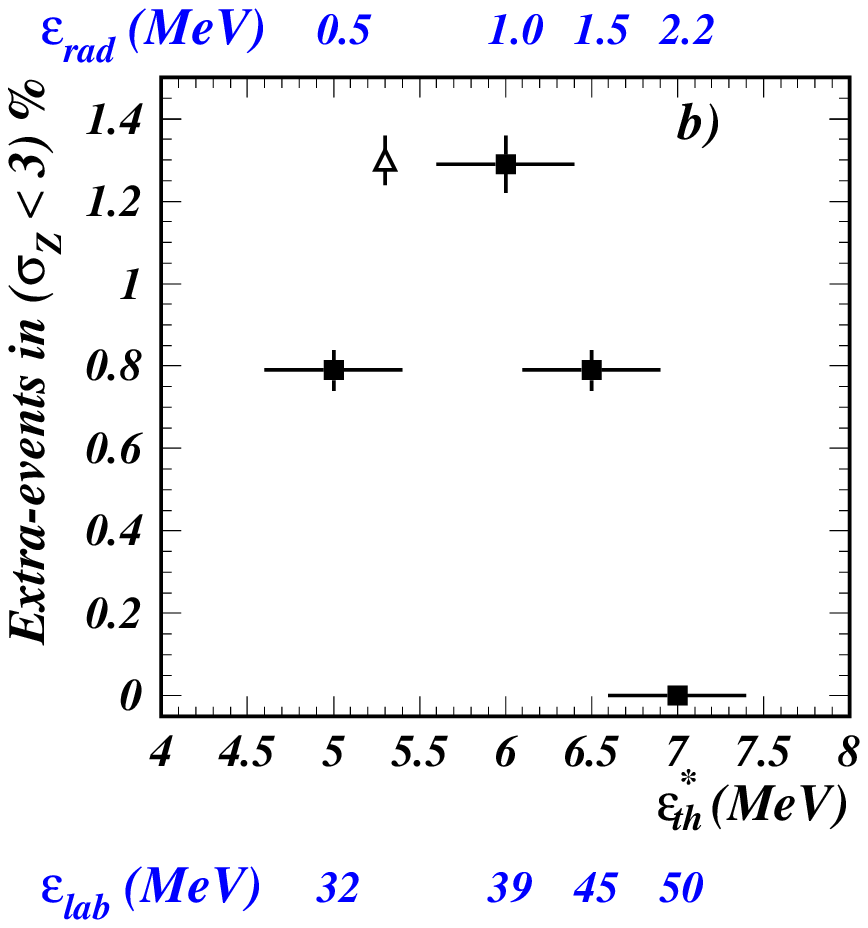}
\caption{Extra production of events with nearly equal-sized fragments
(a: $\sigma_{Z} < 1$ and b: $\sigma_{Z} < 3$) as a function of thermal
excitation energy (full points); the incident and radial
energy scales are also indicated. $\epsilon_{th}^*$ and $\epsilon_{rad}$ 
are deduced from comparisons with SMM. The open point refers to the 
result from BOB simulations; the average thermal excitation energy is 
used. Vertical bars correspond to statistical errors and
horizontal bars refer to estimated uncertainties on the backtraced
quantity, $\epsilon_{th}^*$. (from\cite{I40-Tab03}).}
\label{mfc}
\end{figure*}
It was applied on different data sets corresponding to compact single 
sources produced in central collisions in the incident energy range 
30-50~\AM{}~\cite{Bor06}. For the $^{129}Xe$+ $^{nat}Sn$
system~\cite{I40-Tab03,I31-Bor01} a rise
and fall of the percentage of extra production of events with nearly
equal-sized fragments was observed. The average-charge domains contributing
to the correlation peaks are 12-21 and come mainly, as theoretically 
expected, from fragment multiplicity 3 (dominance of the octupole 
mode)~\cite{Col02}. Figure~\ref{mfc} illustrates the results for two 
different limits on $\sigma_{Z}$; the lower limit corresponds to the
spread estimated as coming from secondary fragment evaporation and the 
higher limit to the observed spread for the correlation function. 
Results obtained 
with BOB simulations (see section~\ref{models}) are also reported. 
For the considered system, incident energies around 35-40~\AM{} appear
as the most favourable to induce spinodal decomposition; it corresponds 
to about 5.5-6~\AM{} thermal excitation energy 
associated to a very gentle expansion energy around 0.5-1~\AM{}. The 
qualitative explanation for those numbers can be well understood in 
terms of a necessary compromise between two time scales.
On one hand the fused systems have to stay in the spinodal region 
$\approx$ 100-150~fm/$c$~\cite{Colo97,Idi94,Nor02}, to allow an 
important amplification 
of the initial fluctuations and thus permit spinodal decomposition; 
this requires a not too high incident energy, high enough however
for multifragmentation to occur. On the other hand, for a
finite system, Coulomb interaction and collective expansion push 
the ``primitive'' fragments apart and reduce the time of their mutual 
interaction, which is efficient to keep a memory of ``primitive'' size 
properties.\\ 
At present confidence levels around 3-4 $\sigma$, observed for charge
correlation peaks, prevent any definitive conclusion. To firmly assess 
or not the validity of this fossil signal new experiments with higher 
statistics are required.\\
If this fossil signal is definitively confirmed the following comments 
have to be done. As we have shown in fig.~\ref{smmexp}, good average 
statistical descriptions of data using for example SMM
are also obtained, which could demonstrate that the dynamics involved 
for finite systems is sufficiently chaotic to finally explore enough 
of the phase space in order
to describe  fragment production through a statistical approach.
It is tempting to associate a part of that chaos with the
coalescence stage which can occur during fragment formation.
Indeed, as shown previously (sections~\ref{frag} and~\ref{FO}),
the peculiar size and position of the largest fragment
could be well understood as resulting from mode beating (size), 
large-wavelength instabilities (size) and the late-stage coalescence 
related to the involved bulk nuclear density (size and position).

\subsection{Spinodal instability and Thermostatistics}
\begin{figure}[htb]
\begin{center}
\includegraphics[width=8cm]{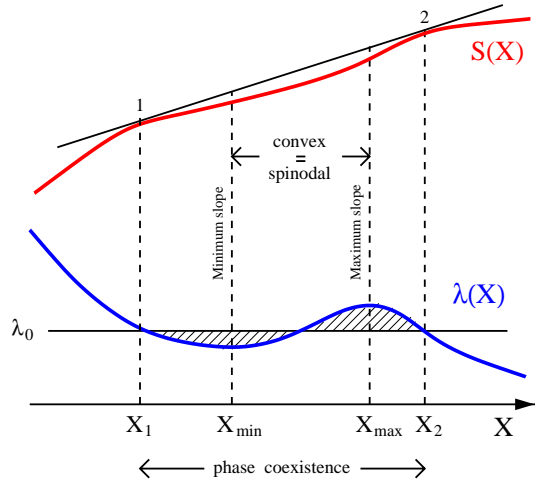}
\end{center}
\caption{
Phase coexistence and spinodal instability. The entropy function $S(X)$ 
is convex between the inflection points $X_{min}$ and $X_{max}$ where
the intensive conjugate variable 
$\lambda(X)$=$\delta(S)$/$\delta(X)$
has a local minimum or maximum, respectively. The infinite system is 
mechanically unstable and entropy is gained by separating the system 
into two phases. The entropy of the corresponding mixed system is 
additive and moves along the common tangent as $X$ increases from $X_1$ 
to $X_2$; the intensive conjugate variable is determined by the 
Maxwell construction requiring that the two hatched
areas are equal.(from~\protect\cite{Cho04}). \label{entrospino}}
\end{figure}

Spinodal instability is intimately related to the occurrence of a first order
phase transition as signaled by a convex anomaly in the entropy function 
$S(X)$. For pedagogical purpose we consider in the following only one
single extensive variable $X$ like energy, volume or number of particles.
Entropy can  be gained by separating the uniform system into two phases 
at equilibrium: it is the spinodal decomposition well known at
the thermodynamical limit for binary solutions and binary alloys.
Figure~\ref{entrospino} illustrates the relationship between entropy 
convexity, spinodal instability and phase coexistence.
\begin{figure}[htb]
\begin{minipage}[t]{0.47\textwidth}
\includegraphics[width=\textwidth]{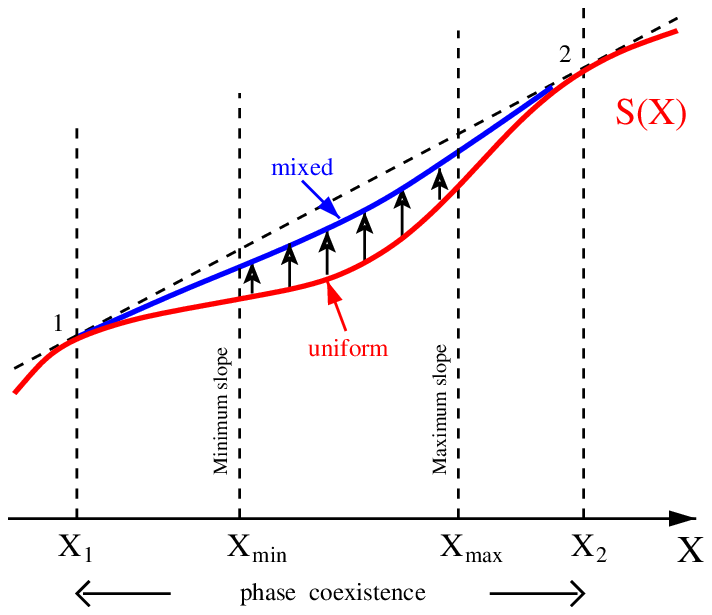}
\caption{
Isolated finite systems. The entropy function for a uniform unstable 
system (lower curve) has a local convexity region and the system gains
entropy reorganizing itself into a mixture of the two phases. The 
resulting equilibrium entropy function (upper curve) will always lie 
below the common tangent (dashed line) (from~\protect\cite{Cho04}). 
\label{convex}}
\end{minipage}%
\hspace*{0.05\textwidth}
\begin{minipage}[t]{0.47\textwidth}
\includegraphics[width=\textwidth]{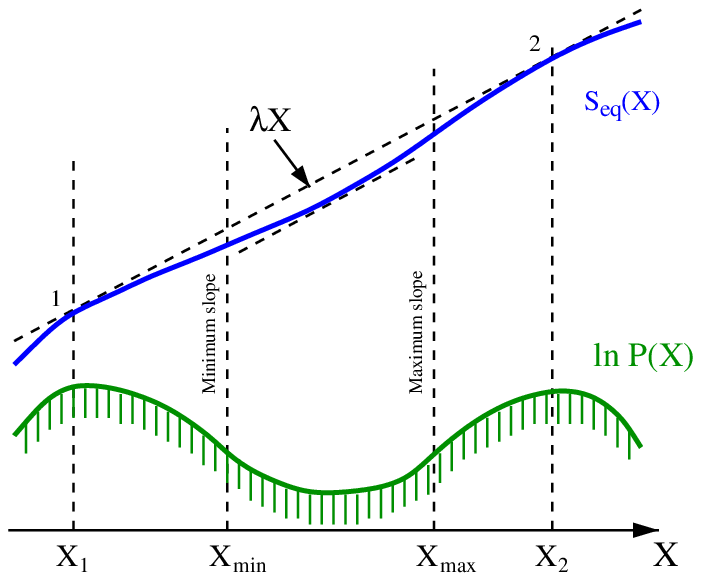}
\caption{
Canonical ensemble of finite systems. The bimodal equilibrium 
distribution is given by $P(X)$$\sim$exp($S(X)$-$\lambda X$). The figure 
shows the case when the Lagrange multiplier $\lambda$  is equal to the 
slope of the common tangent (from~\protect\cite{Cho04}). \label{bimo}}
\end{minipage}%
\end{figure}
Two signatures of a phase transition for hot nuclei
can be added to the previous one (observation of nearly equal-sized
fragments), just starting from the fact that surfaces are no 
more negligible for finite systems. The entropy of the combined system
constituted by the two coexisting phases is not a sum of the individual
subsystem entropies. The uniform unstable finite system
gains entropy by reorganizing itself into a mixture of the two phases
but the resulting equilibrium entropy function keeps a local convexity.
The Maxwell construction is no more valid as illustrated in
fig.~\ref{convex}.\\
The two main consequences of that local convexity are the following.
Using a microcanonical sampling 
(fixed values of the extensive variable E which can be rather well
estimated from experiments)
the first derivative corresponds to the inverse microcanonical temperature
and the bottom curve of fig.~\ref{entrospino} is the inverse of a
caloric curve which presents a backbending. The energetic 
cost paid to produce surfaces is a
decrease of the microcanonical temperature when the excitation energy
increases and
consequently the appearance of a negative microcanonical heat 
capacity~\cite{Gro97,Gro01} as 
a specific signature of a first order phase
transition for finite systems. 
Moreover if a signature of spinodal instability is observed  
one must measure
correlatively a negative microcanonical heat capacity related to the
resulting equilibrium entropy function with local convexity. The reverse 
is not true if spinodal instabilities are not responsible for fragment 
formation. Experimental
observations of correlated signals will be discussed in the next section.\\
If one considers now a finite system  in contact with a
reservoir (canonical sampling), the value of X may fluctuate as the system
explores the phase space; the associated distribution at equilibrium
is $P(X)$$\sim$exp($S(X)$-$\lambda X$). The distribution of $X$ 
acquires a bimodal
character (see fig.~\ref{bimo}): that bimodality is a second specific
signature of a first order phase transition for finite systems.
It gives an
understanding of coexistence as a bimodality of the event distribution, each
component being a phase.
We recall that for infinite systems
two delta function distributions are localized in $X_1$ and $X_2$ and
correspond to spinodal decomposition. The biggest fragment detected in each
multifragmentation event, because of its correlation with the density
which is the natural order parameter of the
liquid-gas phase transition, is a good candidate as a potential order
parameter. A priori that second specific signature can be directly
observed from experiments and could appear as robust. A
difficulty comes however from the absence
of a true canonical sorting in the data. The statistical ensembles    
produced by selecting for example fused systems are
neither canonical nor microcanonical and should be better described in 
terms of the Gaussian ensemble~\cite{Cha88}, which gives a continuous
interpolation between canonical and microcanonical ensembles. Very recently
a simple reweighting of the probabilities associated to each excitation 
energy bin for quasi-projectile events was proposed to allow the comparison
with the canonical ensemble~\cite{Gul07}. 
Signals of bimodality will be discussed and correlated to other
signals in the next section. \emph{Finally one can also
underline a very important consequence of bimodality: for finite systems
large fluctuations of ``order parameters'' are
expected to be observed in the middle of the coexistence region}.

\subsubsection{Caloric curves, negative microcanonical heat capacity and
abnormal energy fluctuations}
\begin{figure}[htb]
\begin{center}
\includegraphics[trim =0 0 137 174,clip,width=0.6\textwidth]{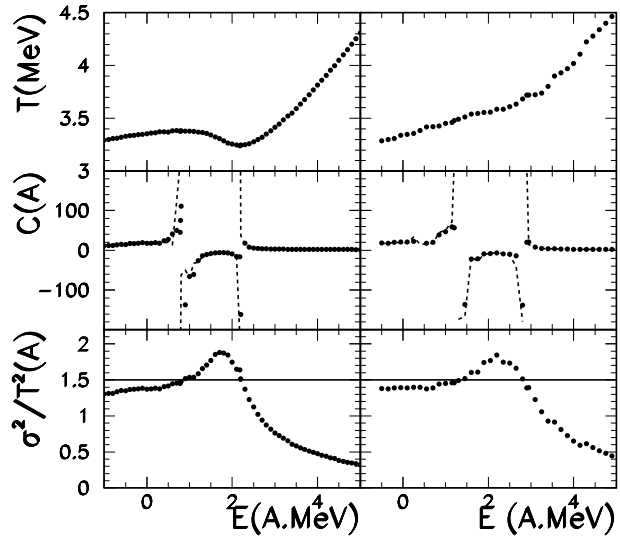}
\end{center}
\caption{Thermodynamic quantities in the microcanonical ensemble for 
a transformation at constant pressure (left part) and at constant 
volume (right part). Caloric curves are displayed in upper panels. 
Normalized kinetic energy fluctuations are compared to the canonical 
expectation (lines) in lower panels. In medium panels microcanonical 
heat capacities (symbols) are compared to 
the estimation (dashed lines) discussed in the text.
(from~\protect\cite{Cho00}). \label{MLGM}}
\end{figure}

\begin{figure}[htb]
\begin{center}
\includegraphics[trim=34 17 68 68,clip,width=0.8\textwidth]{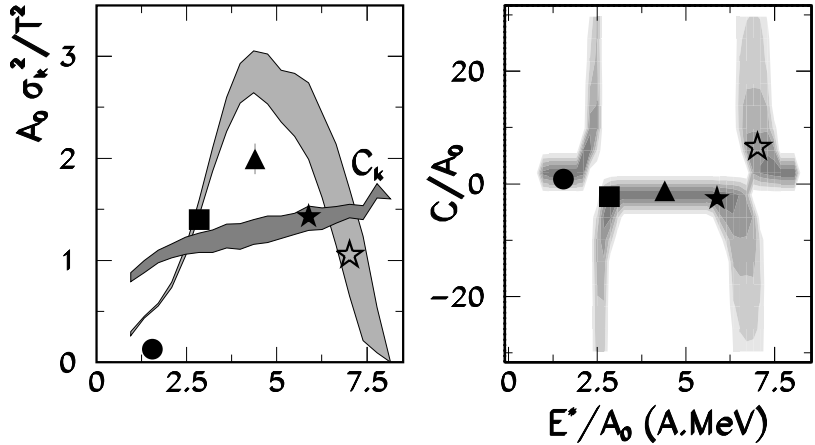}
\end{center}
\caption{Left panel: normalized kinetic energy fluctuations and estimated
canonical fluctuations ($C_k$) for quasi-projectile (QP) events produced 
in Au+Au collisions at 35~\AM{} (grey zones) and for fused systems 
produced in central Au+C (black dots), Au+Cu (squares, triangles) 
and Au+Au reactions before (open stars) and after subtraction of 1~\AM{} 
radial flow (black  stars). Right panel: corresponding microcanonical heat 
capacities per nucleon. (from~\protect\cite{MDA00}). \label{fig:b4}}
\end{figure}

The observation of a plateau in nuclear caloric curves was experimentally 
proposed as a
direct signature of a first-order phase transition~\cite{Poc95}. However,
from a theoretical point of view, a plateau-like shape can not be an
unambiguous signature even if it is a strong indication of a physical 
change and
if its observation can help to better define the energy domain of interest
for phase transition.
Caloric curves for restricted mass regions are presented in
fig.~\ref{calocurve}.
Indeed measured caloric curves can be
misleading because different curve shapes can be generated depending 
on the path
followed in the microcanonical equation of state surface. 
 As examples calculated caloric curves (microcanonical lattice gas 
model-216 particles)  at a constant pressure or a constant average 
volume~\cite{Cho00} are displayed in the upper part of fig.~\ref{MLGM}. 
At constant pressure a backbending is clearly seen whereas at constant 
average volume a smooth behaviour is observed showing that the phase 
transition signal can be hidden in the observation of caloric curves.
In experiments one does not explore a caloric curve at constant pressure 
nor at constant volume, the system follows a path in the excitation 
energy freeze-out volume plane.
For overcoming that situation it was proposed to directly measure the heat
capacity; if an experimental path is intersecting the spinodal region of 
the system phase diagram, it would be signaled by a negative value of the
microcanonical heat capacity.
A method for measuring microcanonical heat capacity using partial 
energy fluctuations was proposed ~\cite{Cho99,Gul00,Cho00}. Abnormal partial
energy fluctuations (as compared to the canonical expectation) are indeed
always seen, independently of the path, if microcanonical negative heat 
capacity is present. 
The prescription is based on the fact that
 for a given total energy of a system, the
average partial energy stored in a part of the system is a good 
microcanonical thermometer, while the associated fluctuations can be
used to construct the heat capacity.
From experiments the most simple decomposition of the total energy $E^{*}$
is in a kinetic part, $E_{k}$, and a potential part, $E_{p}$, 
(Coulomb energy + total mass excess).
However these quantities have to be determined at freeze-out  and
consequently it is necessary to trace back this
configuration on an event by event basis. As discussed in the previous 
section the fragment properties entirely rely on the representation 
of the system at the freeze out stage as non interacting fragments.
The true configuration needs  the knowledge of the freeze-out
volume and of all the particles evaporated from primary hot
fragments including the (undetected) neutrons. Consequently
some working hypotheses are used, possibly constrained by
specific experimental results~\cite{MDA02}.
Then, the experimental correlation between the kinetic energy per nucleon
$E_{k}$/$A$ and
the total excitation energy per nucleon $E^{*}$/$A$ of the considered
system can be
obtained as well as the variance of the kinetic energy
$\sigma_{k}^{2}$. Note that $E_{k}$ is calculated by subtracting 
the potential part $E_{p}$ from the total energy $E^{*}$ and consequently
kinetic energy fluctuations at freeze-out reflect the configurational 
energy fluctuations. An estimator of the
microcanonical temperature of the system can be obtained by inverting the
kinetic equation of state:
$$ < E_{k} > = < \sum_{i=1}^{M} a_i > T^2 + < \frac{3}{2} (M-1) > T $$
The brackets $<>$ indicate the average on events with the same $E^{*}$,
$a_i$ is the level density parameter and M the multiplicity at freeze-out.
An estimate of the total microcanonical heat capacity 
is extracted from the following equations:
\noindent $$ C_k = \frac{\delta <E_k / A >}{\delta T};
\hspace*{0.4cm} A\sigma_{k}^{2} \simeq T^2\frac{C_kC_p}{C_k+C_p};
\hspace*{0.4cm} (\frac{C}{A})_{micro} \simeq C_k+C_p \simeq \frac{C_k^2}
 {C_k - \frac{A \sigma_k^2}{T^2}} $$
The specific microcanonical heat capacity $(C/A)_{micro}$ becomes negative 
if the kinetic/configurational energy
fluctuations $A \sigma_k^2$ overcome  $C_kT^2$.
Figure~\ref{MLGM} (medium and lower panels) illustrates the results of 
such a procedure in the framework of a microcanonical lattice gas model. \\
That procedure was applied by the MULTICS and INDRA collaborations 
on different sources produced in peripheral and central
heavy ion collisions in the incident energy range 
30-80~\AM{}~\cite{MDA00,NLNBorm00,T34Gui02,I59-Bor04,MDA04,I61-Pic06,Gul06}.
Figure~\ref{fig:b4} summarizes the results obtained by the MULTICS
collaboration: on the left side it is seen that normalized kinetic 
energy fluctuations overcome the estimated canonical fluctuations, 
$C_k$; the right part of the figure illustrates the microcanonical 
negative heat capacities observed, the
distances between the poles being associated with the latent heat.\\
These results provide a direct evidence of a first order liquid-gas phase
transition.
They have to be seen as semi-quantitative and correspond to a starting
point for more precise measurements needed for the reconstruction of the
phase diagram. We recall again that the present protocol of measurements
is discussed in details in reference~\cite{MDA02}.
Exact microcanonical formulae, assuming  that a classical treatment 
of the motion of products emitted at freeze-out is appropriate, are also
proposed in~\cite{Radut02} to calculate heat capacity and second order
derivative of the entropy versus energy. They depend on the total kinetic
energy and number of emitted products which have to be estimated event by
event at freeze out. However up to now this method was not used to derive 
information from data.

\subsubsection{Microcanonical heat capacity and open questions}
Before concluding on this part one can indicate that theoretical questions
are still under debate and concern the topology of the system at 
freeze-out and the related order of the phase transition. If the system is
still relatively dense at freeze-out, which seems improbable if we refer to
the previous section, the fragment properties may be very different from the
ones asymptotically measured and the question arises whether the energetic
information measured on ground state properties can be taken for the freeze
out stage~\cite{Sat03}. Classical molecular dynamics calculations have shown
that the ground state Q-value is a very bad approximation of the interaction
energy of fragments in dense systems. This is due to the deformation of
fragments and to the interaction energy when fragment surfaces touch each
other. As a consequence, comparable kinetic energy fluctuations are obtained
in the subcritical and supercritical region of the Lennard-Jones phase
diagram~\cite{Cam05}. On the other hand calculations with a similar model,
the Lattice Gas model, show that even in the supercritical region the correct
fluctuation behaviour can be obtained if both the total energy and the
interaction energy are consistently estimated with the same approximate
algorithm as it is done in the experimental data analysis~\cite{Gulm05}.
Concerning now the order of the transition, recent calculations performed
within a mean-field framework (liquid drop model at finite temperature)
show that the phase transition is seen to be continuous (second order) 
and that a negative heat capacity for finite nuclei is not incompatible with
it~\cite{De06}.

\subsection{Scaling and fluctuations for fragment sizes}
In the early eighties multifragmentation was 
 connected to a
critical phenomenon~\cite{Fin82,Sie83,Cur83,Hir84,Cam84}. 
Percolation models~\cite{Cam84,Sta94,Cam00,Kle02}, the Fisher
Droplet Model~\cite{Fis67,Ell02,Ell03,Mor04,More05} and
more recently the theory of universal
fluctuations~\cite{Bot00,Bot02,Bot01,I51-Fra05,Gul05} have
been employed to describe fragment size distributions and fluctuations and
to tentatively derive information on the critical region of the liquid-gas
transition: in finite systems the critical point becomes a critical
region. The determinations of a scaling function 
and of a consistent set of critical exponents in different 
multifragmentation data also tend to support this 
hypothesis~\cite{Ell98,Sch01,MDA99,MDA03}. However the situation appeared 
as more complex. Indeed signals of a first order phase transition were 
also observed on the same set of
data~\cite{MDA99,MDA03,MDA00,I52-Riv05}, which opened a debate on the 
critical temperature extracted, on the order of the
transition and on the consequences of Coulomb and finite size effects 
on critical behaviours~\cite{Gul99,Gul02,Das02,Ell05,Mor05}.

\subsubsection{The Fisher Droplet Model}
The observation of critical behaviours like power
laws in the charge/mass distribution of multifragmenting systems has been
interpreted as an evidence of a second order phase transition. 
In the Fisher Droplet Model~\cite{Fis67} the vapor coexisting with a 
liquid in the mixed
phase of a liquid-gas phase transition is schematized as an ideal gas of
clusters, which appears as an approximation for non ideal fluids. The
model was recently applied to multifragmentation data~\cite{Ell02} by 
considering all
fragments but the largest as the gas phase, Z$_{max}$ being assimilated
to the liquid part. The yield of a fragment of mass A reads:\\
$\mathrm{d} N / \mathrm{d} A = \eta(A) = 
 q_{0}A^{-\tau} \exp((A\Delta\mu(T)-c_{0}(T)\varepsilon A^{\sigma})/T)$.\\
In this expression, $\tau$ and $\sigma$ are universal critical exponents, 
$\Delta \mu$
is the difference between the liquid and actual chemical potentials,
$c_0(T)\varepsilon A^{\sigma}$ is the surface free energy of a droplet of
size A, $c_0$ being the zero temperature surface energy coefficient;
$\varepsilon = (T_{c}-T)/T_{c}$ is the control parameter and describes 
the distance of the
actual to the critical temperature. At the critical point 
$\Delta\mu=0$ and surface energy vanishes: $\eta(A)$ follows a
power law. Away from the critical point 
but along the coexistence line  $\Delta\mu=0$ and the
cluster distribution is given by:
$\mathrm{d} N / \mathrm{d} A = \eta(A) = 
 q_{0}A^{-\tau} \exp((-c_{0}(T)\varepsilon A^{\sigma})/T)$.\\
\begin{figure}[htp]
\includegraphics[scale=0.65]{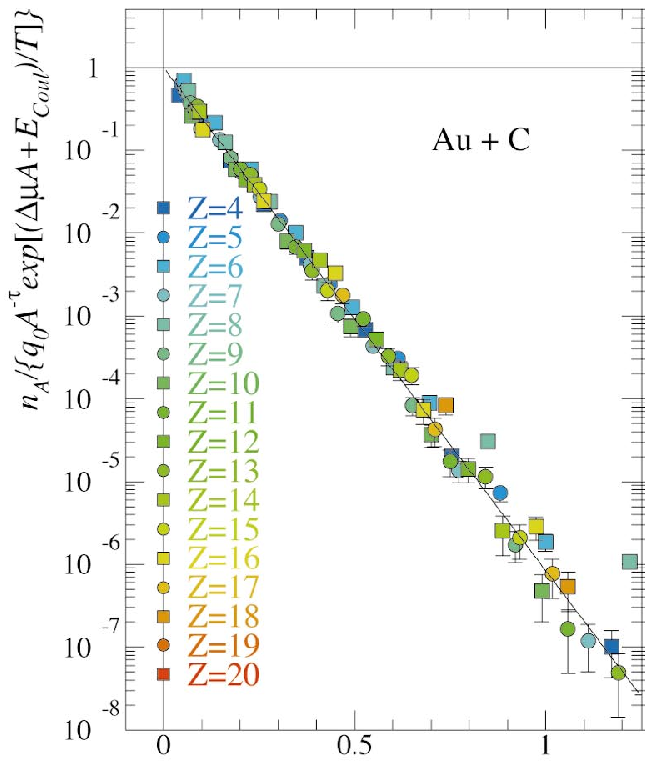} 
\includegraphics[scale=0.65]{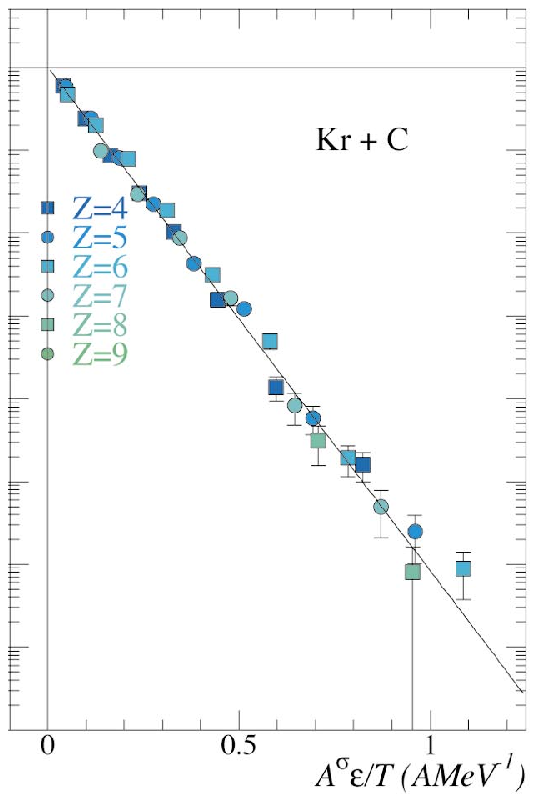} 
\includegraphics[scale=0.6]{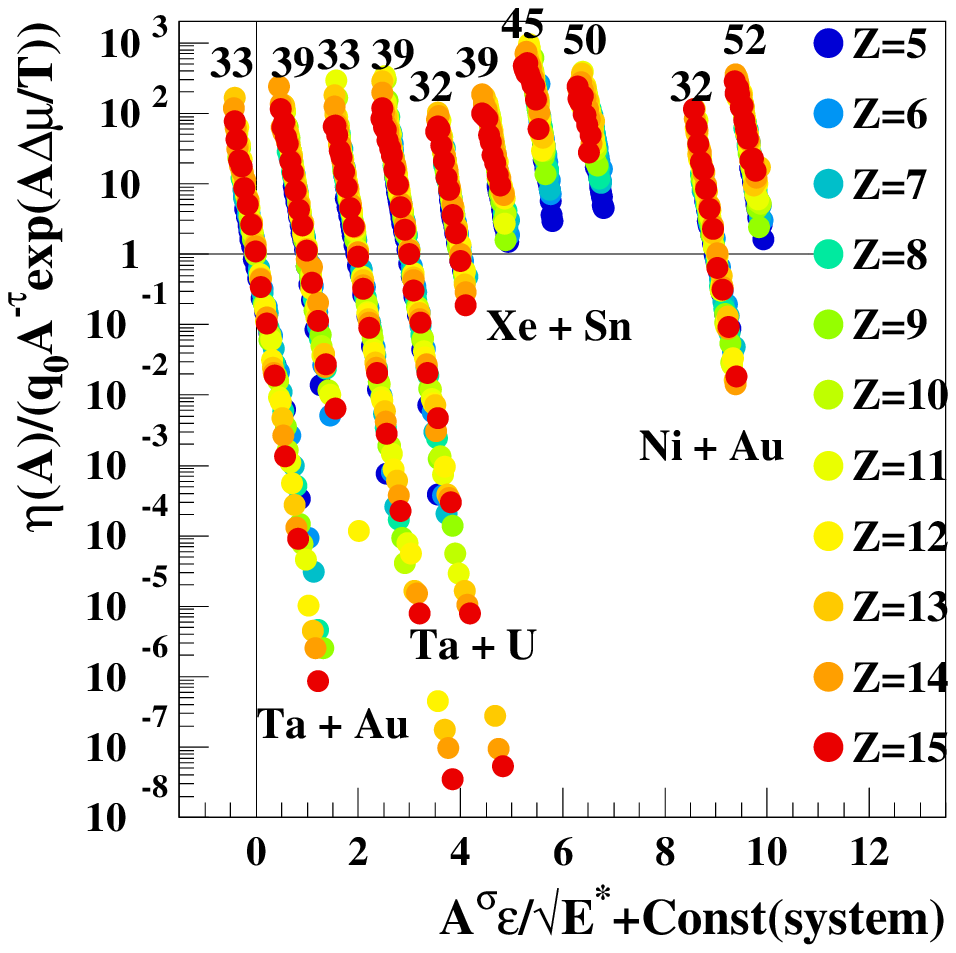} 
\caption{ 
Fragment mass yield distribution scaled by the power law prefactor, the
chemical potential and Coulomb (added to $A\Delta\mu$ for left and 
central panels) terms plotted against the inverse temperature scaled 
by Fisher's parametrization of the surface energy. Left and central 
panels: data correspond to projectile fragmentation
observed in reverse kinematics reactions at 1.0 GeV per nucleon,
from~\protect\cite{Ell03}. Right panel: Data correspond to  semi-peripheral
(Ta+X) or central (Xe+Sn, Ni+Au)
collisions. Bombarding energies in \AM{} are reported on top of
each scaled distribution; they are shifted by one x unit for a better view.
from~\protect\cite{I52-Riv05}.} \label{fisher}
\end{figure}
The temperature $T$ is determined by assuming a degenerate Fermi gas. 
This kind of scaling was found in many
multifragmentation data, in hadron-nucleus as well as in nucleus-nucleus
collisions~\cite{Ell02,Ell03,MDA03,I52-Riv05}; the agreement between data
and theory often holds over orders of
magnitude, and the critical exponents which are deduced are in acceptable
agreement with those expected for the liquid-gas universality class. 
As examples, fig.~\ref{fisher} shows the scaling properties observed
for different systems studied by the EOS and INDRA collaborations.

\subsubsection{Universal fluctuations of an order parameter} \label{unifluc}
The recently developed theory of universal scaling laws of order-parameter
fluctuations 
provides methods to select order parameters, characterize  critical
and off-critical behaviours and determine critical point
without any equilibrium assumption~\cite{Bot00,Bot02}. In this framework, 
universal $\Delta$ scaling laws of one of the order parameters, $m$, 
should be observed:\\ 
$ \langle m \rangle^{\Delta} P(m) = 
\phi ((m - \langle m \rangle )/ \langle m \rangle ^{\Delta}) $ \\
where $\langle m \rangle$ is the mean value of the distribution $P(m)$.
$\Delta$=1/2 corresponds to small fluctuations, 
$\sigma_m^2 \sim \langle m \rangle$, and
thus to an ordered phase. Conversely $\Delta$=1 occurs for the largest
fluctuations nature provides, $\sigma_m^2 \sim \langle m \rangle^2$, 
in a disordered phase. For models of cluster production having a critical 
behaviour there are two possible order parameters~\cite{Bot02}: the 
fragment multiplicity in a fragmentation process or the size of the 
largest fragment in an aggregation process (clusters are
built up from smaller constituents).
\begin{figure}[htp]
\begin{center}
\includegraphics[width=9cm]{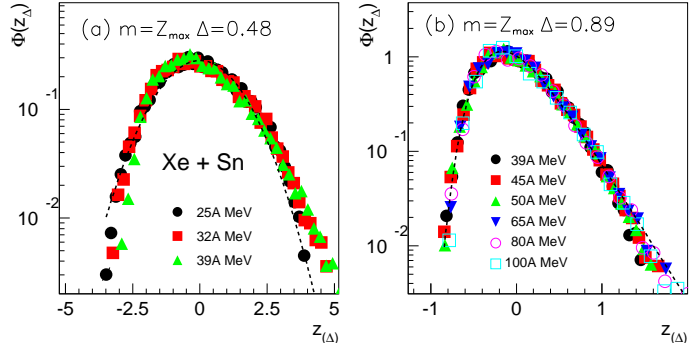}
\end{center}
\caption{ (a) $Z_{max}$ distributions for central Xe+Sn collisions at
25-39~\AM{} bombarding energies, scaled according to $\Delta$ scaling
equation; the dashed curve is a best fit to scaled data using a Gaussian
distribution. (b) As (a) but for bombarding energies 39-100~\AM{}:
the dashed curve is a best fit to scaled data using the Gumbel distribution.
 From~\protect\cite{I51-Fra05}.}
\label{flucuniv1}
\end{figure}
The method was applied to central collision samples (symmetric systems with
total masses $\sim$73-400 at bombarding energies between 25 and 100~\AM{}) 
by the INDRA collaboration~\cite{Bot01,I51-Fra05}.
The  CP multiplicity fluctuations do not show any evolution over the 
whole data set. Conversely the relationship between the
mean value and the fluctuation of the size of
the largest fragment does change as a function of the bombarding energy:
$\Delta \sim$1/2 at low energy, and $\Delta \sim$1 for higher bombarding
energies. The form of the $Z_{max}$ distributions also evolves with
bombarding energy: it is nearly Gaussian in the $\Delta$=1/2 regime and
exhibits for $\Delta$=1 an asymmetric form with a near-exponential tail for
large values of the scaling variable (see fig.~\ref{flucuniv1}). This
distribution is close to that of the modified Gumbel
distribution~\cite{Gum58}, the resemblance increasing with the total mass of
the system studied and being nearly perfect for the Au+Au data. The Gumbel
distribution is the equivalent of the Gaussian distribution in the case of
extreme values: it is obtained for an observable which is an extremum of a
large number of random, uncorrelated, microscopic variables.\\
\begin{figure}[htp]
\begin{center}
\includegraphics[width=9cm]{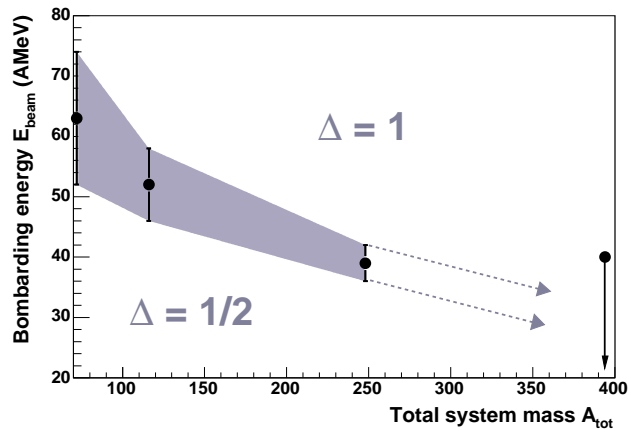}
\end{center}
\caption{ 
Dependence on bombarding energy and total system mass of the two different
regimes of $\Delta$-scaling observed for the size of the largest fragment in
each event, $Z_{max}$, for central collisions of symmetric systems studied
with INDRA. From~\protect\cite{I51-Fra05}.}
\label{flucuniv2}
\end{figure}
The dependence on bombarding energy and total system mass of the
frontier between the two $\Delta$-scalings is shown in 
fig.~\ref{flucuniv2}. Within the developed theory, this
behaviour indicates the transition from an ordered phase to a
disordered phase in the critical region, the fragments being produced 
following some aggregation scenario. Here again, as it was the case
for the Fisher's scaling, the two $\Delta$-scaling regimes
suggest the occurrence of a second-order phase transition. 

\subsubsection{Finite size effects and pseudo-critical behaviours inside 
the coexistence region}

As shown in this section, analyses of given samples of events have 
found at the same time
 scaling laws and evidence for first-order phase transition signals 
(negative microcanonical heat capacity and spinodal instabilities).
At the thermodynamic limit, scalings are general properties of
matter near the critical point and thus typical of second order phase
transition. For finite systems fluctuations correlate points at a distance
which can be comparable to the linear size of the system and can have 
an effect similar
to a diverging correlation length in an infinite system. Therefore the
so called critical behaviours are expected to be observed non only in the
critical zone,  but also deeply in the coexistence region where large 
fluctuations have been observed.
Within a lattice gas model in the canonical ensemble a scaling behaviour,
with critical parameters close to the ones expected for the 
liquid-gas universality class, was observed for finite systems
inside the coexistence region~\cite{Gul99}.
Moreover a good quality scaling of the cluster size distribution was
obtained using the Fisher's formula for the scaling function at 
subcritical as well as supercritical densities with values for the 
critical exponents compatible (within finite size effects) with the 
universality class of the model~\cite{Gul02}. This indicates that the 
observation of scaling does not allow to infer the position of the 
critical zone as it is also compatible with a fragmentation inside the 
coexistence region of a first order phase transition. Concerning the 
order parameter fluctuations, simulations have been performed
in the framework of the Ising model~\cite{Car02} which show that the
distribution of the heaviest fragment approximately obeys the $\Delta$=1 
scaling regime even at subcritical densities where no continuous 
transition takes place.
The observed behaviour was interpreted  as a finite size effect that
prevents the recognition of the order of a transition in a small system.
More recently the distribution of the heaviest fragment was analyzed within
the lattice gas model~\cite{Gul05} and it was shown that the most important
finite size effect comes from conservation laws, the distribution of the
order parameter being strongly deformed if a constraint is applied (mass
conservation) to an
observable that is closely correlated to the order parameter. Moreover the
observation of the $\Delta$=1 scaling regime was indeed observed in the
critical zone but was also confirmed at subcritical densities inside the
coexistence region. 

Very recently a generalization of Fisher's model to extend and
quantify finite size effects was also proposed~\cite{More05}. The Fisher's
formula along the coexistence line is
modified to the case of a vapor in equilibrium with a finite liquid drop
replacing the original formula by 
 $\mathrm{d} N / \mathrm{d} A = \eta(A) = 
 q_{0}[A(A_{d}-A)/A]^{-\tau}
\exp((-c_{0}(T)\varepsilon)/T[A^{\sigma}+(A_d-A)^{\sigma}-A_d^{\sigma}])$.\\
Such a generalization was tested within the canonical lattice gas model. 
Better results are obtained for d=2 (square lattice) than
for d=3 (cubic lattice). Applications to experimental data are needed to
evaluate that potential improvement. 
To conclude  that part one can say that, from both experiments and theory,
many different pseudo-critical behaviours can be observed inside the
coexistence region of a first-order phase transition as far as finite
systems are concerned.

\section{Coherent experimental signals of phase transition}\label{SignPT}

As shown in the previous sections, signals of phase transitions were
observed by many experimental groups for systems with different sizes 
and excitation energies. 
In most cases however only one signal was evidenced, which is not 
sufficient to convince that a phase transition does occur, because 
each of the proposed signals has some intrinsic weakness. 
 The quality of the exclusive measurements nowadays permits to search for 
several different signatures of the transition on a given sample of events; 
indeed the concomitant observation of several signals strongly reinforces
the hypothesis that a phase transition has occurred. Several groups are 
presently looking for different signals in their samples. 
For instance, and if we restrict to reactions below 100~\AM{}, the group 
of Bologna studied QPs from Au+Au collisions at
35~\AM{}, and after having discovered the negative branch of the heat 
capacity~\cite{MDA00}, they showed also Fisher scaling, critical 
exponents~\cite{MDA03} and more recently they examine the possible 
occurrence of a bimodal distribution (see below) of the charge of the 
largest fragment~\cite{BRU07}. The INDRA collaboration evidenced 
several signatures of phase transition for Au QPs, and for
central collisions between Ni and Ni, Ni and Au, Xe and Sn 
(see ref~\cite{I59-Bor04,I52-Riv05,Bor06}. A rather good coherence of 
the  different signatures was found in these cases.
 We chose to focus in this section on three sets of data for which the
most extended analyses were realized. They cover different system sizes
and collision centrality. The findings will only be
summarized, the reader is sent to the quoted references for details: \\
 1) light quasi-projectiles (A$\sim$36), from reactions between
47~\AM{} $^{40}$Ar and $^{27}$Al, $^{48}$Ti and $^{58}$Ni, by Ma
\emph{et al}~\cite{Ma05}; \\
 2)  heavy quasi-projectiles from Au on Au reactions, by the INDRA/ALADIN 
collaboration~\cite{I61-Pic06,T41Bon06,I63-NLN07}; \\
 3) fused systems of mass around 200 formed in central collisions 
between Xe and Sn,  by the INDRA 
collaboration~\cite{Bot01,I46-Bor02,I51-Fra05,I40-Tab03,I63-NLN07}.

Besides the signals mentioned in the previous section, Ma~\emph{et al} 
proposed some other ones as possible signatures of phase transition
or critical behaviour, and  firstly  the nuclear Zipf
law~\cite[and references therein]{MaYG99}. Zipf law, first 
introduced in linguistic to analyse the relative population of words 
in English texts, says that the frequency of a word is inversely
 proportional to its rank, n, in a frequency list (the integer rank n
 is equal to 1 for the most frequently used word). 
It was afterwards evidenced in many different scientific areas and thus 
it was suggested that the Zipf hierarchy is a fingerprint of criticality.
 In nuclear physics, investigations with a lattice gas model showed that 
 cluster distributions follow a Zipf law, where the frequency is replaced 
by the average size (charge or mass) of the clusters:
 $\langle Z_n \rangle \propto n^{-\lambda}$, where n = 1,2
{\ldots}  for the largest, second largest {\ldots}  fragment of 
each partition; $\lambda$ was found equal to 1 (the Zipf law is verified)
at the phase transition temperature.  Ma fitted the average values of the 
 charges of the multifragmentation clusters (starting at Z=1) as a
function of their rank with this type of law for different excitation
energies: $\lambda$ is found to decrease with increasing energy. The
Zipf law is verified ($\lambda=1$) for the energy bin  5-6~\AM{}. 
Consistently with the formulation of the Zipf law, percolation models
suggest that the ratio of the sizes of the second to the first largest 
fragments,
$S_p = \langle Z_{max2} \rangle / \langle Z_{max} \rangle$, 
takes the value 0.5 around the phase separation point~\cite{Cole02}. 
And indeed, in the
sample studied by Ma, $S_p$ exhibits essentially linear behaviours, with 
a change of slope around the energy of 5.2~\AM{}, where it takes the
value of 0.5.

As mentioned in section~\ref{FinitePT}, for finite systems, in the 
transition zone of a first order phase transition, the distribution of 
the order parameter presents a bimodal behaviour in the canonical 
ensemble, due to an anomalous convexity in the underlying 
microcanonical entropy. The transition energy, E$_{tr}^{bimodality}$,
corresponds to the value for which the two bumps of the bimodal 
distribution have the same height. In the case of multifragmentation, 
the size of the heaviest cluster, which is correlated with the 
total energy deposit and the system density/volume, appears as 
a natural potential order parameter (see also the universal scaling laws in
sect.~\ref{unifluc}). 
The bimodal character of the distribution may however be hidden if the
variable is constrained by a conservation law. Bimodality was searched for
by using different observables connected with the charge partition of the 
events; the chosen variable  in cases 1) and 3), where a limited
excitation energy range was covered, has the form: 
$P_b = (\sum_{Z_i \geq Z_{lim}} Z_i - \sum_{Z_i < Z_{lim}} Z_i) / \sum_{Z_i
\geq Z_{th}} $, with $Z_{lim}$ and $Z_{th}$ equal to 4 and 1 for case 1), 
and 13 and 3 for case 3). $P_b$ can be viewed as the difference between 
the liquid and the gas densities. In case 2) the distribution studied 
was firstly that
of the asymmetry between the two largest fragments~\cite{I61-Pic06},
$(Z_{max}-Z_{max2})/(Z_{max}+Z_{max2})$; in a
second step an elaborated analysis  was performed, which consisted in 
weighting  the $Z_{max}$ distribution by a probability given by 
the correlated excitation 
energy distribution on the total range scanned~\cite{T41Bon06,Gul07}. 

The signals studied for the three systems investigated, and the excitation
energy at which they occur are listed in table~\ref{tabPT}; these energies
are those coming from the calorimetry performed, and may thus include not
only thermal energy but also some collective component. Scalings,
fluctuations of the configurational energy and of the largest fragment
charge, 
sudden changes in the evolution of key variables, topological structure 
of the fragment partitions, charge correlations were examined. 

\begin{table}[!h] \centering
\caption{Excitation energies - in \AM{} - at which different signals were
observed for three systems (see text). The energy values were derived from
calorimetry. $Z_s$ is the charge of the considered source. 
E$_{crit}^{ Fisher}$ is the excitation energy corresponding to $T_c$ in
Fisher formula (see sect.~\ref{FinitePT} and see text for the other
definitions of variables). The threshold for radial 
expansion energy is reported for information (see sect.~\ref{frag})
}\label{tabPT}
\begin{tabular}{|l|c|c|c|}
\hline
variable        & QP $A_s \sim$36 & QP $Z_s\sim$68 &monosources $Z_s\sim$82 \\ \hline
E$_{crit}^{ Fisher}$    &   5-6    &   4.2     &  3.8-4.5         \\
$\Delta$ scaling        & 5-6      &  -        &   6.2        \\
max $A_s \sigma_k^2/T^2$ & 4-6.5    & 4-5       &   $\leq $4      \\
max $\sigma^2_{Z_{max}}/ \langle Z_{max} \rangle$ & 5-6 & - & - \\
c $<$ 0                 &   -      & $[$2.5:5.5$]$   &   $[-$:6.5$]$ \\
max $\sigma_{Z_{max}/Z_s}$&  -      & 4-5       &  $\leq$ 5   \\
max $\langle Z_{max2} \rangle$ &   6 & 5        & 4.5-6 \\
E$_{Zipf}$: $\lambda = 1$      &  5.6    & 8.5        & 7.5   \\
S$_p$ = 0.5             & 5.2     & 8.5 and above & 3.2 - 6  \\
change slope S$_p$      & 5.6     &  4         &  -     \\
E$_{tr}^{bimodality}$    & 5.6    &  $[$4.75:5.25$]$ & 7.8    \\
spinodal               &  -     &  $[$5:8$]$        & $[$5:9$]$  \\ \hline
threshold $\varepsilon _{rad}$ & -   & $\sim$5. & $\sim$4.5 \\ \hline
\end{tabular}
\end{table}
\begin{enumerate}
\item For the light $A\sim $36 system, in the Texas A\&M experiment,
both charged products and neutrons were
measured and used for centrality selections. 
A completeness criterion (Z$_{QP} \geq$12) was applied, and QP formed 
in \emph{central collisions} were analysed. The QP excitation energy, 
derived by calorimetry, ranges between 2 and 10~\AM{} with 
a mean value around 4.5~\AM{}. Such a distribution does clearly not 
reflect the expected impact parameter dependence, meaning that at low 
energy all the possible QP partitions are not explored.
In this context, all the variables characterizing a phase transition 
studied present 
a coherent behaviour, with a change in their evolution in the range 
5-6~\AM{}. Note however that, while a maximum of the configurational
energy fluctuations is observed, its value remains very low (0.3), far 
below the canonical expectation of 3/2. The authors give thus no firm
statement about the observation of a negative heat capacity. Their global 
conclusion is that they show a body of evidence  suggesting a liquid-gas 
phase change in an equilibrated system at, or extremely close to, 
the critical point.
\item For Au quasi-projectiles a completeness criterion is also required 
and calorimetry measurements lead to an excitation energy distribution 
decreasing from 2 to 11~\AM{}. For these heavier QP, with a charge 
$\sim$68, a similar consistency between all the signals of phase 
transition emerges. The maxima of fluctuations (fig.~\ref{fluctu_Au}) 
and E$_{crit}^{ Fisher}$ are in
the range 4-5~\AM{}, well inside the domain where the heat capacity is
negative. The changes in the fragment topology (inversion of variation of
$\langle Z_{max2} \rangle$, E$_{tr}^{bimodality}$ ) occur near the second
divergence of the heat capacity, located at $\sim$5~\AM{}, as does 
the maximum of the
extra equal-sized fragment events, possible witnesses
of a spinodal decomposition.
The values of E$_{crit}^{ Fisher}$ and the position of the
energies at which the heat capacity diverges  agree within 10\%
with those obtained from another sample of events obtained with
the Multics/Miniball device~\cite{MDA00,MDA03}.
\begin{figure} 
\begin{minipage}[t]{0.5\textwidth}
\includegraphics[width=\textwidth]{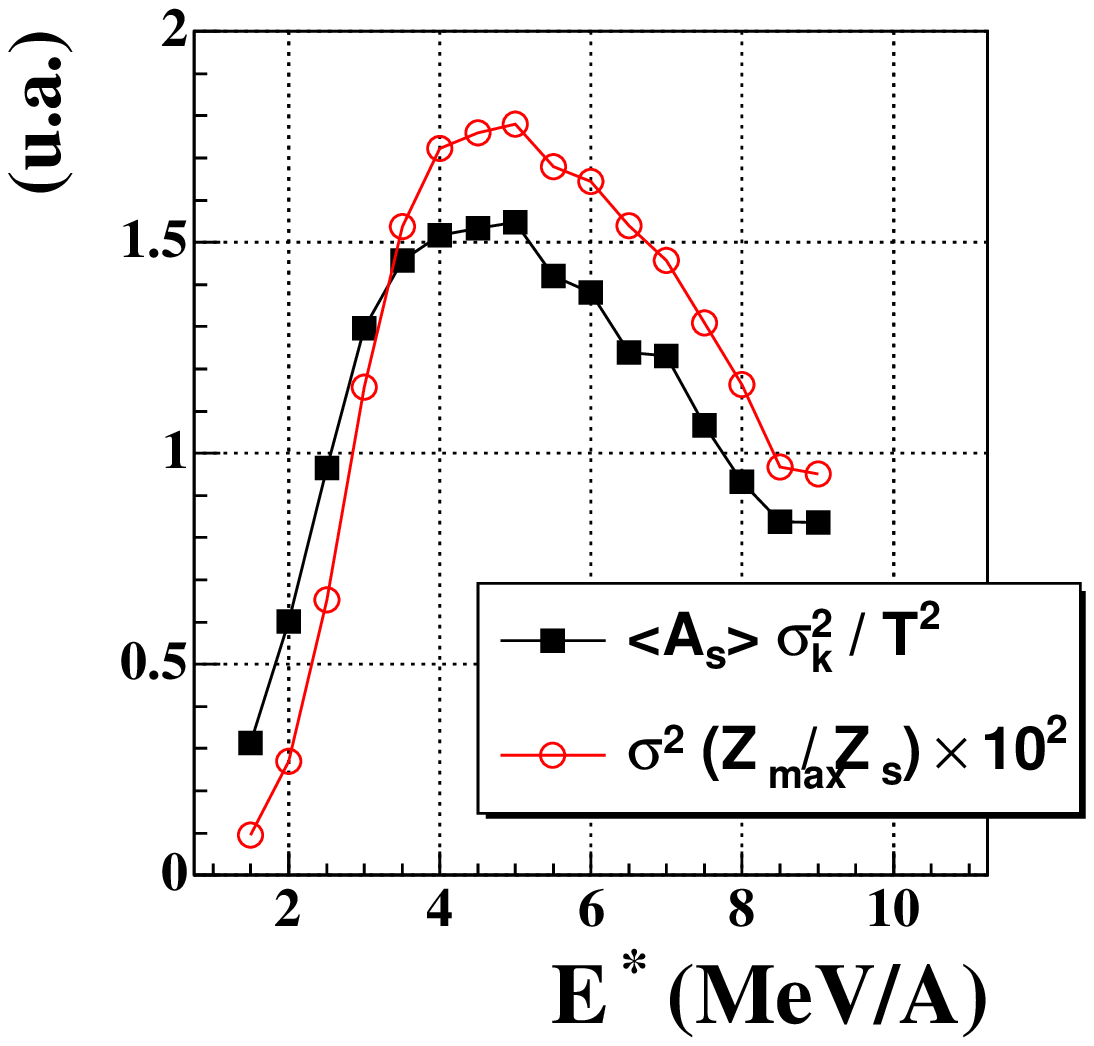}
\caption{Normalized fluctuations as a function of the excitation energy for
Au QP (Au + Au 80~\AM{}). Open circles correspond to fluctuations of the
largest fragment normalized by the source size $Z_1/Z_s$ and full squares
refer to normalized configurational energy fluctuations.
(from~\protect\cite{BonBorm07}).} \label{fluctu_Au}
\end{minipage}
\hspace*{0.02\textwidth}
\begin{minipage}[t]{0.48\textwidth}
\includegraphics[width=1.07\textwidth]{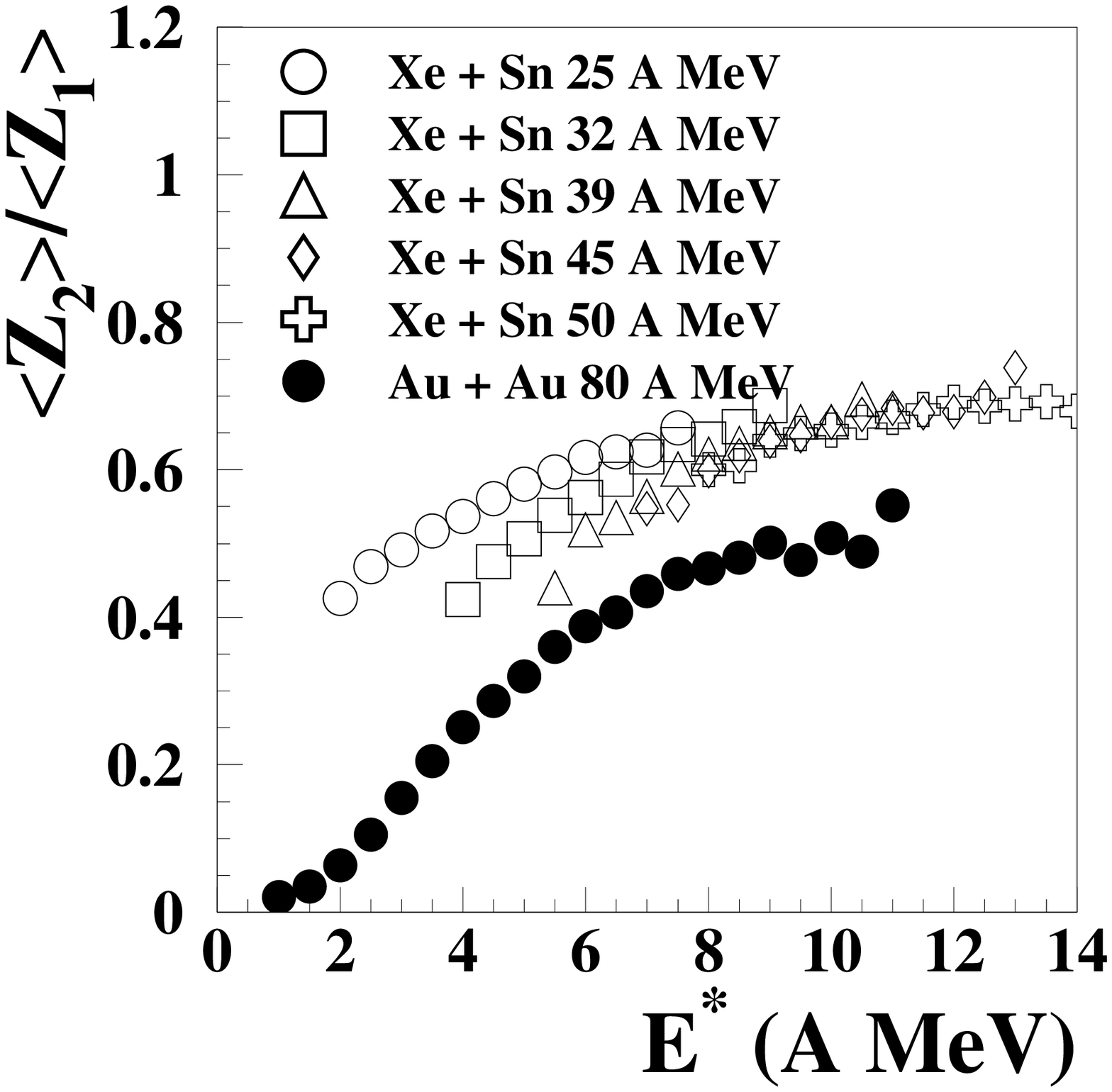}
\caption{Evolution of the variable $S_p$ versus the excitation energy for Au
quasi-projectiles and Xe+Sn compact sources from central collisions. data
are from the INDRA collaboration.}\label{Sp}
\end{minipage}%
\end{figure}
Conversely, while the variable $S_p$ changes slope in the same domain, it
reaches a value of 0.5 - where it saturates - at a much higher energy
(8.5~\AM{}, see fig.~\ref{Sp}); consistently the Zipf law is 
verified at the same high value (only fragments with Z$\geq$3 were 
taken into account for the heavy systems).
In that case, the observations plead in favor of the occurrence of a first
order phase transition of the liquid-gas type. The transition energy is
lower than that derived for the lighter QP; the Zipf law and related
signals can not be taken as a sign of transition.
\item What can be said for the monosources formed in central Xe+Sn
collisions  at 25, 32, 39, 45 and 50~\AM{}, which are slightly heavier 
than the QP of case 2)? They were isolated  by a
detected-charge completeness criterion, and by selecting events with 
a flow angle greater than 60$^o$ (compact event shapes in velocity 
space). The average excitation energies determined by calorimetry 
lie between $\sim$5 and $\sim$10~\AM{}.
In agreement with the previous case, the fluctuations of both the 
configurational energy and the largest fragment 
charge, and the average charge of the second largest 
fragment, present a maximum 
in the region 4-5~\AM{} where lies also E$_{crit}^{ Fisher}$. 
Only the second divergence of the heat capacity is seen in the explored 
energy range, at a slightly higher than for Au QP's. 
The change of $\Delta$ scaling and the
maximum of extra equal-sized fragment events occur near the second
divergence, while the transition energy indicated by the bimodality signal
is slightly higher (however the answer in that case is somewhat dependent on
the chosen value of $Z_{lim}$). Note that the $Z_{max}$ distributions are
never bimodal in monosources, due to the strong constraints on the energy.
Finally the Zipf law is also verified above the second
divergence, while the variable $S_p$ takes the value of 0.5 inside the
coexistence region, but at an excitation energy varying with the incident
energy (fig~\ref{Sp}). Let us underline that monosources with close 
mass and charge were
also formed with an asymmetric entrance channel, Ni+Au, and that 
negative heat capacities and signal of spinodal decomposition were 
observed in an energy range consistent with that quoted for 
Xe+Sn~\cite{Bor06}. \\
The consistency of all signals for central collisions is however not 
as good as in the Au QP case. It must be noted that for QP's radial 
expansion energy appears just above
the higher limit of the coexistence region. Conversely the threshold of
expansion lies inside the coexistence region for the monosources and may
have a different influence on the signals, particularly those
related to the fragment topology: indeed as explained in 
sect.~\ref{FinitePT},
radial expansion prevents coalescence between the nascent fragments,
favouring the observation of the spinodal signal.
\end{enumerate}
   In conclusion of this section, it may be said that a large body 
 of signals of
phase transition has been obtained in the three cases examined here. The
transition energy is higher when the system is lighter. The radial energy
seems to influence the energy at which some signals occur, leading to a
better overall consistency for QP's than for monosources formed in violent
collisions. Finally there is no indication that the Zipf law is a signal 
of phase transition or critical behaviour for the two
heavy systems studied by the INDRA collaboration. 
The excitation energy, E$_{Zipf}$, at which the Zipf law is 
verified was however shown to vary with the minimal value of the charge 
of the considered fragments: raising this value shifts E$_{Zipf}$ 
downwards~\cite{I63-NLN07}. One may thus suspect a connection between 
the system size and the range of charge to consider in order to establish 
the Zipf law.
While Ma \emph{et al.}
conclude that ``the transition occurs close to the critical point''
for their light system, for heavy systems the ensemble of results 
inclines towards the occurrence of a first order phase transition.

\section{Conclusions}

With the advent of powerful multidetectors dedicated to the study of
multifragmentation, which allow in particular the
detection of more complete events, important improvements
have been realized in the sorting of data and the
construction of global observables. New correlation methods were 
developed (higher order charge correlations, fragment-particle) to 
investigate fragment formation mechanism and primary fragment excitation 
energy. Yield scaling as well as various fluctuation properties were 
also studied in relation with phase transition.
Detailed and instructive comparisons with dynamical and
statistical models have been reported.\\
Fragment emission exhibits both dynamical and statistical aspects. For
example, heavy-ion collisions and hadron-nucleus collisions were used 
to clearly show that fragment radial expansion energy is higher for 
central heavy-ion collisions for which a compression-expansion phase is 
predicted to occur in dynamical simulations; deduced radial expansion 
energies from experiments are around 4-5~\AM{} at 
the maximum of fragment production. Fragment multiplicity appears as 
mainly governed by thermal energy whereas fragment partitions seem 
to be sensitive to radial expansion, which could be an indication of 
different trajectories in the phase diagram. Note that statistical
models predict the sensitivity of partitions to break-up density/volume.
Fragment-particle correlation studies strongly suggest
that primary fragments are, on average, in thermal equilibrium. 
That result is supported by both the success of statistical models in 
describing rather well average multifragmentation properties and the 
information derived from dynamical simulations concerning the different 
time scales involved along the collisions. A saturation of the average 
excitation energy of primary fragments around 3~\AM{} is also deduced. 
Moreover a few recent examples show a rather coherent link 
between dynamical and statistical descriptions.\\
What is the mechanism of fragment formation? It is still an open
question. In stochastic mean field approaches the
fragmentation process follows the spinodal decomposition scenario whereas
in molecular dynamics many-body correlations play a stronger role and
pre-fragments appear at earlier times.
New experiments with larger statistics are needed 
to eventually demonstrate through fragment charge correlations that
spinodal instabilities could be responsible of fragmentation. 
On the theoretical side, although large progresses have been
made, the development of quantal transport theory is still a challenge to
formally describe the basic quantal nature of hot nuclear systems and of
their fragmentation.\\
The specific role of the largest fragment in each event was also clearly 
revealed. Firstly,  at a given excitation energy, its charge distribution 
is independent of the total mass of the multifragmenting system 
when the latter exceeds A $\sim$~180-190; this behaviour is also
observed in statistical models down to A$\sim$~150. 
Then, its specific position close to the centre
of mass of the multifragmenting systems produced in central heavy-ion 
collisions, and more randomly positioned for $^3$He-nucleus collisions,
informs on event topology at freeze-out. Finally, its scaling properties 
and bimodal distributions can indicate that the largest fragment 
may be considered as a good candidate as order parameter of a phase 
transition in nuclei.  

Since the early eighties multifragmentation has been tentatively
associated with the liquid-gas phase transition theoretically predicted for
nuclear matter at sub-saturation densities and temperature below 16-18 MeV.
However, when dealing with the breaking of a nucleus, statistical 
mechanics of finite systems appeared as a key issue to progress and 
propose new first-order phase transition signatures related to
thermodynamic anomalies like negative microcanonical heat capacity and
bimodalities. Those anomalies are generic features in non-extensive systems
and have been reported in other fields such as atomic cluster physics and
self-gravitating systems.\\
A big effort to accumulate experimental indications of a phase
transition has been presently made, including scaling and fluctuations 
for fragment sizes. At the same time progresses have been obtained to 
experimentally better localize nuclear multifragmentation in the nuclear 
phase diagram thanks to improvements in calorimetry, thermometry and 
estimates of density at the break-up stage within working hypotheses 
and approximations. Today a rather coherent picture has been reached 
for few exhaustive studies and some semi-quantitative estimates of the 
transition region become available. However the order and the nature of 
the transition is still subject to debate. For a large part this debate 
is related to finite-size effects, which remain an important challenge.

How to progress further ? Experimentally new advances can be
made \emph{by identifying charge but also mass of fragments with 
the construction of a new generation of 4$\pi$ multidetectors}. 
With such new devices some hypotheses would be reduced or suppressed 
to evaluate freeze-out properties from intra-event correlations and 
it would become possible to disentangle radial expansion from Coulomb 
repulsion. On the theoretical side, 
by largely varying the proportions of neutrons and protons, a new 
phenomenon related to the phase transition for nuclei is predicted: 
the distillation, which makes the gas phase
more asymmetric than the liquid phase (even more asymmetric for spinodal
decomposition as compared to phase equilibrium). 
By measuring the isotopic composition of all the fragments it would become 
possible to identify the two phases and determine by a
new way the dynamics of the transition.



\begin{thebibliography}{100}
\expandafter\ifx\csname url\endcsname\relax
  \def\url#1{\texttt{#1}}\fi
\expandafter\ifx\csname urlprefix\endcsname\relax\def\urlprefix{URL }\fi
\providecommand{\eprint}[2][]{\url{#2}}

\bibitem{Boh36}
N.~Bohr, Nature 137 (1936) 351.

\bibitem{Sou06}
R.~T. {de Souza} et~al., P.~Chomaz et~al. (eds.) Dynamics and Thermodynamics
  with nuclear degrees of freedom, Springer, 2006, vol.~30 of \emph{Eur. Phys.
  J. A}, 275--291.

\bibitem{Fuc06}
C.~Fuchs et~al., P.~Chomaz et~al. (eds.) Dynamics and Thermodynamics with
  nuclear degrees of freedom, Springer, 2006, vol.~30 of \emph{Eur. Phys. J.
  A}, 5--22.

\bibitem{Ono06}
A.~Ono et~al., P.~Chomaz et~al. (eds.) Dynamics and Thermodynamics with nuclear
  degrees of freedom, Springer, 2006, vol.~30 of \emph{Eur. Phys. J. A},
  109--120.

\bibitem{Sie83}
P.~J. Siemens, Nature 305 (1983) 410.

\bibitem{Bug00}
K.~A. Bugaev et~al., Phys. Rev. C 62 (2000) 044320.

\bibitem{Bug01}
K.~A. Bugaev et~al., Phys. Lett. B 498 (2001) 144.

\bibitem{Jaq83}
H.~Jaquaman et~al., Phys. Rev. C 27 (1983) 2782.

\bibitem{Jaq84}
H.~R. Jaqaman et~al., Phys. Rev. C 29 (1984) 2067.

\bibitem{Cse86}
L.~P. Csernai et~al., Phys. Rep. 131 (1986) 223.

\bibitem{Mul95}
H.~Müller et~al., Phys. Rev. C 52 (1995) 2072.

\bibitem{Ber83}
G.~F. Bertsch et~al., Phys. Lett. B 126 (1983) 9.

\bibitem{Hei88}
H.~Heiselberg et~al., Phys. Rev. Lett. 61 (1988) 818.

\bibitem{Lope89}
J.~L\'opez et~al., Phys. Lett. B 219 (1989) 215.

\bibitem{I46-Bor02}
B.~Borderie, J. Phys. G: Nucl. Part. Phys. 28 (2002) R217.

\bibitem{De99}
J.~N. De et~al., Phys. Rev. C 59 (1999) 1.

\bibitem{Col00}
A.~J. Cole, Statistical Models for Nuclear Decay, Institute of Physics
  Publishing, Bristol, 2000.

\bibitem{Beau96}
L.~Beaulieu et~al., Phys. Rev. C 54 (1996) 973.

\bibitem{Sch96}
A.~Schuttauf et~al. ({ALADIN Collaboration}), Nucl. Phys. A 607 (1996) 457.

\bibitem{T25NLN99}
N.~{Le Neindre}, thèse de doctorat, Universit\'e de Caen (1999),
  http://tel.archives-ouvertes.fr/tel-00003741.

\bibitem{I37-Bel02}
N.~Bellaize et~al. (INDRA Collaboration), Nucl. Phys. A 709 (2002) 367.

\bibitem{Cha90}
R.~J. Charity et~al., Nucl. Phys. A 511 (1990) 59.

\bibitem{Bou89}
R.~Bougault et~al., Phys. Lett. B 232 (1989) 291.

\bibitem{Lou94}
M.~Louvel et~al., Phys. Lett. B 320 (1994) 221.

\bibitem{Lou95}
M.~Louvel et~al., preprint LPCC 95-01 (1995).

\bibitem{Gel95}
C.~J. Gelderloos et~al., Phys. Rev. Lett. 75 (1995) 3082.

\bibitem{He00}
Z.~Y. He et~al., Phys. Rev. C 63 (2000) 011601.

\bibitem{He01}
Z.~Y. He et~al., Phys. Rev. C 65 (2001) 014606.

\bibitem{Beau01}
L.~Beaulieu et~al., Phys. Rev. C 64 (2001) 064604.

\bibitem{Kar06}
V.~A. Karnaukhov, Phys. of Particles and Nuclei. 37 (2006) 165.

\bibitem{Tsa93}
M.~B. Tsang et~al., Phys. Rev. Lett. 71 (1993) 1502.

\bibitem{I2-Bac95}
C.~O. Bacri et~al. (INDRA Collaboration), Phys.\ Lett. B 353 (1995) 27.

\bibitem{I6-Riv96}
M.~F. Rivet et~al. (INDRA Collaboration), Phys.\ Lett. B 388 (1996) 219.

\bibitem{I7-Bor96}
B.~Borderie et~al. (INDRA Collaboration), Phys.\ Lett. B 388 (1996) 224.

\bibitem{Pie00}
L.~Pienkowski et~al., Phys. Lett. B 472 (2000) 15.

\bibitem{I15-Bor99}
B.~Borderie et~al. (INDRA Collaboration), Eur. Phys. J. A 6 (1999) 197.

\bibitem{Gul97}
F.~Gulminelli et~al., Nucl. Phys. A 615 (1997) 117.

\bibitem{I60-Lau06}
P.~Lautesse et~al. (INDRA Collaboration), Eur. Phys. J A 27 (2006) 349.

\bibitem{Gal05}
W.~Gawlikowicz et~al., Prog. Rep. DE/ER/40414-18 (2005),
  \eprint{http://nuchem.chem.rochester.edu/webreports.html}.

\bibitem{Bow91}
D.~R. Bowman et~al., Phys. Rev. C 67 (1991) 1527.

\bibitem{Tok01}
J.~Tõke et~al., Phys. Rev. C 63 (2001) 024604.

\bibitem{Ma05}
Y.~G. Ma et~al., Phys. Rev. C 71 (2005) 054606.

\bibitem{Kre93}
P.~Kreutz et~al., Nucl. Phys. A 556 (1993) 672.

\bibitem{I30-Dor01}
D.~Doré et~al. (INDRA Collaboration), Phys.\ Rev. C 63 (2001) 034612.

\bibitem{Gal07}
E.~Galichet et~al.  (2007), to be published.

\bibitem{Mor97}
L.~G. Moretto et~al., Phys. Rep 287 (1997) 249.

\bibitem{I10-Luk97}
J.~{\L}ukasik et~al. (INDRA Collaboration), Phys. Rev. C 55 (1997) 1906.

\bibitem{Jou98}
B.~Jouault et~al., Nucl. Phys. A 628 (1998) 119.

\bibitem{Yan03}
R.~Yanez et~al., Phys. Rev C 68 (2003) 011602.

\bibitem{Pia06}
S.~Piantelli et~al., Phys. Rev. C 74 (2006) 034609.

\bibitem{Bar02}
V.~Baran et~al., Nucl. Phys. A 703 (2002) 603.

\bibitem{Llo95}
W.~J. Llope et~al., Phys. Rev. C 51 (1995) 1325.

\bibitem{Colon98}
M.~Colonna et~al., Nucl. Phys. A 642 (1998) 449.

\bibitem{Buc84}
G.~Buchwald et~al., Phys. Rev. Lett. 52 (1984) 1594.

\bibitem{Sch93}
W.~Schmidt et~al., Phys. Rev. C 47 (1993) 2782.

\bibitem{Kam93}
B.~Kämpfer et~al., Phys. Rev. C 48 (1993) 955.

\bibitem{I18-Lec00}
J.~F. Lecolley et~al. (INDRA Collaboration), Nucl. Instr. and Meth. in Phys.
  Res. A 441 (2000) 517.

\bibitem{Des95}
P.~Désesquelles, Ann. Phys. Fr. 20 (1995) 1.

\bibitem{Des96}
P.~Désesquelles et~al., Phys. Rev. C 53 (1996) 2252.

\bibitem{Dav95}
C.~David et~al., Phys. Rev. C 51 (1995) 1453.

\bibitem{Bas96}
S.~A. Bass et~al., Phys. Rev. C 53 (1996) 2358.

\bibitem{Had97}
F.~Haddad et~al., Phys. Rev. C 55 (1997) 1371.

\bibitem{I22-Des00}
P.~Désesquelles et~al. (INDRA Collaboration), Phys. Rev. C 62 (2000) 024614.

\bibitem{I50-Lau05}
P.~Lautesse et~al. (INDRA Collaboration), Phys. Rev. C 71 (2005) 034602.

\bibitem{T39Mou04}
R.~Moustabchir, thèse de doctorat, Université Claude Bernard - Lyon I et
  Université Laval Québec (2004), http://tel.archives-ouvertes.fr/tel-00008654.

\bibitem{H4Lau05}
P.~Lautesse, Habilitation à diriger des recherches, Université Claude Bernard
  Lyon (2005), {http://tel.archives-ouvertes.fr/tel-00127989}.

\bibitem{Cav90}
C.~Cavata et~al., Phys. Rev. C 42 (1990) 1760.

\bibitem{Ogi91}
C.~A. Ogilvie et~al., Phys. Rev. Lett. 67 (1991) 1214.

\bibitem{Sis01}
D.~Sisan et~al., Phys. Rev. C 63 (2001) 027602.

\bibitem{Ins00}
A.~Insolia et~al. ({EOS Collaboration}), Phys. Rev. C 61 (2000) 044902.

\bibitem{Rei04}
W.~Reisdorf et~al. ({FOPI Collaboration}), Phys. Lett. B 595 (2004) 118.

\bibitem{I40-Tab03}
G.~T\u{a}b\u{a}caru et~al., Eur. Phys. J. A 18 (2003) 103.

\bibitem{Pea94}
G.~F. Peaslee et~al., Phys. Rev. C 49 (1994) 2271.

\bibitem{T41Bon06}
E.~Bonnet, thèse de doctorat, Université Paris-XI Orsay (2006),
  {http://tel.archives-ouvertes.fr/tel-00121736}.

\bibitem{Tam06}
B.~Tamain, P.~Chomaz et~al. (eds.) Dynamics and Thermodynamics with nuclear
  degrees of freedom, Springer, 2006, vol.~30 of \emph{Eur. Phys. J. A},
  71--79.

\bibitem{I12-Riv98}
M.~F. Rivet et~al. (INDRA Collaboration), Phys.\ Lett. B 430 (1998) 217.

\bibitem{T32Hud01}
S.~Hudan, thèse de doctorat, Université de Caen (2001), {GANIL T 01 07}.

\bibitem{I57-Tab05}
G.~T\u{a}b\u{a}caru et~al. (INDRA Collaboration), Nucl. Phys. A 764 (2006) 371.

\bibitem{Viol04}
V.~E. Viola et~al., Phys. Rev. lett. 93 (2004) 132701.

\bibitem{Rad05}
A.~R. Raduta et~al., Phys. Lett. B 623 (2005) 43.

\bibitem{Wad04}
R.~Wada et~al. ({NIMROD Collaboration}), Phys. Rev. C 69 (2004) 044610.

\bibitem{T29Lav01}
F.~Lavaud, thèse de doctorat, Université Louis Pasteur Strasbourg I (2001),
  http://tel.archives-ouvertes.fr/tel-00004100.

\bibitem{T30Bou01}
B.~Bouriquet, thèse de doctorat, Université de Caen (2001),
  http://tel.archives-ouvertes.fr/tel-00003803.

\bibitem{Pak96}
R.~Pak et~al., Phys. Rev. C 54 (1996) 1681.

\bibitem{Wil97}
C.~Williams et~al., Phys. Rev. C 55 (1997) 2132.

\bibitem{MDA96}
M.~D'Agostino et~al., Phys. Lett. B 371 (1996) 175.

\bibitem{Dur96}
D.~Durand et~al., Preprint LPCC 96-02 (1996).

\bibitem{BBro96}
B.~Borderie, G.~Giardina et~al. (eds.) Proc. Int. Symposium on large-scale
  collective motions of atomic nuclei, Brolo, Italy, World scientific, 1997, 1.

\bibitem{ReiR97}
W.~Reisdorf et~al., Ann. Rev. Nucl. Part. Sci. 47 (1997) 663.

\bibitem{T18ADN98}
{Anh-Dung Nguyen}, thèse de doctorat, Université de Caen (1998), {LPCC T
  98-02}.

\bibitem{I29-Fra01}
J.~D. Frankland et~al. (INDRA Collaboration), Nucl. Phys. A 689 (2001) 940.

\bibitem{I47-Lef04}
A.~{Le Fèvre} et~al. (INDRA and ALADIN collaborations), Nucl. Phys. A 735
  (2004) 219.

\bibitem{Das04}
C.~B. Das et~al., Phys. Rev. C 70 (2004) 064610.

\bibitem{Gul04}
F.~Gulminelli et~al., Nucl. Phys. A 734 (2004) 581.

\bibitem{T16Sal97}
S.~Salou, thèse de doctorat, Université de Caen (1997),
  http://tel.archives-ouvertes.fr/tel-00003688.

\bibitem{Jeo96}
S.~C. Jeong et~al., Nucl. Phys. A 604 (1996) 208.

\bibitem{MDA99}
M.~D'Agostino et~al., Nucl. Phys. A 650 (1999) 329.

\bibitem{Bea00}
L.~Beaulieu et~al. ({ISIS Collaboration}), Phys. Rev. Lett. 84 (2000) 5971.

\bibitem{T12LeFe97}
A.~{Le Fèvre}, thèse de doctorat, Université Paris 7 - Denis Diderot (1997),
  {GANIL T 97 03}.

\bibitem{Des98}
P.~Désesquelles et~al. (Multics/Miniball Collaboration), Nucl. Phys. A 633
  (1998) 547.

\bibitem{I9-Mar97}
N.~Marie et~al. (INDRA Collaboration), Phys. Lett. B 391 (1997) 15.

\bibitem{Mek78}
A.~Z. Mekjian, Phys. Rev. C 17 (1978) 1051.

\bibitem{Fai83}
G.~Fái et~al., Nucl. Phys. A 404 (1983) 551.

\bibitem{Koo87}
S.~E. Koonin et~al., Nucl. Phys. A 474 (1987) 173.

\bibitem{Gro90}
D.~H.~E. Gross, Rep. Prog. Phys. 53 (1990) 605.

\bibitem{Lee92}
S.~J. Lee et~al., Phys. Rev. C 45 (1992) 1284.

\bibitem{Hah88}
D.~Hahn et~al., Phys. Rev. C 37 (1988) 1048.

\bibitem{Hahn88}
D.~Hahn et~al., Nucl. Phys. A 476 (1988) 718.

\bibitem{Kon94}
J.~Konopka et~al., Phys. Rev. C 50 (1994) 2085.

\bibitem{Bon95}
J.~Bondorf et~al., Phys. Rep. 257 (1995) 133.

\bibitem{Rad97}
A.~H. Raduta et~al., Phys. Rev. C 55 (1997) 1344.

\bibitem{Radu00}
A.~H. Raduta et~al., Phys. Rev. C 61 (2000) 034611.

\bibitem{Pei89}
G.~Peilert et~al., Phys. Rev. C 39 (1989) 1402.

\bibitem{Aic91}
J.~Aichelin, Phys. Rep. 202 (1991) 233.

\bibitem{Luk93}
J.~{\L}ukasik et~al., Acta Phys. Polonica B 24 (1993) 1959.

\bibitem{Ono93}
A.~Ono et~al., Phys. Rev. C 47 (1993) 2652.

\bibitem{Fel90}
H.~Feldmeier, Nucl. Phys. A 515 (1990) 147.

\bibitem{Ono96}
A.~Ono et~al., Phys. Rev. C 53 (1996) 2598.

\bibitem{Sug99}
Y.~Sugawa et~al., Phys. Rev. C 60 (1999) 064607.

\bibitem{I16-Neb99}
R.~Nebauer et~al. (INDRA Collaboration), Nucl. Phys. A 658 (1999) 67.

\bibitem{Pra95}
S.~Pratt et~al., Phys. Lett. B 349 (1995) 261.

\bibitem{Ayi88}
S.~Ayik et~al., Phys. Lett. B 212 (1988) 269.

\bibitem{Ayi90}
S.~Ayik et~al., Nucl. Phys. A 513 (1990) 187.

\bibitem{Ran90}
J.~Randrup et~al., Nucl. Phys. A 514 (1990) 339.

\bibitem{Cho91}
P.~Chomaz et~al., Phys. Lett. B 254 (1991) 340.

\bibitem{Rei92}
P.~Reinhard et~al., Ann. Phys. 213 (1992) 204.

\bibitem{Rein92}
P.~Reinhard et~al., Ann. Phys. 216 (1992) 98.

\bibitem{Reinh92}
P.~Reinhard et~al., Nucl. Phys. A 545 (1992) 59c.

\bibitem{Cho94}
P.~Chomaz et~al., Phys. Rev. Lett. 73 (1994) 3512.

\bibitem{Gua96}
A.~Guarnera et~al., Phys. Lett. B 373 (1996) 267.

\bibitem{Gua97}
A.~Guarnera et~al., Phys. Lett. B 403 (1997) 191.

\bibitem{Cho96}
P.~Chomaz, Ann. Phys. Fr. 21 (1996) 669.

\bibitem{Mat00}
F.~Matera et~al., Phys. Rev. C 62 (2000) 044611.

\bibitem{Lee97}
S.~J. Lee et~al., Phys. Rev C 56 (1997) 2621.

\bibitem{Par00}
A.~S. Parvan et~al., Nucl. Phys. A 676 (2000) 409.

\bibitem{Das01}
C.~B. Das et~al., Phys. Rev. C 64 (2001) 017601.

\bibitem{Cho00}
P.~Chomaz et~al., Phys. Rev. Lett. 85 (2000) 3587.

\bibitem{Fri90}
W.~A. Friedman, Phys. Rev. C 42 (1990) 667.

\bibitem{Bow95}
D.~R. Bowman et~al., Phys. Rev. C 52 (1995) 818.

\bibitem{Char88}
R.~J. Charity et~al., Nucl. Phys. A 483 (1988) 371.

\bibitem{Sch01}
R.~P. Scharenberg et~al. ({EOS Collaboration}), Phys. Rev. C 64 (2001) 054602.

\bibitem{Sri02}
B.~K. Srivastava et~al. ((EOS Collaboration)), Phys. Rev. C 65 (2002) 054617.

\bibitem{I62-Zbi07}
K.~Zbiri et~al. (ALADIN and INDRA collaborations), Phys. Rev. C 75 (2007)
  034612.

\bibitem{Kru85}
H.~Kruse et~al., Phys. Rev. C 31 (1985) 1770.

\bibitem{Gre87}
C.~Grégoire et~al., Nucl. Phys. A 465 (1987) 317.

\bibitem{Ber88}
G.~F. Bertsch et~al., Phys. Rep. 160 (1988) 189.

\bibitem{Bon94}
A.~Bonasera et~al., Phys. Rep. 243 (1994) 1.

\bibitem{Str99}
A.~Strachan et~al., Phys. Rev. C 59 (1999) 285.

\bibitem{Cus02}
D.~Cussol, Phys. Rev. C 65 (2002) 054614.

\bibitem{Che04}
A.~Chernomoretz et~al., Phys. Rev. C 69 (2004) 034610.

\bibitem{Mul93}
W.~Müller et~al., Phys. Lett. B 298 (1993) 27.

\bibitem{Ono99}
A.~Ono, Phys. Rev. C 59 (1999) 853.

\bibitem{Ayi94}
S.~Ayik et~al., Phys. Rev. C 50 (1994) 2947.

\bibitem{T28Tab00}
G.~T\u{a}b\u{a}caru, thèse de doctorat, Université Paris-XI Orsay (2000),
  http://tel.archives-ouvertes.fr/tel-00007912.

\bibitem{Rad06}
A.~H. Raduta et~al., Phys. Rev. C 74 (2006) 034604.

\bibitem{Lle93}
A.~Lleres et~al., Phys. Rev. C 48 (1993) 2753.

\bibitem{Pla01}
R.~Planeta et~al., Eur. Phys. J. A 11 (2001) 297.

\bibitem{I61-Pic06}
M.~Pichon et~al. (INDRA and ALADIN collaborations), Nucl. Phys. A 779 (2006)
  267.

\bibitem{H5Vie06}
E.~Vient, Habilitation à diriger des recherches, Université de Caen (2006),
  {http://tel.archives-ouvertes.fr/tel-00141924}.

\bibitem{Hau96}
J.~A. Hauger et~al. (EOS Collaboration), Phys. Rev. Lett. 77 (1996) 235.

\bibitem{Vio06}
V.~E. Viola et~al., Phys. Rep. 434 (2006) 1.

\bibitem{Jan05}
M.~Jandel et~al., J. Phys. G: Nuclear and Particle Physics 31 (2005) 29.

\bibitem{Hud05}
S.~Hudan et~al., Phys. Rev. C 71 (2005) 054604.

\bibitem{Sum00}
K.~Sümmerer et~al., Phys. Rev. C 61 (2000) 034607.

\bibitem{Cha98}
R.~J. Charity, Phys. Rev. C 58 (1998) 1073.

\bibitem{Lef01}
T.~Lefort et~al., Phys. Rev. C 64 (2001) 064603.

\bibitem{I27-Ste01}
J.~C. Steckmeyer et~al. (INDRA Collaboration), Nucl. Phys. A 686 (2001) 537.

\bibitem{DasG01}
S.~{Das Gupta} et~al., Adv. Nucl. Phys. 26 (2001) 91.

\bibitem{Kel06}
A.~{Keli\'c} et~al., P.~Chomaz et~al. (eds.) Dynamics and Thermodynamics with
  nuclear degrees of freedom, Springer, 2006, vol.~30 of \emph{Eur. Phys. J.
  A}, 203--213.

\bibitem{Gon90}
M.~Gonin et~al., Phys. Rev. C 42 (1990) 2125.

\bibitem{Ort06}
R.~Ortega et~al., Eur. Phys. J. A 28 (2006) 161.

\bibitem{Alb85}
S.~Albergo et~al., Nuovo Cimento 89 A (1985) 1.

\bibitem{Radu99}
A.~H. Raduta et~al., Phys. Rev. C 59 (1999) 1855.

\bibitem{Tsa97}
M.~B. Tsang et~al., Phys. Rev. Lett. 78 (1997) 3836.

\bibitem{Rad00}
A.~H. Raduta et~al., Nucl. Phys. A 671 (2000) 600.

\bibitem{Sch02}
K.~H. Schmidt et~al., Nucl. Phys. A 710 (2002) 157.

\bibitem{Nato02}
J.~B. Natowitz et~al., Phys. Rev. C 65 (2002) 034618.

\bibitem{Poc95}
J.~Pochodzalla et~al. ({ALADIN Collaboration}), Phys. Rev. Lett. 75 (1995)
  1040.

\bibitem{Bon84}
P.~Bonche et~al., Nucl. Phys. A 427 (1984) 278.

\bibitem{Bon86}
P.~Bonche et~al., Nucl. Phys. A 436 (1986) 265.

\bibitem{Ent02}
D.~G. {d'Enterria} et~al., Phys. Lett. B 538 (2002) 27.

\bibitem{Par05}
M.~Pârlog et~al., Eur. Phys. J. A 25 (2005) 223.

\bibitem{KimY92}
Y.~D. Kim et~al., Phys. Rev. C 45 (1992) 387.

\bibitem{Zaj84}
W.~A. Zajc et~al., Phys. Rev. C 29 (1984) 2173.

\bibitem{Lis91}
M.~A. Lisa et~al., Phys. Rev. C 44 (1991) 2865.

\bibitem{Fox93}
D.~Fox et~al., Phys. Rev. C 47 (1993) 421.

\bibitem{Bau93}
E.~Bauge et~al., Phys. Rev. Lett. 70 (1993) 3705.

\bibitem{Bow93}
D.~R. Bowman et~al., Phys. Rev. Lett. 70 (1993) 3534.

\bibitem{Zhi00}
{Zhi-Yong He} et~al., Phys. Rev. C 63 (2000) 011601.

\bibitem{San95}
T.~C. Sangster et~al., Phys. Rev. C 51 (1995) 1280.

\bibitem{Wan99}
G.~Wang et~al., Phys. Rev. C 60 (1999) 014603.

\bibitem{I11-Mar98}
N.~Marie et~al. (INDRA Collaboration), Phys. Rev. C 58 (1998) 256.

\bibitem{I39-Hud03}
S.~Hudan et~al. (INDRA Collaboration), Phys. Rev. C 67 (2003) 064613.

\bibitem{Sta01}
P.~Staszel et~al., Phys. Rev. C 63 (2001) 064610.

\bibitem{Frit99}
S.~Fritz et~al., Phys. Lett. B 461 (1999) 315.

\bibitem{Natow02}
J.~B. Natowitz et~al., Phys. Rev. C 66 (2002) 031601.

\bibitem{I28-Fra01}
J.~D. Frankland et~al. (INDRA Collaboration), Nucl.\ Phys. A 689 (2001) 905.

\bibitem{I58-Pia05}
S.~Piantelli et~al. (INDRA Collaboration), Phys. Lett. B 627 (2005) 18.

\bibitem{Radu05}
A.~R. Raduta et~al., Phys. Rev. C 72 (2005) 057603.

\bibitem{Gro01}
D.~H.~E. Gross, Microcanonical Thermodynamics - Phase Transitions in ``small''
  systems, World Scientific, Singapore, 2001.

\bibitem{Gro95}
D.~H.~E. Gross et~al., Z. Phys. D 35 (1995) 27.

\bibitem{Gro97}
D.~H.~E. Gross, Phys. Rep. 279 (1997) 119.

\bibitem{Gul99}
F.~Gulminelli et~al., Phys. Rev. Lett. 82 (1999) 1402.

\bibitem{Cho99}
P.~Chomaz et~al., Nucl. Phys. A 647 (1999) 153.

\bibitem{Gro00}
D.~H.~E. Gross et~al., Eur. Phys. J. B 15 (2000) 115.

\bibitem{Bor00}
P.~Borrmann et~al., Phys. Rev. Lett. 84 (2000) 3511.

\bibitem{Bot00}
R.~Botet et~al., Phys. Rev. E 62 (2000) 1825.

\bibitem{Jel00}
J.~Jellinek et~al., J. Chem. Phys. 113 (2000) 2570.

\bibitem{Mul01}
O.~Mülken et~al., Phys. Rev. C 63 (2001) 024306.

\bibitem{Cam01}
X.~Campi et~al., Nucl. Phys. A 681 (2001) 458.

\bibitem{Cho01}
P.~Chomaz et~al., Phys. Rev. E 64 (2001) 046114.

\bibitem{Car01}
J.~M. Carmona et~al., Eur. Phys. J A 11 (2001) 87.

\bibitem{Radut01}
A.~H. Raduta et~al., Phys. Rev. Lett. 87 (2001) 202701.

\bibitem{Radut02}
A.~H. Raduta et~al., Nucl. Phys. A 703 (2002) 876.

\bibitem{Mor02}
L.~G. Moretto et~al., Phys. Rev. C 66 (2002) 041601.

\bibitem{Gulm04}
F.~Gulminelli et~al., Phys. Rev. C 69 (2004) 069801.

\bibitem{Rad03}
A.~H. Raduta et~al., Nucl. Phys. A 724 (2003) 233.

\bibitem{Cho03}
P.~Chomaz et~al., Physica A 330 (2003) 451.

\bibitem{Gul03}
F.~Gulminelli, Ann. Phys. Fr. 29 (2004) N$^{o}$ 6.

\bibitem{Cho04}
P.~Chomaz et~al., Phys. Rep. 389 (2004) 263.

\bibitem{Mor04}
L.~G. Moretto et~al., Prog. Part. Nucl. Phys. 53 (2004) 101.

\bibitem{Das05}
C.~B. Das et~al., Phys. Rep. 406 (2005) 1.

\bibitem{Cam05}
X.~Campi et~al., Phys. Rev. C 71 (2005) 041601.

\bibitem{Gulm05}
F.~Gulminelli et~al., Phys. Rev. C 72 (2005) 064618.

\bibitem{Gul05}
F.~Gulminelli et~al., Phys. Rev. C 71 (2005) 054607.

\bibitem{More05}
L.~G. Moretto et~al., Phys. Rev. Lett. 94 (2005) 202701.

\bibitem{De06}
J.~N. De et~al., Phys. Rev. C 73 (2006) 034602.

\bibitem{Colo97}
M.~Colonna et~al., Nucl. Phys. A 613 (1997) 165.

\bibitem{Jac96}
B.~Jacquot et~al., Phys. Lett. B 383 (1996) 247.

\bibitem{Ayi95}
S.~Ayik et~al., Phys. Lett. B 353 (1995) 417.

\bibitem{Idi94}
D.~Idier et~al., Ann.\ Phys.\ Fr. 19 (1994) 159.

\bibitem{Nor00}
W.~Nörenberg et~al., Eur. Phys. J. A 9 (2000) 327.

\bibitem{JacT96}
B.~Jacquot, thèse de doctorat, Université de Caen (1996), gANIL T 96 05.

\bibitem{Bru92}
M.~Bruno et~al., Phys. Lett. B 292 (1992) 251.

\bibitem{Mor96}
L.~G. Moretto et~al., Phys. Rev. Lett. 77 (1996) 2634.

\bibitem{I41-Cha03}
J.~L. Charvet et~al., Nucl. Phys. A 730 (2004) 431.

\bibitem{Bor06}
B.~Borderie et~al., P.~Chomaz et~al. (eds.) Dynamics and Thermodynamics with
  nuclear degrees of freedom, Springer, 2006, vol.~30 of \emph{Eur. Phys. J.
  A}, 243--251.

\bibitem{I31-Bor01}
B.~Borderie et~al. (INDRA Collaboration), Phys.\ Rev.\ Lett. 86 (2001) 3252.

\bibitem{Col02}
M.~Colonna et~al., Phys. Rev. Lett. 88 (2002) 122701.

\bibitem{Nor02}
W.~Nörenberg et~al., Eur. Phys. J. A 14 (2002) 43.

\bibitem{Cha88}
M.~Challa et~al., Phys. Rev. Lett. 60 (1988) 77.

\bibitem{Gul07}
F.~Gulminelli, Nucl. Phys. A 791 (2007) 165.

\bibitem{Gul00}
F.~Gulminelli et~al., Europhys. Lett. 50 (2000) 434.

\bibitem{MDA02}
M.~D'Agostino et~al., Nucl. Phys. A 699 (2002) 795.

\bibitem{MDA00}
M.~D'Agostino et~al., Phys. Lett. B 473 (2000) 219.

\bibitem{NLNBorm00}
N.~{Le Neindre} et~al. ({INDRA collaboration}), I.~Iori et~al. (eds.) Proc.
  XXXVIII Int. Winter Meeting on Nuclear Physics, Bormio, Italy, Ricerca
  scientifica ed educazione permanente, 2000, 404.

\bibitem{T34Gui02}
B.~Guiot, thèse de doctorat, Université de Caen (2002),
  http://tel.archives-ouvertes.fr/tel-00003753.

\bibitem{I59-Bor04}
B.~Borderie et~al., Nucl. Phys. A 734 (2004) 495.

\bibitem{MDA04}
M.~D'Agostino et~al., Nucl. Phys. 734 (2004) 512.

\bibitem{Gul06}
F.~Gulminelli et~al., P.~Chomaz et~al. (eds.) Dynamics and Thermodynamics with
  nuclear degrees of freedom, Springer, 2006, vol.~30 of \emph{Eur. Phys. J.
  A}, 253--262.

\bibitem{Sat03}
N.~Sator, Phys. Rep. 376 (2003) 1.

\bibitem{Fin82}
J.~E. Finn et~al., Phys. Rev. Lett. 49 (1982) 1321.

\bibitem{Cur83}
M.~W. Curtin et~al., Phys. Lett. 123 (1983) 289.

\bibitem{Hir84}
A.~S. Hirsch et~al., Phys. Rev. C 29 (1984) 508.

\bibitem{Cam84}
X.~Campi et~al., Phys. Lett. B 138 (1984) 353.

\bibitem{Sta94}
D.~Stauffer et~al., Introduction to Percolation Theory, Taylor \& Francis
  Publishers, London, 1994.

\bibitem{Cam00}
X.~Campi et~al., Eur. Phys. J. D 11 (2000) 233.

\bibitem{Kle02}
M.~{Kleine Berkenbusch} et~al., Phys. Rev. Lett. 88 (2002) 022701.

\bibitem{Fis67}
M.~E. Fisher, Physics 3 (1967) 255.

\bibitem{Ell02}
J.~B. Elliott et~al., Phys. Rev. Lett. 88 (2002) 042701.

\bibitem{Ell03}
J.~B. Elliott et~al., Phys. Rev. C 67 (2003) 024609.

\bibitem{Bot02}
R.~Botet et~al., Universal fluctuations, World Scientific, 2002, vol.~65 of
  \emph{World scientific Lecture Notes in Physics}.

\bibitem{Bot01}
R.~Botet et~al., Phys. Rev. Lett. 86 (2001) 3514.

\bibitem{I51-Fra05}
J.~D. Frankland et~al. (INDRA and ALADIN collaborations), Phys. Rev. C 71
  (2005) 034607.

\bibitem{Ell98}
J.~B. Elliott et~al. ({EOS Collaboration}), Phys. Lett. B 418 (1998) 34.

\bibitem{MDA03}
M.~D'Agostino et~al., Nucl. Phys. 724 (2003) 455.

\bibitem{I52-Riv05}
M.~F. Rivet et~al. (INDRA and ALADIN collaborations), Nucl. Phys. A 749 (2005)
  73.

\bibitem{Gul02}
F.~Gulminelli et~al., Phys. Rev. C 65 (2002) 051601R.

\bibitem{Das02}
C.~B. Das et~al., Phys. Rev. C 66 (2002) 044602.

\bibitem{Ell05}
J.~B. Elliot et~al., Phys. Rev. C 71 (2005) 024607.

\bibitem{Mor05}
L.~G. Moretto et~al., Phys. Rev. C 72 (2005) 064605.

\bibitem{Gum58}
E.~J. Gumbel, Statistics of extremes, Columbia University Press, 1958.

\bibitem{Car02}
J.~M. Carmona et~al., Phys. Lett. B 531 (2002) 71.

\bibitem{BRU07}
M.~Bruno et~al., Eur. Phys. J A (2007) submitted, \eprint{nucl-ex/0612030}.

\bibitem{I63-NLN07}
N.~{Le Neindre} et~al. (INDRA and ALADIN collaborations), Nucl. Phys. A 795
  (2007) 44.

\bibitem{MaYG99}
Y.~G. Ma, Eur. Phys. J. A 6 (1999) 367.

\bibitem{Cole02}
A.~J. Cole, Phys. Rev. C 65 (2002) 031601.

\bibitem{BonBorm07}
E.~Bonnet et~al. ({INDRA and ALADIN collaborations}), I.~Iori et~al. (eds.)
  Proc. XLV Int. Winter Meeting on Nuclear Physics, Bormio, Italy, Ricerca
  scientifica ed educazione permanente, 2007,
  \eprint{hal.archives-ouvertes.fr/hal-00141043}.

\end{thebibliography}

\end{document}